\documentclass[journal]{IEEEtran}
\usepackage{color}
\usepackage{amsmath,amsfonts}
\usepackage{algorithmic}
\usepackage{algorithm}
\usepackage{subfigure}
\usepackage{textcomp}
\usepackage{multirow}
\usepackage{colortbl}
\usepackage{makecell}
\usepackage{verbatim}
\usepackage{stfloats}
\usepackage{subfloat}
\usepackage{url}
\usepackage{braket}
\usepackage{booktabs}
\usepackage{verbatim}
\usepackage{graphicx}
\usepackage{physics}
\usepackage{cite}
\usepackage{CJK}
\usepackage{siunitx}
\usepackage{array}
\graphicspath{{figures/}{photos/}}

\begin{document}
\title{Entanglement-Assisted Quantum Networks: Mechanics, Enabling Technologies, Challenges, and Research Directions}	
	
\author{Zhonghui Li, \IEEEmembership{Member,~IEEE}, Kaiping Xue, \IEEEmembership{Senior Member,~IEEE}, Jian Li, \IEEEmembership{Member,~IEEE}, \\ Lutong Chen, \IEEEmembership{Graduate Student Member,~IEEE}, Ruidong Li, \IEEEmembership{Senior Member,~IEEE}, Zhaoying Wang, \\ Nenghai Yu, David S.L. Wei,~\IEEEmembership{Life Senior Member,~IEEE}, Qibin Sun, \IEEEmembership{Fellow, IEEE}, Jun Lu
		
\thanks{Z. Li, K. Xue, J. Li, L. Chen, Z. Wang, N. Yu and Q. Sun are with the School of Cyber Science and Technology, University of Science and Technology of China, Hefei 230027, China.}	
\thanks{R. Li is with the College of Science and Engineering, Kanazawa University, Kakuma, Kanazawa 920-1192, Japan.}		
\thanks{J. Lu is also with the Department of Electronic Engineering and Information Science, University of Science and Technology of China, Hefei 230027 China.}
\thanks{D. Wei is with the Department of Computer and Information Science, Fordham University, Bronx, NY 10458, USA.}
\thanks{Corresponding author:K. Xue~(email: kpxue@ustc.edu.cn).}
}			
\maketitle
		
\begin{abstract}
Over the past few decades, significant progress has been made in quantum information technology, from theoretical studies to experimental demonstrations. Revolutionary quantum applications are now in the limelight, showcasing the advantages of quantum information technology and becoming a research hotspot in academia and industry. To enable quantum applications to have a more profound impact and wider application, the interconnection of multiple quantum nodes through quantum channels becomes essential. Building an entanglement-assisted quantum network, capable of realizing quantum information transmission between these quantum nodes, is the primary goal. However, entanglement-assisted quantum networks are governed by the unique laws of quantum mechanics, such as the superposition principle, the no-cloning theorem, and quantum entanglement, setting them apart from classical networks. Consequently, fundamental efforts are required to establish entanglement-assisted quantum networks. While some insightful surveys have paved the way for entanglement-assisted quantum networks, most of these studies focus on enabling technologies and quantum applications, neglecting critical network issues. In response, this paper presents a comprehensive survey of entanglement-assisted quantum networks. Alongside reviewing fundamental mechanics and enabling technologies, the paper provides a detailed overview of the network structure, working principles, and development stages, highlighting the differences from classical networks. Additionally, the challenges of building wide-area entanglement-assisted quantum networks are addressed. Furthermore, the paper emphasizes open research directions, including architecture design, entanglement-based network issues, and standardization, to facilitate the implementation of future entanglement-assisted quantum networks.
\end{abstract}

	
\section{Introduction}\label{Introduction}
\IEEEPARstart{o}{ver} the past two decades, there has been dramatic development in classical information technology. However, such tremendous progress poses various challenges for the current classical Internet. For example, the security of public-key crypto-systems relys on the hardness of integer factorization and discrete logarithmic problems, which are no longer effective under a quantum computer using Shor's algorithm~\cite{shor1994algorithms}. Meanwhile, the existing computing power struggles to cope with increasingly complex computing tasks due to the limitations of Moore's Law \cite{schaller1997moore}. Such problems are extremely difficult to address using traditional information technologies, significantly hindering the further development of the Internet. Fortunately, quantum information technology provides new solutions to these problems. Moreover, quantum applications show excellent advantages over classical solutions owing to the unique characteristics of quantum mechanics with no counterpart in classical information technology.

\begin{figure}[t]
	\centering
	\includegraphics[width=1.0\linewidth]{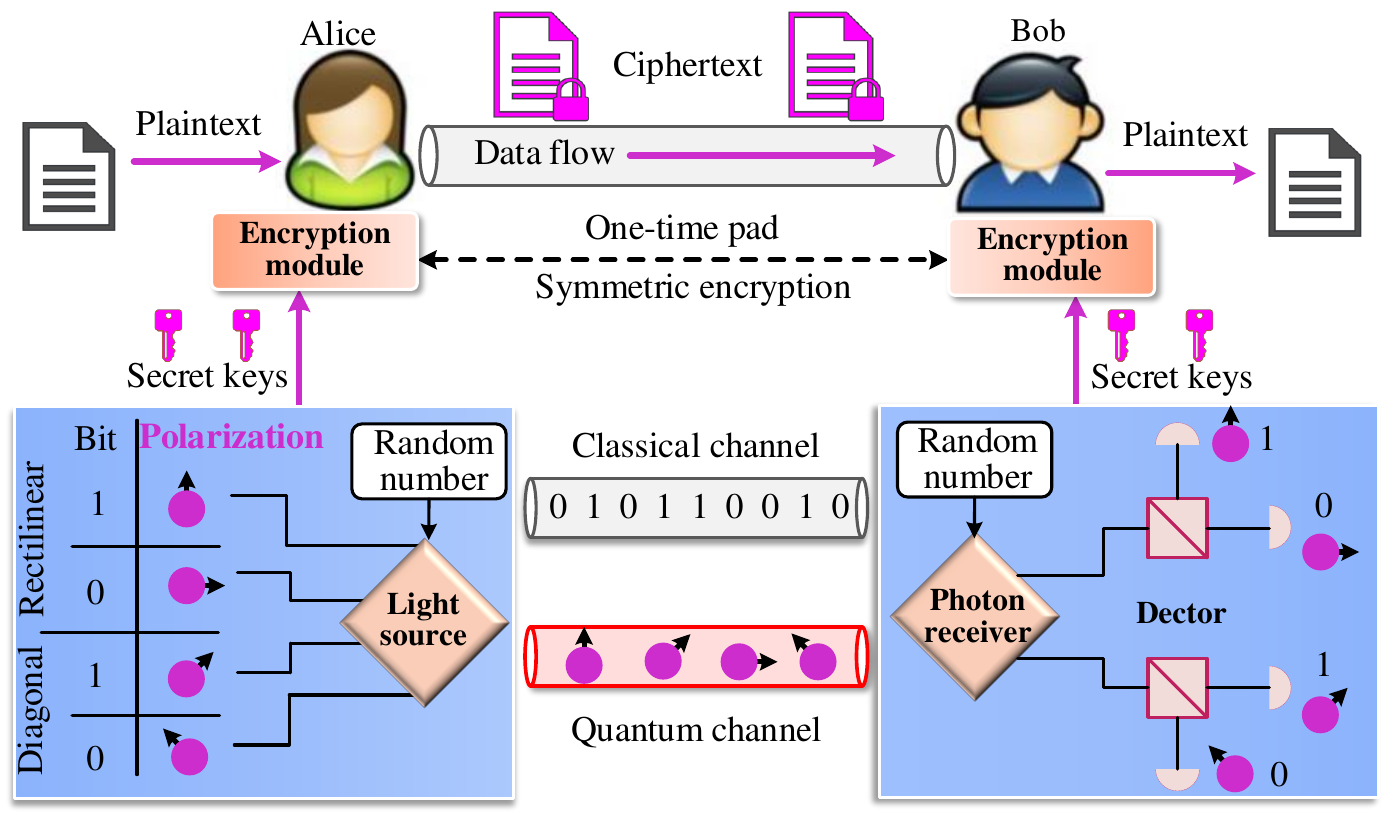}
	\caption{QKD-based secure classical communication.}
	\label{QKD}
\end{figure}

Quantum key distribution (QKD) \cite{bennett1984proceedings,ekert1991quantum,lo2012measurement,lo2014secure,lucamarini2018overcoming} is the most well-known and successful application of quantum information technology and has evolved from experimental demonstration to commercial service. QKD aims to securely distribute random keys between two communicating parties (hereafter referred to as Alice and Bob), exploiting the unique properties of quantum mechanics. \textbf{Fig.~\ref{QKD}} shows the implementation of QKD-based secure communication between Alice and Bob. Generally, QKD-based secure communication follows these pivotal steps. Firstly, Alice and Bob negotiate a rule for encoding particles, e.g., they encode horizontal polarization in rectilinear polarization as binary 0. Then Alice randomly chooses the base (rectilinear or diagonal polarization) to encode photons generated by the light source and sends them to Bob via quantum channels. Thirdly, Bob randomly selects measurement bases to process the received photons. If Bob uses the same base as the decoding base, he can successfully obtain the encoded information. Lastly, they can share secure keys after base comparison and some post-processing operations, such as private amplification. According to the fundamental principles of quantum mechanics, eavesdropping on quantum channels inevitably changes the state of encoded photons, introducing bit errors into coded information. Hence, eavesdroppers can be detected by estimating the bit error rate, which guarantees the security of shared keys~\cite{lo1999unconditional,pirandola2020advances}. Consequently, Alice and Bob can realize unconditional secure communication using the one-time pad strategy and symmetric encryption algorithms~\cite{vernam1926cipher}.

Currently, some QKD networks have been deployed, such as DARPA QKD network \cite{elliott2005current}, SECOQC QKD network \cite{peev2009secoqc}, Tokyo QKD network \cite{sasaki2011field}, SwissQuantum \cite{stucki2011long}, Beijing-Shanghai QKD network \cite{zhang2018large}, and Cambridge QKD network \cite{dynes2019cambridge}, to provide security services for the fields of finance, banking, and defense. Additionally, satellite-based intercontinental QKD network \cite{liao2018satellite} and integrated space-to-ground QKD network \cite{chen2021integrated} have been demonstrated in recent years.
While QKD has matured as a technology capable of providing small-scale commercial services, and valuable surveys \cite{scarani2009security,sasaki2017quantum,hosseinidehaj2018satellite,gyongyosi2019survey,xu2020secure,cavaliere2020secure,mehic2020quantum,cao2022evolution} have paved the way for building large-scale QKD networks, it is worth noting that QKD networks are not the final form of quantum networks. Although QKD technology applies quantum channels to transmit quantized particles between adjacent quantum nodes, the information transmitted via quantum channels is still classical, essentially random classical 0 or 1 bits.

Nevertheless, many quantum applications require quantum information to be transmitted between distant quantum nodes. For example, distributed quantum computing requires quantum information to be transmitted between multiple quantum computers \cite{Denchev2008Distributed}, so as to cooperate to complete specific complex computing tasks that are challenging to solve by distributed classical computing, and quantum sensing usually requires quantum nodes to transmit quantum information by establishing entangled systems, thus realizing high-precision and high-sensitivity measurement. Notably, the no-cloning theorem in quantum mechanics illustrates that it is not feasible to transmit quantum information by measuring microscopic particles and then transmitting them through other channels. Therefore, QKD networks, the network infrastructure built for enhancing classical secure communication by distributing random secret keys, cannot achieve quantum information transmission between communicating parties because measurement operations performed by QKD nodes cause microscopic particles to collapse and destroy quantum information. As a result, QKD networks cannot support various quantum applications, such as quantum imaging \cite{brida2010experimental}, blind quantum computing (a quantum system that allows clients to outsource their computing tasks to quantum servers that do the job for them with blinding information) \cite{barz2012demonstration}, and improved sensing~\cite{degen2017quantum}. Notably, quantum applications, especially quantum computing and quantum communication, have made tremendous progress in recent years and are gradually maturing, which drives the large-scale application of quantum information technology. To fully exploit the potential of quantum information technology, a complete network infrastructure that can support the transmission of quantum information between quantum nodes is required to serve various quantum applications.

\begin{table*}[htbp]
	\centering
	\caption{Comparison of this survey with other related surveys}
	\label{Comparison}
	\begin{tabular}{m{1.0cm}<{\centering}m{0.4cm}<{\centering}m{1.2cm}<{\centering}m{1.6cm}<{\centering}m{1.3cm}<{\centering}m{1.3cm}<{\centering}m{1.2cm}<{\centering}m{1.1cm}<{\centering}m{1.3cm}<{\centering}m{1.4cm}<{\centering}m{1.75cm}<{\centering}}
		\toprule
		\textbf{Refs} & \textbf{Year} &\textbf{Quantum Mechanics} & \textbf{Enabling Technologies} & \textbf{Development Stages}&\textbf{Quantum Elements} &\textbf{Network Structure}& \textbf{Challenges} &\textbf{Architecture Design} &\textbf{ Network Issues} & \textbf{Standardization} \\			
		\midrule
		\textbf{\cite{dur2017towards}}&2017&\checkmark & \checkmark & && &\checkmark & && 	\\
		\textbf{\cite{wehner2018quantum}}&2018&&\checkmark &\checkmark &\checkmark& &\checkmark &\checkmark& &	 	\\
		\textbf{\cite{caleffi2018quantum}} &2018& &\checkmark & && &\checkmark &			& &	 	\\
		\textbf{\cite{kozlowski2019towards}} &2019&\checkmark &\checkmark  & &\checkmark	& &\checkmark  &\checkmark	& &	 	\\
		\textbf{\cite{cacciapuoti2019quantum}}&2019&\checkmark &\checkmark&& 	&   &\checkmark & &	& 	\\
		\textbf{\cite{cacciapuoti2020entanglement}} &2020&\checkmark &\checkmark&&&&\checkmark& & &	\\	
		\textbf{\cite{caleffi2020rise} } &2020& &\checkmark&\checkmark && & \checkmark&	&	 &	\\	
		\textbf{\cite{singh2021quantum}}& 2021&\checkmark &\checkmark & &\checkmark&\checkmark&\checkmark & & &		\\
		\textbf{\cite{li2021building}}& 2021 &&\checkmark& &\checkmark&&\checkmark && &	\\
		\textbf{\cite{sandilya2021quantum}}&2021&\checkmark &\checkmark& &\checkmark &	& 	& 	&     &	\\
		\textbf{\cite{illiano2022quantum}}&2022&\checkmark  &\checkmark& & 	&\checkmark	&\checkmark & & 	&\checkmark	\\
		\textbf{\cite{chehimi2022physics}}&2022 & && \checkmark		& & 	& 	\checkmark		&	&	&	\\
		
		\textbf{This Survey}               & &\checkmark &\checkmark &\checkmark  &\checkmark &\checkmark &\checkmark &\checkmark &\checkmark&\checkmark \\
		\bottomrule
	\end{tabular}
\end{table*}

\subsection{Motivation}	

Entanglement plays an important role in realizing the potential of quantum information technology. On the one hand, the non-local correlation of entangled systems is one of the essential cornerstones of quantum information transmission between distant quantum nodes. On the other hand, most quantum applications require quantum nodes to share entangled systems. Therefore, entanglement-assisted quantum networks are the promising platform for supporting various quantum applications. An entanglement-assisted quantum network is a network interconnecting numerous quantum nodes capable of generating, storing, transmitting, and processing quantum information (i.e., quantum bits, also known as qubits) in addition to classical information. In entanglement-assisted quantum networks, any quantum node can establish entanglement connections with others by sharing entangled systems to realize quantum information transmission, thus effectively supporting various quantum applications. However, the fact that entangled systems are susceptible to environmental noise significantly prevents entangled qubits from being distributed over long distances. Fortunately, the second quantum revolution has activated the development of quantum devices, especially quantum memory and quantum repeaters. With the assistance of complete quantum devices, quantum nodes can effectively overcome distance limitations to establish distant entanglement connections. Hence, it is conceivable that large-scale entanglement-assisted quantum networks can be realized in the near future.

Nevertheless, entanglement-assisted quantum networks follow the fundamental principles of quantum mechanics and are essentially different from classical networks in many aspects, such as information resources, enabling technologies, and upper-layer applications. Hence, entanglement-assisted quantum networks are not the product of the iterative development of classical networks. As a result, the network design generally adopted in classical networks cannot be directly applied to entanglement-assisted quantum networks. Therefore, the fundamental but pivotal studies on network designs are considerably required to pave the way for entanglement-assisted quantum networks to fully demonstrate the advantages of quantum information technology.

The research on entanglement-assisted quantum networks, which offer effective and efficient support for various quantum applications, has gained significant attention. In recent years, valuable research works have been undertaken to facilitate the development of these networks and the broader concept of a quantum internet.
\begin{itemize}
	\item \cite{dur2017towards} reviews the potential of a quantum internet for the secure transmissions of classical and quantum information, along with the theoretical and experimental approaches and recent advances in realizing them.
	
	\item \cite{wehner2018quantum} categorizes the different stages of developing quantum internet and outlines the technological advances required for reaching these stages.
	
	\item \cite{caleffi2018quantum} discusses the exponential computing speed-up achievable by interconnecting numerous quantum computers through a quantum internet and identifies key future research challenges for quantum internet deployment.
	
	\item \cite{kozlowski2019towards} provides a gentle introduction to quantum networking targeted at computer scientists, surveys the state of the art, and discusses the key challenges related to computer science in order to make such quantum networks a reality.
	
	\item \cite{cacciapuoti2019quantum} reviews some basic knowledge of quantum mechanics, introduces quantum teleportation as the key strategy for transmitting quantum information, and discusses some pivotal research challenges in designing future quantum communication networks.
	
	\item \cite{cacciapuoti2020entanglement} reviews some preliminaries on quantum mechanics to show the fundamental differences between the transmission of classical information and the teleportation of quantum information, introduces the communications functionalities underlying quantum teleportation, and addresses the challenges of the practical deployment of these functionalities for the upcoming quantum internet.
	
	\item \cite{caleffi2020rise} introduces two essential quantum operations, quantum teleportation and entanglement swapping, and envisions roughly three subsequent necessary steps toward the envisioned quantum internet, whose complexity varies with time and the level of platform heterogeneity.
	
	\item \cite{singh2021quantum} surveys quantum internet functionalities, technologies, applications, and open challenges to help readers gain a basic understanding of the infrastructure required for realizing a high-performance quantum internet.
	
	\item \cite{li2021building} reviews the enabling technologies required for building entanglement-assisted quantum networks and provides a novel design of a cluster-based structure and an OSI-alike layering model for high-performance entanglement-assisted quantum networks.	
	
	\item \cite{sandilya2021quantum} discusses more of the technologies that make up the quantum internet and its concept and shows that an entanglement-assisted quantum network is a collaboration of various technologies forming a network of networks.
	
	\item \cite{illiano2022quantum} presents a review of the relevant literature about quantum internet protocol stack and discusses the open problems and efforts required for the design of an effective and complete quantum internet protocol stack.
	
	\item \cite{chehimi2022physics} identifies physics-informed performance metrics and controls that enable entanglement-assisted quantum networks to leverage state-of-the-art advancements in quantum technologies to enhance their performance. It also analyzes multiple challenges and open research directions that must be addressed using a physics-informed approach to generate practically viable results.	
\end{itemize}

These valuable surveys provide insights into different perspectives on quantum information technology and entanglement-assisted quantum networks. Nevertheless, none pays attention to networking and internetworking details. For example, many of them focus on the fundamental features of quantum mechanics, enabling technologies, and research challenges, with little attention paid to the problem of how quantum nodes connect and interact with each other in a large-scale entanglement-assisted quantum network, especially pivotal network designs like routing, request scheduling, and resource allocation. Thus, there is a paucity of literature on entanglement-assisted quantum networks. \textbf{Table~\ref{Comparison}} explicitly compares this survey against the existing works mentioned above. In addition to presenting quantum mechanics and enabling technologies, we also take an overview of entanglement-assisted quantum networks, including definition, development stages, differences from classical networks, network elements, network structure, and working principles. We conclude that the dramatic development of quantum information technology makes it possible to build entanglement-assisted quantum networks, and the study of network issues is required. Hence, we further present some research directions, including architecture design and entanglement-based network problems, as well as standardization, thus paving the way for constructing large-scale and wide-area entanglement-assisted quantum networks that enable iterative development and perform well in quality of service (QoS). In summary, this survey is the first to provide a comprehensive and up-to-date review of entanglement-assisted quantum networks, from the unique features of quantum mechanics to networking studies.

\subsection{Contributions}	
More concretely, the major contributions of this survey can be summarized as follows:
\begin{itemize}
\item [1)] We describe the basic concepts and unique principles of quantum mechanics required to understand entanglement-assisted quantum networks. Besides, a comprehensive comparison between bits and qubits, as well as between classical gates and quantum gates used to process quantum information, is presented.

\item [2)] We discuss the enabling technologies required to build an entanglement-assisted quantum network, following the logic of function, principle, implementation process, and the state of development. Specifically, we discuss the implementation steps of each technology with the help of quantum circuits and then present the theoretical and experimental development of each enabling technology.

\item [3)] Based on the detailed comparison between classical communication and quantum communication, we comprehensively compare classical networks with entanglement-assisted quantum networks, ranging from physical resources to protocol stacks, to demonstrate that entanglement-assisted quantum networks are fundamentally different from classical networks.

\item [4)] We conclude the development stages of entanglement-assisted quantum networks. Besides, considering the implementation of a future entanglement-assisted quantum network, network elements and the requirements of network elements are discussed.

\item [5)] We present a general structure of entanglement-assisted quantum networks and describe how entanglement-assisted quantum networks work with the assistance of the enabling technologies discussed in this survey to support various quantum applications.

\item [6)] We discuss the challenges of building a large-scale entanglement-assisted quantum network from three perspectives: the inherently imperfect nature of quantum systems, the vast variability of different physical resources, and the convergence of entanglement-assisted quantum networks and classical networks.

\item [7)] Finally, we summarize a range of detailed research directions, particularly concerning network issues such as routing, scheduling, and resource allocation. Specially, we present the profound effects of these issues on the interaction of quantum nodes and then discuss possible solutions to critical problems to pave the way for building high-performance entanglement-assisted quantum networks.
\end{itemize}

\begin{figure}[t]
	\centering
	\includegraphics[width=0.98\linewidth]{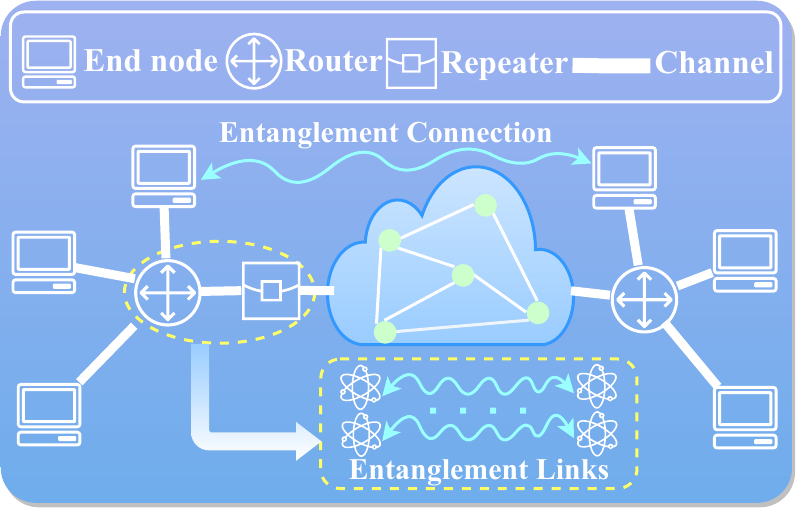}
	\caption{An abstract structure of a large-scale and wide-area entanglement-assisted quantum network.}
	\label{network}
\end{figure}

\subsection{Entanglement-Assisted Quantum Networks}	

In terms of physical structure, an entanglement-assisted quantum network can be regarded as a mesh consisting of three types of network elements: physical channels, networking devices, and quantum end nodes. Firstly, physical channels are utilized to transmit microscopic particles between adjacent quantum nodes. Secondly, networking devices, e.g., quantum repeaters and quantum routers, are pivotal in building large-scale and wide-area entanglement-assisted quantum networks. Quantum repeaters utilize the unique feature of entanglement to overcome the distance limitation caused by the inherent loss of physical channels, thus extending the communication range. Quantum routers aim at converging numerous quantum nodes to expand networks' scale. Quantum end nodes work by transmitting and processing quantum information to support the top-level quantum applications that run on them. \textbf{Fig.~\ref{network}} depicts the abstract structure of an entanglement-assisted quantum network. Concretely, a small number of quantum end nodes are converged to form a small-scale and local-area quantum network with the assistance of networking devices and physical channels, and the networking devices are interconnected through physical channels to form a wide-area core quantum network in a mesh topology, i.e., the cloud icon. In this way, the core quantum network can connect many local-area quantum networks together, thus forming a large-scale and wide-area entanglement-assisted quantum network. Any pair of adjacent quantum nodes can build entanglement links, the essential resource for quantum information transmission, by sharing entangled qubit pairs. With the help of quantum routers and quantum repeaters, any pair of quantum end nodes can establish long-distance entanglement connections by ``coupling" multiple entanglement links along a selected path and thus achieve remote quantum information transmission.

\subsection{Paper Organization}	
An entanglement-assisted quantum network acts as the fundamental platform formed by numerous quantum nodes and physical channels to realize the transmission of quantum information between arbitrary quantum end nodes, thus supporting various quantum applications. The research on entanglement-assisted quantum networks mainly focuses on four pivotal problems: what are the unique properties of entanglement-assisted quantum networks? how to interconnect numerous quantum nodes to form the network infrastructure? how to realize quantum information transmission between quantum end nodes? how to realize an entanglement-assisted quantum network with high performance and QoS guarantee through network design? To better understand entanglement-assisted quantum networks, we guide readers through questions decomposed by the four aforementioned problems. Here, we summarize the question sets in the following categories:

\textbf{QS1.} What is to be transmitted (Qubit)? What are the characteristics of auxiliary tools used for transmitting quantum information (Quantum Entanglement)? What are the characteristics of quantum devices used for qubit transmissions (Decoherence and Fidelity)? How to operate on the transmitted qubit (Quantum Gates)?

\textbf{QS2.} How to generate entangled qubits (Entanglement Preparation)? How to code and decode qubits? How to enable quantum information transmission (Quantum Teleportation)? How to improve transmission performance (Entanglement Purification)? How to correct errors caused by quantum decoherence (Quantum Error Correction)? How to enable multi-hop transmissions (Entanglement Swapping)? How to store/cache qubits (Quantum Memory)?

\begin{figure}[t]
	\centering
	\includegraphics[width=0.92\linewidth]{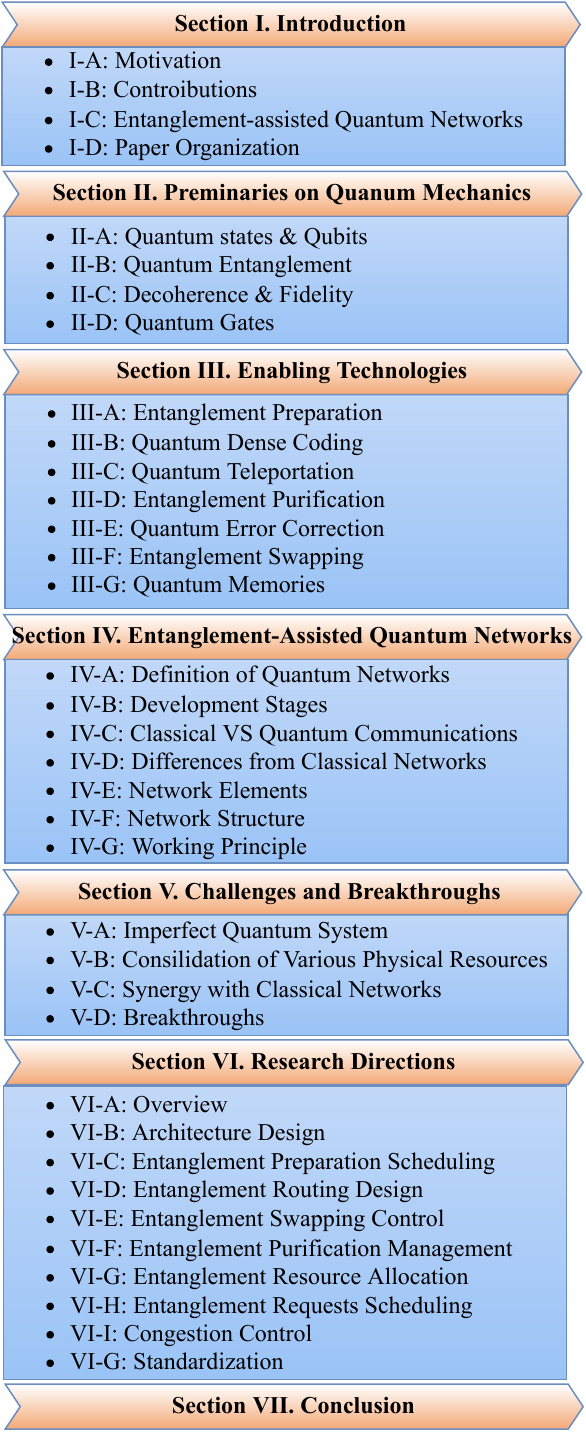}
	\caption{The structure of this paper and topic classification.}
	\label{organization}
\end{figure}

\textbf{QS3.} What are the development stages of entanglement-assisted quantum networks (Development Stages)? What is the difference between an entanglement-assisted quantum network and a classical network (Differences from Classical Networks)? What is the network structure for an entanglement-assisted quantum network (Network Elements and Network Structure)? How does an entanglement-assisted quantum network work (Working Principles)?

\textbf{QS4.} What impedes the implementation of a large-scale and wide-area entanglement-assisted quantum network (Challenges)? How to effectively and efficiently realize quantum information transmission between quantum end nodes in concurrent communication scenarios (Research Directions)?

The rest of this paper is organized as shown in \textbf{Fig.~\ref{organization}}. We answer the first three questions mentioned above in Section~\ref{Sec2} preliminaries on quantum mechanisms (QS1), Section~\ref{Sec3} enabling technologies (QS2), Section~\ref{Sec4} entanglement-assisted quantum networks (QS3). Based on the answers to the above questions, we introduce the challenges in Section~\ref{Sec5} and elaborate on the research directions in Section~\ref{Sec6} to answer the fourth question (QS4). Finally, Section~\ref{Sec7} draws the conclusion of this survey.

\section{Preliminaries of Quantum Mechanics}\label{Sec2}
This section primarily outlines the fundamental but crucial knowledge of quantum mechanics. Firstly, the concept of ``qubits'' used in the rest of the paper and two features of qubits, i.e., uncertainty and no-cloning, are introduced. Then, some unique features of quantum mechanics with no counterpart in classical networks are elaborated, including quantum entanglement, quantum decoherence, and fidelity. Finally, we introduce some quantum gates commonly adopted to manipulate qubits. In particular, we make a comprehensive comparison between qubits and bits, as well as between quantum gates and classical gates, to demonstrate that quantum information technology is fundamentally different from classical information technology.

\subsection{Quantum State and Qubits}\label{Sec2.1}
\textit{Quantum physics} is a fundamental theory used to describe the microscopic physical world. Generally, quantum physics differs from classical physics in three aspects. Firstly, the energy, momentum, angular momentum, and other quantities of a closed quantum system are restricted to discrete values, i.e., quantization. Secondly, microscopic objects are characterized by particles and waves simultaneously, i.e., wave-particle duality. Finally, it is hard to accurately predict the value of a physical quantity before measurement in the quantum world, i.e., the uncertainty principle. The fundamental feature of quantum physics is that it usually cannot predict with certainty what will happen to a closed quantum system but can only give probabilities of some possible outcomes. In other words, the quantum world presents a probabilistic feature.
		
As an essential concept of quantum physics, ``quantum'' represents the minimum amount of any physical entity or physical property involved in an interaction. An isolated or closed quantum system is characterized by its micro-state, known as the quantum state. Quantum state is a mathematical quantity used to describe the probability distribution for each possible measurement outcome of a quantum system \cite{messiah2014quantum}. In quantum physics, a quantum system is usually described by a wave function associated with a probability at each point in space. Mathematically, the probability of each possible outcome of a quantum system is found by taking the square of the absolute value of a complex number, known as the \textit{probability amplitude}. Besides, each quantum system has a corresponding Hilbert space, which is a generalized complete inner vector product space usually used to explore quantum physics~\cite{young1988introduction}. Hence, the state of a quantum system can also be represented by a vector of its corresponding Hilbert space, i.e., the state vector. For example, a quantum state is denoted as $\ket{\psi}$, where the Dirac symbol ``$\ket{}$" is called \textsl{``ket"} and represented by a $N\times1$ matrix. Suppose that the quantum system described by $\ket{\psi}$ exhibits two possible observed outcomes, and we denote these two possibilities as ``0'' and ``1''. Then, the quantum system can be represented by the state $\ket{0}$ if it is in ``0'' and represented by the state $\ket{1}$ if it is in ``1''. Therefore, quantum states $\ket{0}$ and $\ket{1}$ display different possible states of $\ket{\psi}$.
	
\begin{table}[htbp]
	\centering
	\caption{The comparison between Bits and Qubits.}
	\label{bitvsqubit}
	\begin{tabular}{m{2.0cm}<{\centering}m{2.6cm}<{\centering}m{2.8cm}<{\centering}}
	\toprule
	\textbf{Topics} &\textbf{Classical Bits} & \textbf{Qubits}  \\
	\midrule
	\textbf{Existing Form}  &Current, Voltage &Ions, Atoms, Photons, Electrons, Superconductors \\
	\textbf{State} & Deterministic, 0 or 1   & Superposition, $\ket{0}$ and $\ket{1}$ and $\alpha\ket{0}+\beta\ket{1}$ \\
	\textbf{Coherence} & Incoherent    & Coherent  \\
	\textbf{Information Capacity} & Polynomial increase & Exponential increase \\
	\textbf{Impact of Measurement}  & Not affected  & The state of a quantum system is changed  \\
	\textbf{Clonability} & Yes &  No \\
	\bottomrule
	\end{tabular}
\end{table}

In classical information theory, the data or information is encoded as binary bit strings. Similarly, quantum information theory introduces the concept of quantum bits (\textit{qubits}) \cite{braunstein1992maximal}, to ``encode'' the quantum information of a quantum system. Qubits describe the state of quantum systems mathematically. As shown in \textbf{Table~\ref{bitvsqubit}}, qubits essentially vary from classical bits. We elaborate on these differences as follows.
	
Classical physics is a fundamental theory that provides a description of the macroscopic physical world. In the macroscopic world, all observed physical objects can be determined with infinite precision in theory. Hence, classical bits usually exist in the form of deterministic signals such as current and voltage. However, qubits can only be represented by different physical resources with these unique features, including quantization, wave-particle duality, and the uncertainty principle. Generally, qubits can be represented by ions \cite{schmidt2003realization,leibfried2003experimental,randall2015efficient,ballance2016high}, atoms~\cite{isenhower2010demonstration,saffman2010quantum}, photons \cite{kok2010introduction,hacker2016photon,qiang2018large}, spinning electrons \cite{kane1998silicon,van2012decoherence,van2018readout}, and superconductors~\cite{yamamoto2003demonstration,chow2012universal,wendin2017quantum}.

A classical bit is determined, i.e., a classical bit can represent only one state, either 0 or 1. In contrast to the deterministic properties of classical bits, the peculiarity of qubits is that a single qubit can be in a \textit{``superposition''} of multiple states before being measured. That is, a single qubit can be in different states at the same time. In two-dimensional Hilbert space, a single qubit is an arbitrary linear combination of two possible states, $\ket{0}$ and $\ket{1}$, with two probability amplitudes. Mathematically, in two-dimensional Hilbert space, we can represent the state of a single qubit $\ket{\psi}$ by
\begin{equation}
	\ket{\psi}=\alpha\ket{0}+\beta\ket{1},
	\label{qubit}
\end{equation}
where $\ket{0}$ and $\ket{1}$ form a set of standard orthonormal states in two-dimensional Hilbert space, also known as the computational base states. For example, they can be written as
\begin{equation}
	\ket{0}=
	\begin{bmatrix}
		1\\
		0
	\end{bmatrix}, \quad \text{and} \quad
	\ket{1}=
	\begin{bmatrix}
		0\\
		1
	\end{bmatrix}.
\end{equation}
The complex coefficients $\alpha$ and $\beta$ are the probability amplitudes of the observed outcomes of $\ket{\psi}$. The two complex coefficients need to satisfy the condition
\begin{equation}
	{\left| \alpha \right|}^{2} + {\left| \beta \right|}^{2} = 1.
\end{equation}
	
As discussed above, classical bits are deterministic and independent, so classical bits are not coherent. However, qubits follow the superposition principle, and the probability amplitude of all possible measured outcomes determines the state of a closed quantum system. Hence, different from classical bits, qubits present quantum coherence, a special correlation between qubits in a quantum system. Besides, the information capacity of a classical system polynomially increases with the number of classical bits. The superposition state of qubits facilitates the exponential improvement of computing power, which demonstrates quantum superiority, i.e., a quantum computer can solve some challenging tasks that none classical computer can do in an acceptable amount of time using any known algorithm.

Moreover, the state of a classical bit is unaffected by measurement or observation. Hence, classical bits can be completely cloned and recovered during classical communication. However, measurement operations significantly affect the superposition state of a single qubit. For example, by measuring $\ket{\psi}$, we will obtain the state $\ket{0}$ with probability $\left|\alpha\right|^{2}$ or $\ket{1}$ with probability $\left|\beta\right|^{2}$. Measurement operations destroy the initial state of a quantum system, and this phenomenon is often called \textit{``collapse after measurement''}. As a result, qubits follow the no-cloning theorem, and we describe this theorem in more detail below.

\begin{figure}[t]
	\centering
	\includegraphics[width=0.75\linewidth]{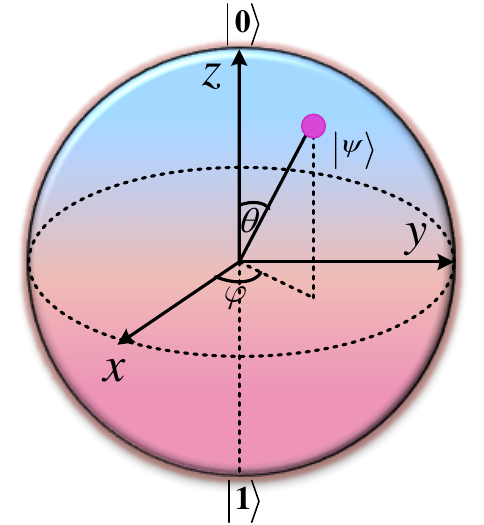}
	\caption{The Bloch sphere representation of a single qubit.}
	\label{Qubit}
\end{figure}

Qubits can also be geometrically represented by a Bloch sphere~\cite{glendinning2005bloch}. Next, we elaborate on the derivation of Bloch spherical coordinates corresponding to a single qubit. In quantum mechanics, a quantum system can be described by a wave function associated with a probability amplitude at each point in space. Hence, $\ket{\psi}$ can be formulated as a wave function using polar coordinates:
\begin{equation}
    \ket{\psi}=a e^{i\varphi_{a}} \ket{0} + b e^{i\varphi_{b}} \ket{1},
    \label{wave-function}
\end{equation}
where $a$, $b$, $\varphi_{a}$, and $\varphi_{b}$ are real parameters and $i$ is an imaginary unit. The global phase factor can be removed from Eq.~\eqref{wave-function}, so we can get

\begin{equation}
    \ket{\psi}=a \ket{0} + b e^{i(\varphi_{b}-\varphi_{a})} \ket{1}=a\ket{0}+b e^{i\varphi}\ket{1},
    \label{wave-function1}
\end{equation}
where $\varphi=\varphi_{b}-\varphi_{a}$ is a real parameter. Eq.~\eqref{wave-function1} can be further normalized. Let $b e^{i\varphi}=x+iy$, where $x$ and $y$ are real parameters, we can get
\begin{equation*}
    {\left| a \right|}^{2} + {\left| b e^{\varphi}  \right|}^{2} = a^{2}+{\left| x+iy  \right|}^{2}=a^{2}+x^{2}+y^{2}=1.
\end{equation*}
Notably, the spherical coordinates of any point in Euclidean space can be further expressed as
\begin{equation}
    x=r\sin{\theta^{'}}\cos{\varphi}, y=r\sin{\theta^{'}}\sin{\varphi}, z=r\cos{\theta^{'}},
\end{equation}
where $r$ is the distance from the point to the center of the sphere, $\theta^{'}$ is the angle between the line formed by the connection between the point and the center of the sphere and the Z-axis, and $\varphi$ is the angle between the line (i.e., the projection of the line between the point and the center of the sphere onto the X-Y plane) and the X-axis. Therefore, the wave function of a quantum system can be written as
\begin{equation}
	\begin{aligned}
	\ket{\psi}=&z\ket{0}+(x+iy)\ket{1} \\
	&=\cos{\theta^{'}}\ket{0}+\sin{\theta^{'}}(\cos{\varphi}+i\sin{\varphi})\ket{1}\\
	&=\cos{\theta{'}}\ket{0}+e^{i\varphi}\sin{\theta^{'}}\ket{1}.
	\end{aligned}
\label{Bloch-function}
\end{equation}
To make the wave function correspond to Bloch spherical coordinates, we introduce $\theta=2\theta{'}$. As a result, Eq.~\eqref{Bloch-function} can be further expressed as
\begin{equation}
\ket{\psi}=\cos\frac{\theta}{2}\ket{0}+e^{i\varphi}\sin{\frac{\theta}{2}}\ket{1}).
\end{equation}
In this, any point on the Bloch sphere can be defined by a function related to $\theta$ and $\varphi$. Similar to the probability amplitude of the wave function of $\ket{\psi}$, these two parameters also determine the state of $\ket{\psi}$. As depicted in \textbf{Fig.~\ref{Qubit}}, the qubit $\ket{\psi}$ exhibits a point on the surface of the Bloch sphere with spherical coordinates of $\theta$ as the polar angle and $\varphi$ as the azimuth angle. For example, when $\theta=\ang{0}$ and $\varphi=\ang{0}$, $\ket{\psi}=\ket{0}$, that is, the point is at the vertex of the Bloch sphere's Z-axis. Besides, the point is at the vertex Bloch sphere's X-axis when $\theta=\ang{90}$ and $\varphi=\ang{0}$, i.e., $\ket{\psi}=\frac{1}{\sqrt{2}}(\ket{0}+\ket{1})$.
		
A quantum system can be extended from a single qubit to multiple qubits. Due to to the fact that the size of a classical bit's state space equals two, the amount of information a classical system can present increases only by a polynomial degree when one more classical bit is added to this system. However, the amount of information that a quantum system can present increases exponentially with the number of qubits. A multi-qubit quantum system shows a much larger state space than a classical system consisting of multiple classical bits. Due to the superposition phenomenon, the state of a quantum system consisting of $n$ qubits is a linear combination of $2^{n}$ base states and is determined by $2^{n}$ probability amplitudes. That is, each possible state of the quantum system can be represented by a state vector in $2^{n}$-dimensional space. Formally, the description of a $n$-qubit quantum system can be expressed as
\begin{equation}
	\ket{\psi}=\sum_{i=0}^{2^{n}-1}\alpha_{i}\ket{i}, \\
	\forall i \neq j, \Braket{i|j}=0,\Braket{i|i}=1.
\end{equation}
$\alpha_{i}$ is the probability amplitude of the state in which $\ket{\psi}$ equals $\ket{i}$, and $\sum |\alpha_{i}|^{2}=1$. $\ket{i}$ is the $i$-th base state of $\ket{\psi}$, and $\Braket{i|j}$ is the inner product of two state vectors, $\bra{i}$ and $\ket{j}$, where $\bra{i}$ is the dual state vector of $\ket{i}$. When $n=500$, the number of possible states in this quantum system exceeds the estimated total number of atoms in the entire universe, and storing all these complex numbers in a classical computer is unimaginable. Hence, quantum mechanics holds information processing capabilities beyond those of classical systems, and we can use this property to achieve efficient quantum computing.

In addition to the unique feature of the superposition state, qubits also follow other principles that have no counterpart in the classical world. Other properties, including the uncertainty principle and the no-cloning theorem, are described as follows.

\textbf{Uncertainty Principle.} In quantum mechanics, observers usually cannot predict with certainty what will happen to a closed quantum system, but only give probabilities of some possible outcomes. This phenomenon was first discussed by Heisenberg in 1927 \cite{heisenberg1985anschaulichen} and then named the uncertainty principle (also called \textit{``Heisenberg uncertainty principle''}). The uncertainty principle suggests that the position and momentum of a particle cannot be determined simultaneously. There is an opposite relationship between the determinism of position and momentum, i.e., the more precisely the position is determined, the less precisely the momentum is known, and vice versa. This principle results from the ``probability interpretation'' in quantum mechanics, e.g., the superposition state and the phenomenon of collapse after measurement.

\begin{figure}[t]
	\centering
	\includegraphics[width=1.0\linewidth]{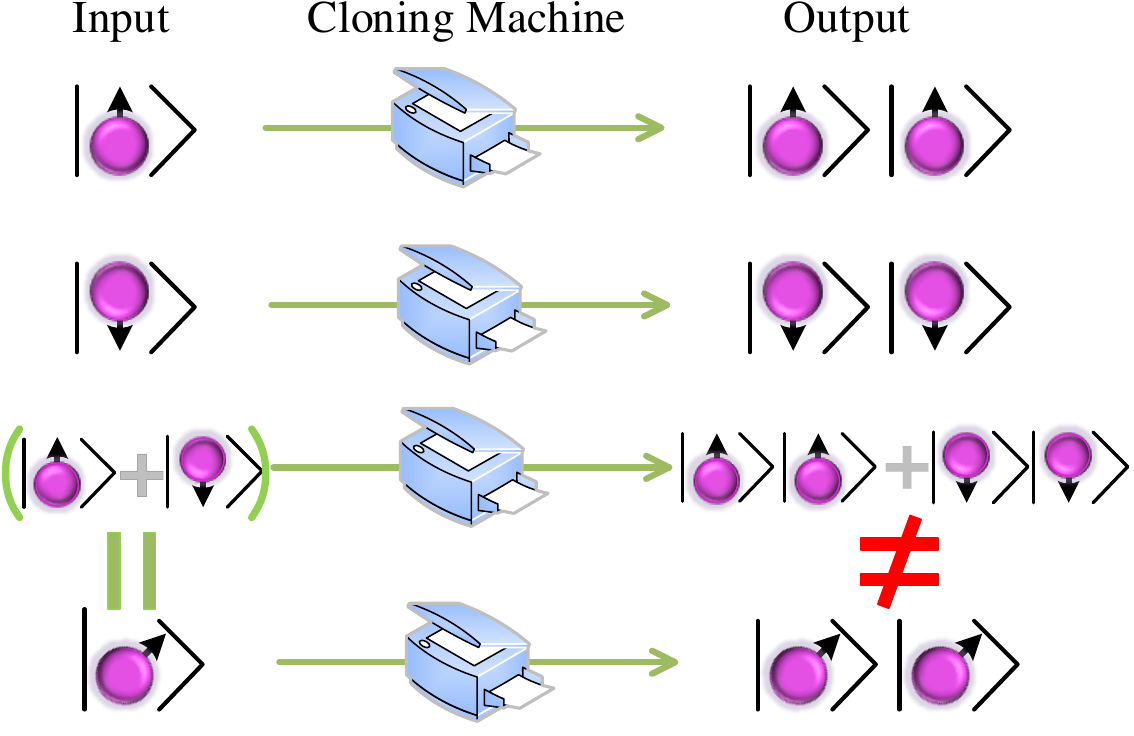}
	\caption{The illustration of the no-cloning theorem.}
	\label{no-cloning}
\end{figure}

\textbf{No-cloning Theorem.} Quantum systems are subject to a restriction known as the \textit{``no-cloning theorem''}. Wotters and Zurek first proposed the no-cloning theorem in 1982 \cite{Wootters1982A}, which indicates that a third party cannot copy and steal quantum information without interfering with the quantum system when qubits are transmitted in quantum channels. The linear feature of quantum mechanics, i.e., any linear combination of all possible states of a quantum system is still a possible state of the system, is the fundamental reason for the no-cloning theorem \cite{park1970concept,dieks1982communication,scarani2005quantum}. We assume that there is a cloning machine that can copy the quantum information from a photon or an electron. As shown in \textbf{Fig.~\ref{no-cloning}}, when the original state enters this machine, two copies come out, each having the same state as the original. If such a machine is successful, it will convert the state $\ket{0}$ to $\ket{00}$ ($\ket{00}$ is also often written as $\ket{0}\ket{0}$) and $\ket{1}$ to $\ket{11}$, where $\ket{00}=\ket{0}\otimes\ket{0}$ is the tensor product of two state vectors $\ket{0}$ and $\ket{0}$. However, the problem arises when we send a linear combination of $\ket{0}$ and $\ket{1}$, i.e., $\ket{\psi}=\alpha\ket{0}+\beta\ket{1}$, through a hypothetical cloning machine. If $\ket{0}$ and $\ket{1}$ can be cloned correctly, the output for their superposition must be the superposition of the outputs, i.e., $\alpha\ket{00}+\beta\ket{11}$, due to the linearity of quantum mechanics. Whereas the cloning machine will output $\ket{\psi}\ket{\psi}=\alpha^{2}\ket{00}+\alpha\beta\ket{01}+\alpha\beta\ket{10}+\beta^{2}\ket{11}$. Obviously, the output is not what we expected, i.e., the superposition state $\ket{\psi}$ is not copied exactly. Hence, cloning machines do not exist in the quantum physics world. The no-cloning theorem prevents the application of signal regeneration and amplification techniques in quantum information transmission, which is essential for ensuring the unconditional security of quantum information transmission.

\subsection{Quantum Entanglement}\label{Sec2.2}

Lying at the center of the interest in quantum physics of the 21st century, \textit{quantum entanglement} was first regarded as a \textit{``spooky action at a distance''} phenomenon of quantum mechanics by Einstein, Podolsky, and Rosen (EPR) and Schr{\"o}dinger in 1935~\cite{einstein1935can}. This phenomenon implies that the global states of an entangled system composed of multiple entangled qubits cannot be factored as a product of the states of its local constituents. That is, they are not individual particles but are an inseparable whole, i.e., one constituent cannot be fully described without considering the other(s) in an entangled system.

\begin{figure}[t]
	\centering
	\includegraphics[width=1.0\linewidth]{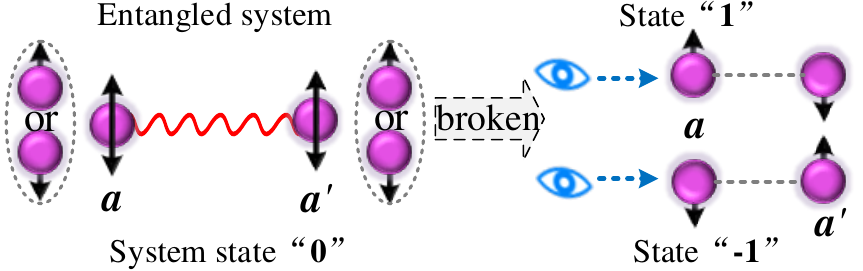}
	\caption{The unique property of an entangled system.}
	\label{quantum-entanglement}
\end{figure}
	
As shown in \textbf{Fig.~\ref{quantum-entanglement}}, an entangled system consists of two photons, $a$ and $a'$, with opposite spin directions, i.e., up-spin and down-spin photons. Hence, the entangled system's angular momentum, a metric to measure particles' spin intensity, is equal to zero, that is, the entangled system is a spin-zero system (denoted as the ``0'' state). In this entangled system, an observer only knows that each spin photon is in one of the up-spin (denoted as the ``1'' state) and down-spin (denoted as the ``-1'' state) states. According to the law of conservation of angular momentum, i.e., the total angular momentum of the isolated system remains constant (``0''=``1''+``-1'') \cite{kanarev2002law}, there is a spin anti-correlated case when the entangled system is broken. If the spin photon $a$ is measured to be the up-spin or the ``1'' state, $a'$ spins down or the ``-1'' state. If $a$ spins down or is the ``-1'' state, $a'$ is measured to be the up-spin or the ``1'' state. The essence of entanglement is a non-local correlation in quantum mechanics. When two or more quantum systems form an entangled system, their states are interdependent and cannot be described independently. This means that the measurement result of one system will immediately affect the state of the other quantum systems, even if they are far apart, and this correlation is superluminal. In summary, an entangled state refers to a special correlation between multiple quantum systems, such that their states cannot be described individually but can only be understood through a collective description.

In 1964, Bell proposed the Bell inequality based on the EPR conclusion and showed that the probabilities for the results obtained by measuring some entangled systems violated the Bell inequality \cite{bell1964einstein}. This result demonstrates that entanglement is a fundamentally non-classical phenomenon, i.e., it is impossible to simulate quantum behavior in classical systems. Next, we introduce entangled systems by dividing them into two categories according to the number of entangled qubits they contain.

Let $N$ and $V$ denote two two-dimensional Hilbert spaces, and their base state vectors are $\ket{0}$ and $\ket{1}$. For the four-dimensional Hilbert space formed by the tensor product of $N$ and $V$, its base state vectors are $\ket{00}$, $\ket{11}$, $\ket{01}$, and $\ket{10}$. In this four-dimensional Hilbert space, any state vector $\ket{\Psi}$ can be written as the tensor product of two qubits' state vectors $\ket{\psi}=\alpha\ket{0}+\beta\ket{1}$ and $\ket{\phi}=\gamma\ket{0}+\delta\ket{1}$:
\begin{equation}
	\begin{aligned}
		\ket{\Psi}&=\ket{\psi}\otimes\ket{\phi}\\
		&=(\alpha\ket{0}+\beta\ket{1})\otimes(\gamma\ket{0}+\delta\ket{1})\\
		&=\alpha\gamma\ket{00}+\alpha\delta\ket{01}+\beta\gamma\ket{10}+\beta\delta\ket{11},
	\end{aligned}
\end{equation}
where $\alpha$, $\beta$, $\gamma$, and $\delta$ are the four probability amplitudes that determine the states of $\ket{\psi}$ and $\ket{\phi}$, and they need to satisfy the condition  $\left|\alpha\gamma\right|^2+\left|\alpha\delta\right|^2+\left|\beta\gamma\right|^2+\left|\beta\delta\right|^2=1$. However, not each two-qubit quantum system can be decomposed into the sum product of $\ket{\psi}$ and $\ket{\phi}$. For example, $\ket{\Psi}=\ket{00}+\ket{11}$. A necessary and sufficient condition for the equation $\ket{00}+\ket{11}=\ket{\psi}\otimes\ket{\phi}$ is that $\alpha\gamma=\beta\delta=1$ and $\alpha\delta=\beta\gamma=0$. However, there are no four real or complex parameters that can satisfy this condition. According to the entanglement definition discussed above, the state $\ket{00}+\ket{11}$ is an entangled state.

There are four famous entangled states composed by a pair of entangled qubits, often referred to as \textit{Bell state} or \textit{EPR pair}, which are written as
\begin{equation}
	\begin{aligned}
		\ket{\Phi^{\pm}}&=\frac{1}{\sqrt{2}}(\ket{00}\pm\ket{11}), \\
		\ket{\Psi^{\pm}}&=\frac{1}{\sqrt{2}}(\ket{10}\pm\ket{01}), \\
	\end{aligned}
\label{Bell-state}	
\end{equation}
where $\ket{\Phi^{+}}$, $\ket{\Phi^{-}}$, $\ket{\Psi^{+}}$, and $\ket{\Phi^{-}}$ are four Bell bases of two one-half spin particle systems, and each Bell base carries non-local two-bit information: a parity bit and a phase bit. According to Eq.~\eqref{Bell-state}, we can get that each component of the Bell state is determined by two qubits' states, i.e., $\ket{0}$ and $\ket{1}$. Thus, these two qubits in the entangled system show a common non-local correlation: once we measure one of them to obtain quantum information about its state, we can know the state of the another qubit no matter how far apart they are. There is a measurement operation applied to the Bell state, known as Bell state measurement (BSM). BSM is the measurement over Bell bases. For example, by performing the BSM operation on a single qubit and an entangled qubit separated from an entangled system in the Bell state, the quantum system composed of the single qubit and an entangled qubit will be transformed into one of the four Bell states. The transformation process is essential to the implementation of the entanglement-based quantum teleportation. In this survey, we present some representative schemes for preparing EPR pairs and the application of the BSM operation Section~\ref{entanglement-generation}.

As discussed in Section~\ref{Sec2.1}, a quantum system can be represented by multiple qubits. Hence, the entangled system can also contain multiple entangled qubits. Typically, there are two maximum entangled states in a composite quantum system containing three or more entangled qubits: the GHZ state and the W state. The GHZ state was proposed in 1989, and the landmark event was that Greenberger, Horne, and Zeilinger (GHZ) went beyond Bell inequality, demonstrating that entanglement of more than two quantum particles leads to a contradiction with the local hidden variable model in 1989~\cite{greenberger1989going}. The W state was proposed by D{\"u}r \textit{et al.} in 2000~\cite{dur2000three}. Here, we use the entangled system containing three entangled qubits as an example to describe the GHZ state and the W state, i.e., the three-qubit GHZ state and the three-qubit W state, the state vectors of them are written as
\begin{equation}
	\begin{aligned}
	&\ket{GHZ}=\frac{1}{\sqrt{2}}(\ket{000}+\ket{111}),\\
	&\ket{W}=\frac{1}{\sqrt{3}}(\ket{001}+\ket{010}+\ket{100}).\\
	\end{aligned}
\label{three-entangled}	
\end{equation}
Currently, three-qubit entangled systems have been successfully implemented experimentally \cite{eibl2004experimental,bourennane2004experimental}. With the development of quantum hardware, high-dimensional entangled systems will show vast advantages in the application of quantum information technology in the future.

Entanglement is one of the most interesting properties in quantum mechanics and characterizes strong correlations between multiple local constituents. The strong correlation prompts entangled states to be used as a physical ``resource'', i.e., something costly to allows the implementation of valuable transformations. The valuable transformation plays a pivotal role in many of the most interesting quantum applications, such as quantum computing and quantum communication. In a nutshell, most ground-breaking quantum applications require distant quantum nodes to establish entangled correlations by sharing EPR pairs~\cite{pirandola2016physics}. Besides, entanglement provides crucial tools for networking numerous quantum information processors due to non-local correlation~\cite{santra2021quantum,van2012quantum}. Hence, entanglement is essential to an entanglement-assisted quantum network. In this survey, we mainly discuss how numerous quantum information processors are interconnected using entanglement resources to effectively support various quantum applications and how to utilize limited entanglement resources (or control entanglement-based quantum operations) to improve the performance of quantum networks.

\subsection{Quantum Decoherence and Fidelity}
Quantum systems are very fragile: unstable excited atoms transformed from stable atoms absorbing energy will decay spontaneously, and a single atom will spontaneously flip and change its state. An open quantum system will inevitably interact with the noisy environment, which contributes to the irreversible effect on the state of the quantum system \cite{zeh1970interpretation}. Specifically, the interaction between a quantum system and the noisy environment introduces perturbations to the quantum system, leading to partial leakage of the system's state information to the environment, i.e., the quantum properties of the quantum system are lost. In quantum mechanics, the gradual decay of the state of a quantum system is known as \textit{quantum decoherence}, sometimes also called \textit{dynamical decoherence or environment-induced decoherence} \cite{zurek1982environment,joos2013decoherence}. As discussed in Section \ref{Sec2.1}, the state of a quantum system is determined by its probability amplitudes. Hence, the probability amplitudes will change when a quantum system undergoes quantum decoherence in a noisy environment~\cite{zurek2003decoherence}. As a result, the measurement of a decoherent quantum system will not generate the desired results, thereby causing errors in quantum information.

We can quantify the degree of quantum decoherence by using \textit{fidelity} to measure the information overlap between two quantum states, especially the decoherent quantum state and the initial quantum state. In this survey, fidelity is mainly used to describe the degree of information overlap between the decoherent and initial states of an entangled system. Generally, fidelity (denoted as $F$) can be defined as \cite{jozsa1994fidelity}
\begin{equation}
	F=\sqrt{\bra{\psi}\rho\ket{\psi}},
\end{equation}
where $0\leq F\leq 1$ and $\rho$ is the density matrix of the quantum state $\ket{\psi}$. For a multi-qubit quantum system, its density matrix can be written as
\begin{equation}
	\rho=\sum_{i}p_{i}\ket{\psi}_{i}\bra{\psi}_{i},
\end{equation}
where $p_{i}$ is the probability amplitude of observing the quantum system in the pure state $\ket{\psi}_{i}$, and  $\ket{\psi}_{i}\bra{\psi}_{i}$ is the outer product of two state vectors $\ket{\psi}_{i}$ and $\bra{\psi}_{i}$. Most notably, quantum states can be either ``pure'' or ``mixed''. Pure state means that it is possible to write the quantum state in state vector form rather than the quantum state containing only one term of the superposition state. For example, $\ket{\psi}=\ket{0}$ and $\ket{\psi}=\frac{1}{\sqrt{2}}(\ket{0}+\ket{1})$ are both pure states, and the mixed Bell state is usually called Werner state~\cite{werner1989quantum}. According to the definition of fidelity, we can know that a quantum system's fidelity will gradually decay from 1.0 to 0 as it interacts with the noisy environment, including transmission channels, quantum memory, and measurement devices.

To better understand the decay process of quantum systems, we use two sets to describe quantum decoherence and fidelity: we abstractly denote the initial quantum state and the decoherent quantum state as the set $A$ containing $m$ elements and the set $A'$, respectively. The decoherence of the initial quantum state can be regarded as the process in which elements in the set $A$ are gradually deleted. After a certain period of quantum decoherence, the set $A$ becomes the set $A'$ containing $k$ elements. We can use the union of two sets to correspond to the information overlap between two quantum states. That is, the ratio of $k$ to $m$, i.e., $0 \leq \frac{k}{m} \leq 1$, corresponds to the quantum system's fidelity. The number of deleted elements in $A$ can be regarded as quantum information lost by the quantum system during quantum decoherence. The more quantum information is lost, the greater the error rate introduced by quantum decoherence in quantum systems.

\begin{figure}[t]
	\centering
	\includegraphics[width=1.0\linewidth]{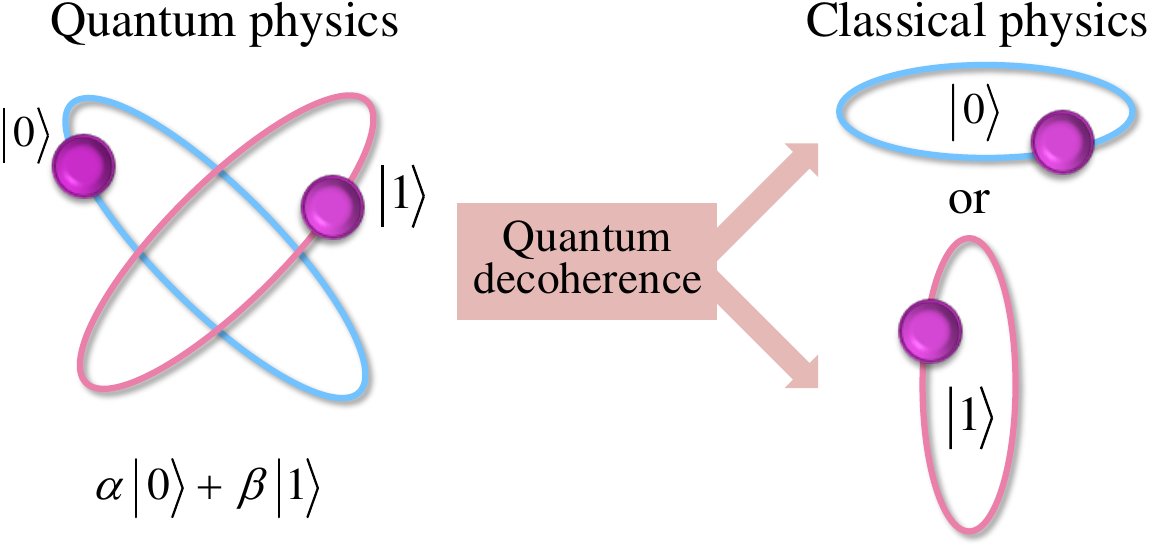}
	\caption{Quantum to classical transition caused by quantum decoherence.}
	\label{decoherence}
\end{figure}

Quantum decoherence is one of the main obstacles to the application of quantum information technology. This is because quantum decoherence decays the probabilistic features of quantum systems. When fidelity equals zero, a quantum system's probabilistic features are completely lost. In other words, quantum decoherence drives the quantum behavior of a quantum system to become classical, the process of which is called \textit{``quantum-to-classical transition''} (as shown in \textbf{Fig.~\ref{decoherence}}). The higher the fidelity, the less information about the initial quantum system is lost during the interaction between the quantum system and the noisy environment. Similar to classical information theory, quantum information theory also shows that the loss of system information can lead to errors. Hence, high fidelity means fewer errors are introduced in quantum computing and quantum communication. The decoherence phenomenon significantly destroys the unique properties of quantum mechanics, thus negatively affecting the performance of quantum applications. Hence, a quantum technology that can mitigate the negative influence of quantum decoherence is required in entanglement-assisted quantum networks, and we will discuss this technology in Section \ref{entanglement-purification}.

\begin{table}[htbp]
	\centering
	\caption{The comparison between classical gates and Quantum  gates.}
	\label{gatecomparision}
	\begin{tabular}{m{2.5cm}<{\centering}m{2.5cm}<{\centering}m{2.5cm}<{\centering}}
		\toprule
		\textbf{Topics}      &  \textbf{Classical Gates} & \textbf{Quantum Gates}  \\
		\midrule
		\textbf{Fundamental Unit}           &  Bit                & Qubit             \\
		\textbf{Gates}                      &  Logic gates        & Unitary gates     \\
		\textbf{Universal Gates (Example)}  &  NAND, NOR          & H, CNOT  \\
		\textbf{Algebra}                    &  Boolean            & Linear            \\
		\textbf{Church-Turing Thesis}       &  Strong supports    & Possibly violates \\
		\textbf{Gates Reversible}           &  Sometimes          & Always           \\
		\bottomrule
	\end{tabular}
\end{table}

\subsection{Quantum Gates}
The processing of quantum information requires controlling, manipulating, and measuring qubits. The basic operations applied to qubits are called quantum gates \cite{hey1999quantum}. Quantum gates are usually embedded in quantum circuits, the model for quantum computation, in which a computation is a sequence of quantum gates, measurements, initializations of qubits to known values, and possibly other actions, to manipulate qubits. Mathematically, a quantum gate is an operator or transformation matrix for the state vector of a quantum system. That is, the state vector $\ket{\psi_{i}}$ can be transformed to $\ket{\psi_{f}}$ through a unitary operator $\widehat{U}$, i.e., $input: \ket{\psi_{i}} \xrightarrow{\widehat{U}} output: \ket{\psi_{f}}$. \textbf{Table \ref{gatecomparision}} summarizes some differences between classical gates and quantum gates \cite{kitaev2002classical}. Classical gates are logic gates applied to binary bit strings and are intrinsically Boolean that strongly adhere to the Church-Turing thesis \cite{copeland1997church}. Besides, only some classical gates are reversible, e.g., the classical NOT gate. However, quantum gates violate the Church-Turing thesis and are linear unitary operations applied to qubits. Quantum gates manipulate superposition states (or qubits) and follow the principle of unitary operation. In other words, a quantum gate is a linear transformation that maintains the total probability of the quantum system as one. For example, there is a quantum gate, i.e., the identity matrix $I$, and the quantum state $\ket{G}=\ket{0}+\ket{1}$. Applying $I$ to $\ket{G}$, we can get the generated state $\ket{G'}$
\begin{equation*}
	\ket{G'}=I\ket{G}=
	\begin{bmatrix}
		1 & 0 \\
		0 & 1
	\end{bmatrix}
	\begin{bmatrix}
	1 \\
	1
	\end{bmatrix}
=
	\begin{bmatrix}
	1  \\
	1
\end{bmatrix}.
\end{equation*}
Obviously, $\ket{G'}$ is a linear transformation of $\ket{G}$ and they have the same inner product, that is, a quantum gate is essentially a reversible unitary transformation.

\begin{table}[htbp]
\centering
\caption{Single-qubit Gates}
\label{single-qubit}
\begin{tabular}{m{2.5cm}<{\centering} m{2.5cm}<{\centering} m{2.5cm}<{\centering}}
\toprule
\textbf{Quantum Gates}  &\textbf{Notation}& \textbf{Matrix}  \\
\midrule
\textbf{Pauli-X (X)}  & \begin{minipage}[b]{0.22\columnwidth}
    \centering
    {\includegraphics[width=0.8\textwidth]{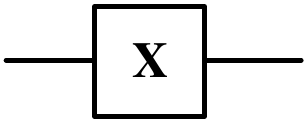}}
\end{minipage}   &
\begin{center}
    $\left[
   \begin{array}{cc}
    0 & \ 1 \\
    1 & \ 0
   \end{array}
\right]$
\end{center}
\\
\textbf{Pauli-Y (Y)}   & \begin{minipage}[b]{0.22\columnwidth}
    \centering
    {\includegraphics[width=0.8\textwidth]{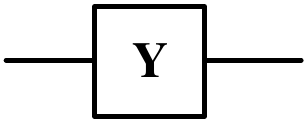}}
\end{minipage}   &
\begin{center}
   $\left[
   \begin{array}{cc}
   0 & -i \\
   i & 0 \\
   \end{array}
   \right]$
\end{center}
\\
\textbf{Pauli-Z (Z)}  & \begin{minipage}[b]{0.22\columnwidth}
    \centering
    {\includegraphics[width=0.8\textwidth]{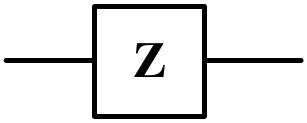}}
\end{minipage}   &
\begin{center}
   $\left[
   \begin{array}{cc}
   1 & 0 \\
   0 & -1\\
   \end{array}
   \right]$
\end{center}
\\
\textbf{Hardamard (H)}   & \begin{minipage}[b]{0.22\columnwidth}
    \centering
    {\includegraphics[width=0.8\textwidth]{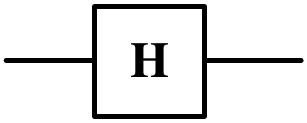}}
\end{minipage}   &
\begin{center}
   $\frac{1}{\sqrt{2}}\left[
   \begin{array}{cc}
   1 & 1 \\
   1 & -1 \\
   \end{array}
   \right]$
\end{center}
\\
\bottomrule
\end{tabular}
\end{table}

\begin{table}[htbp]
	\centering
	\caption{Truth table of sing-qubit gates.}
	\label{Gate-Truth}
	\begin{tabular}{m{0.7cm}<{\centering}m{1.15cm}<{\centering}m{1.55cm}<{\centering}m{1.2cm}<{\centering}m{2.2cm}<{\centering}}
		\toprule
		\textbf{Topic} & \textbf{Pauli-X} & \textbf{Pauli-Y} & \textbf{Pauli-Z} & \textbf{H} \\
		\midrule
		\textbf{Input}  & $\ket{0}$ or $\ket{1}$ & $\ket{0}$ or $\ket{1}$ & $\ket{0}$ or $\ket{1}$ & $\ket{0}$ or $\ket{1}$ \\
		\textbf{Output} & $\ket{1}$ or $\ket{0}$ & $i\ket{1}$ or -$i\ket{0}$ & $\ket{0}$ or -$\ket{1}$ & $\frac{1}{\sqrt{2}}(\ket{0}+\ket{1})$ or $\frac{1}{\sqrt{2}}(\ket{0}-\ket{1})$\\
		\bottomrule
	\end{tabular}
\end{table}

There are various quantum gates that can be used to process quantum information. However, according to quantum information theory, any unitary operation applied to a quantum system can be constructed from the combination of the basic single-qubit gates and the two-qubit \textit{controlled-NOT} (known as \textit{CNOT}) gate \cite{wilde2013quantum}. Therefore, in this survey, we only discuss some typical single-qubit gates and the CNOT gate.

\textbf{Single-qubit Gates.} The single-qubit gates mainly include \textit{Pauli} gates and the \textit{Hardamard} gate (also called \textit{H} gate). The notations and matrix presentations of single-qubit gates are shown in \textbf{Table \ref{single-qubit}}. Similar to classical gates, a quantum gate represents a quantum information processing unit with input and output. A quantum gate can change the state vector of qubits. Hence, the output of the processing unit will exhibit a quantum state change compared to the input. Specifically, the truth table of single-qubit gates is shown as \textbf{Table \ref{Gate-Truth}}. Based on this truth table, we can get the transformation of a qubit. As shown in Eq.~\eqref{transformation}, we use a single qubit described in two-dimensional Hilbert space as the input of quantum gates to present the function of single-qubit quantum gates as follows:
\begin{equation}
	\centering
	\begin{aligned}
		\alpha\ket{0}+\beta\ket{1} &\xrightarrow{X} \beta\ket{0}+\alpha\ket{1}, \\
		\alpha\ket{0}+\beta\ket{1} &\xrightarrow{Y} i(\beta\ket{0}-\alpha\ket{1}),  \\
		\alpha\ket{0}+\beta\ket{1} &\xrightarrow{Z} \alpha\ket{0}-\beta\ket{1},\\
		\alpha\ket{0}+\beta\ket{1} &\xrightarrow{H} \alpha\frac{\ket{0}+\ket{1}}{\sqrt{2}}+\beta\frac{\ket{0}-\ket{1}}{\sqrt{2}}.\\
	\end{aligned}
	\label{transformation}	
\end{equation}

\begin{figure}[t]
	\centering
	\subfigure[Pauli-X gate]{
		\includegraphics[width=0.95\linewidth]{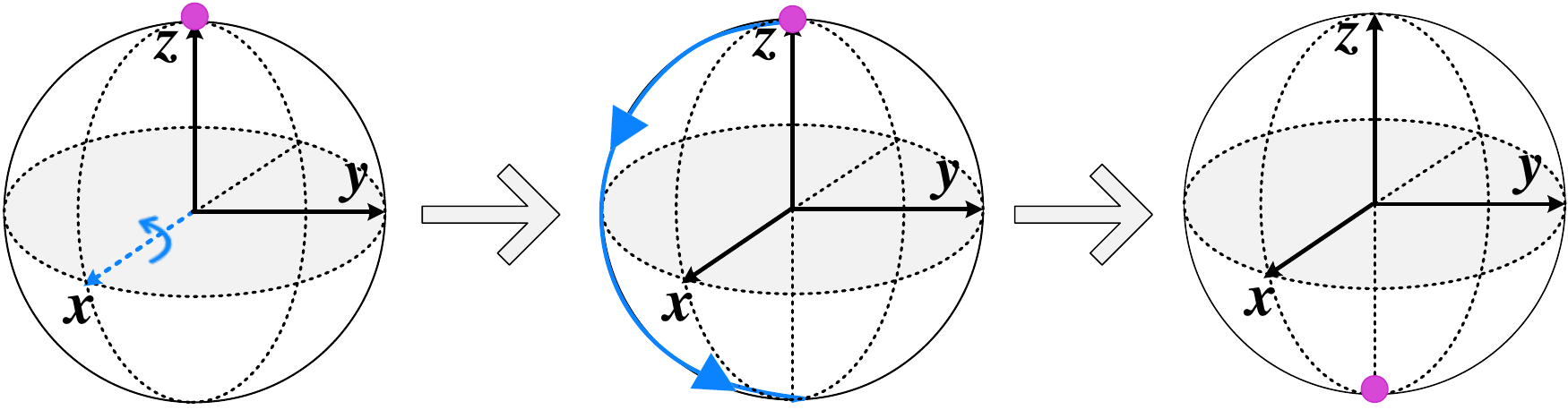}}
	\quad
	\subfigure[Pauli-Y gate.]{
		\includegraphics[width=0.95\linewidth]{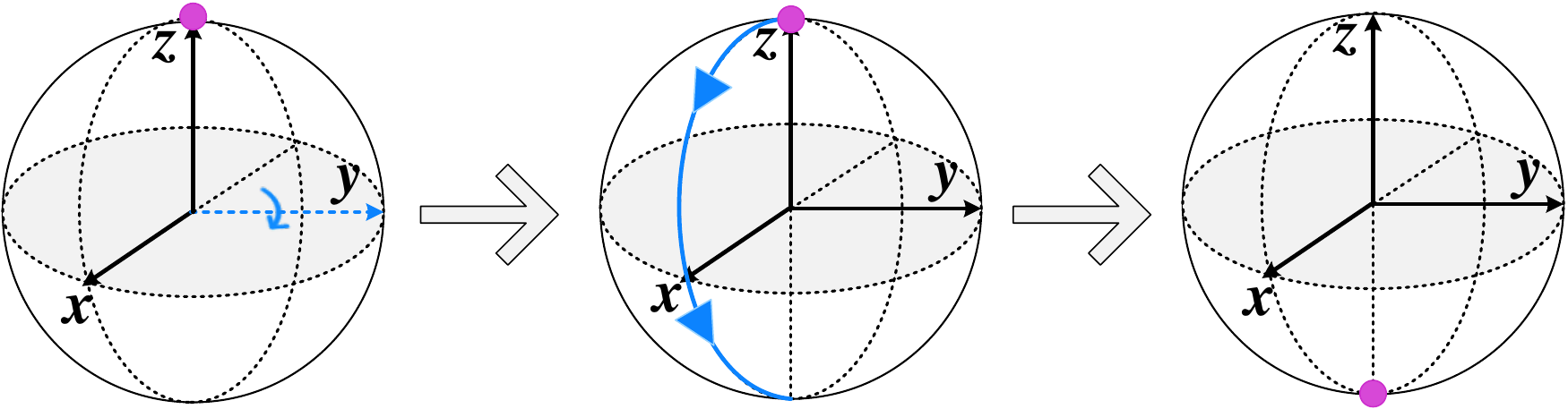}}
	\quad
	\subfigure[Pauli-Z gate.]{
		\includegraphics[width=0.95\linewidth]{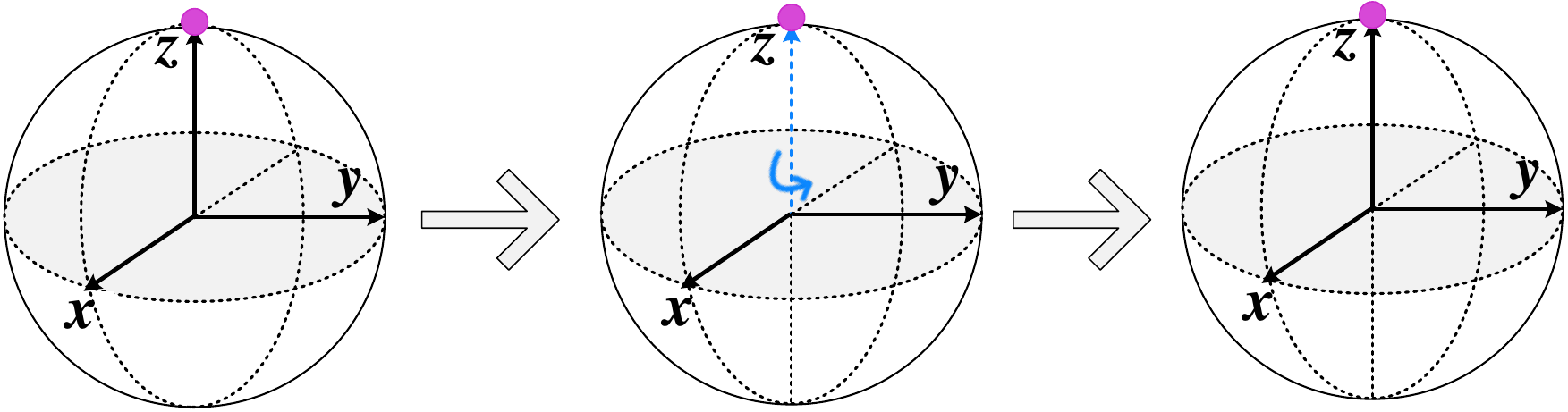}}
	\quad
	\subfigure[H gate.]{
		\includegraphics[width=0.95\linewidth]{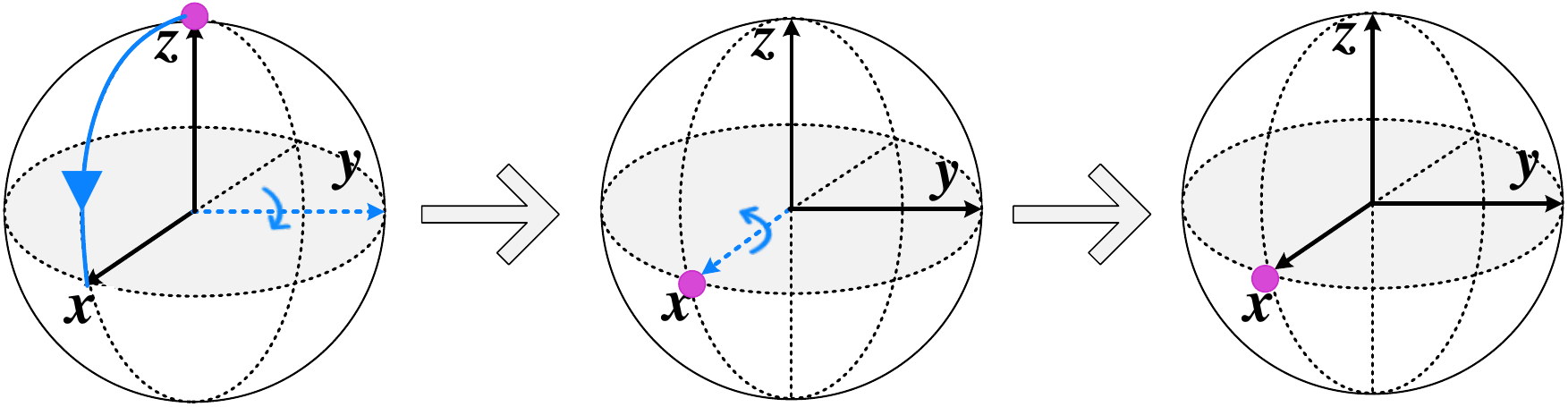}}
	\caption{The rotation effects of different single-qubit gates on a Bloch sphere.}
	\label{Gate-Bloch}
\end{figure}

In addition, since a single qubit can also be geometrically represented by a Bloch sphere, the influence of applying a single-qubit gate to a single qubit can be regarded as the rotation of the point on the Bloch sphere. More concretely, the \textit{Pauli-X} gate, also called the NOT gate, represents a transformation with an \ang{180} rotation along the X-axis. The \textit{Pauli-Y} gate means the \ang{180} rotation along the Y-axis. The \textit{Pauli-Z} gate represents the \ang{180} rotation along the Z-axis. The H gate is defined by a combination of two rotational actions, i.e., a \ang{90} rotation along the Y-axis first and then \ang{180} rotation along the X-axis. For example, by applying different single-qubit gates to the qubit $\ket{0}$, the rotations of $\ket{0}$ in the Bloch sphere are shown in \textbf{Fig.~\ref{Gate-Bloch}}.

\begin{table}[t]
	\centering
	\caption{controlled-NOT Gate}
	\label{CNOT-Gate}
	\begin{tabular}{m{2.5cm}<{\centering} m{2.5cm}<{\centering} m{2.5cm}<{\centering}}
		\toprule
		\textbf{Quantum Gates}  & \textbf{Notation} & \textbf{Matrix} \\
		\midrule
		\textbf{CNOT}  & \begin{minipage}[b]{0.22\columnwidth}
			\centering
			{\includegraphics[width=0.8\textwidth]{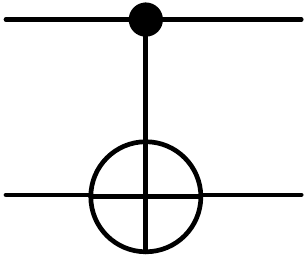}}
		\end{minipage}   &
		\begin{center}
			$\left[
			\begin{array}{cccc}
				1 & 0 & 0 & 0 \\
				0 & 1 & 0 & 0 \\
				0 & 0 & 0 & 1 \\
				0 & 0 & 1 & 0
			\end{array}
			\right]$
		\end{center}
		\\
		\bottomrule
	\end{tabular}
\end{table}

\textbf{Two-qubit Gates.} The most well-known two-qubit gate is the CNOT gate. The CNOT gate can be utilized to create a strong correlation between two incoherent qubits. The notation and matrix representation of the CNOT gate are shown in \textbf{Table ~\ref{CNOT-Gate}}. The CNOT gate's matrix is a tensor product of two $2\times2$ matrices and can convert one four-dimensional vector into another. As a quantum information processing unit, the CNOT gate requires two qubits, a control qubit and a target qubit, as input. Here, we assume that the control and target qubits are $x$ and $y$, respectively, i.e., the input of the CNOT gate is $\ket{xy}$. The feature of the CNOT gate is that the target qubit is inverted as the output only when the control qubit is $\ket{1}$. Concretely, if the input of the CNOT gate is $\ket{00}$ or $\ket{01}$, the output is the original state. Otherwise, $\ket{10}$ and $\ket{11}$ will be transformed to $\ket{11}$ and$\ket{10}$, respectively. According to the above discussion, we can use the CNOT gate to generate entanglement between two incoherent qubits with the help of the H gate. For example, \textbf{Table~\ref{Bell_generation}} presents the implementation of generating four Bell states using the H and CNOT gates. Here, we use two qubits, $\ket{0}$ and $\ket{0}$, as the input of the quantum circuit to illustrate the production of a Bell state. Firstly, we apply the H gate to $\ket{0}$. As a result, $\ket{0}$ is transformed into $\frac{1}{\sqrt{2}}(\ket{0}+\ket{1})$, and the quantum system composed by $\frac{1}{\sqrt{2}}(\ket{0}+\ket{1})$ and the second qubit $\ket{0}$ is presented by $\frac{1}{\sqrt{2}}(\ket{00}+\ket{10})$. Then we apply the CNOT gate to two qubits by regarding $\frac{1}{\sqrt{2}}(\ket{0}+\ket{1})$ as the control qubit and $\ket{0}$ as the target qubit. Consequently, the quantum system, $\frac{1}{\sqrt{2}}(\ket{00}+\ket{10})$, is transformed into $\frac{1}{\sqrt{2}}(\ket{00}+\ket{11})$, i.e., the Bell state $\ket{\Phi^{+}}$.

\begin{table}[t]
	\centering
	\caption{The generation of Bell states.}
	\label{Bell_generation}
	\begin{tabular}{m{1.7cm}<{\centering} m{2.8cm}<{\centering} m{3.0cm}<{\centering}}
		\toprule
		\textbf{Input}  &\textbf{Quantum Circuit} & \textbf{Output} \\
		\midrule
		$\ket{0}$ and $\ket{0}$  & \multirow{4}{*}{
			\begin{minipage}[b]{0.25\columnwidth}
				\centering
				{\includegraphics[width=1.0\textwidth]{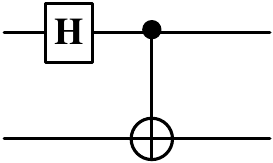}}
			\end{minipage} }  &
		$\ket{\Phi^{+}}=\frac{1}{\sqrt{2}}(\ket{00}+\ket{11})$\\
		$\ket{0}$ and $\ket{1}$& &$\ket{\Psi^{+}}=\frac{1}{\sqrt{2}}(\ket{01}+\ket{10})$\\
		$\ket{1}$ and $\ket{0}$& &$\ket{\Phi^{-}}=\frac{1}{\sqrt{2}}(\ket{00}-\ket{11})$\\
		$\ket{1}$ and $\ket{1}$& &$\ket{\Psi^{-}}=\frac{1}{\sqrt{2}}(\ket{01}-\ket{10})$\\
		\bottomrule
	\end{tabular}
\end{table}
	                                                                	
\section{Enabling Technologies}\label{Sec3}
This section first reviews some enabling technologies that support the interconnection and intercommunication of quantum nodes in entanglement-assisted quantum networks, ranging from point-to-point entanglement distribution to qubit transmission, e.g., quantum dense coding, quantum teleportation, entanglement purification, quantum error correction, entanglement swapping, and quantum memories. Besides, the functions of these enabling technolog are discussed. In particular, for each enabling technology, we first describe its working principle, and then the implementation steps are provided with the help of a quantum circuit. Moreover, we present the development status of these enabling technologies based on the principles of theory and experimental results to demonstrate that entanglement-assisted quantum networks will be built not too far from now.
	
\subsection{Entanglement Preparation and Distribution}\label{entanglement-generation}
As discussed in Section \ref{Sec2.2}, entanglement plays a crucial role in quantum information transmission since communicating parties need to share entangled qubit pairs. Hence, a quantum technology that enables quantum nodes to be entangled is required to act as the building block for entanglement-assisted quantum networks. In general, establishing entanglement between adjacent quantum nodes is realized by two pivotal operations: entanglement preparation and entanglement distribution \cite{cirac1997quantum}. Especially, entanglement preparation aims at generating entangled qubits, and entanglement distribution enables the prepared entangled qubits to be shared by spatially separated quantum nodes with the assistance of quantum channels. Here, we introduce three typical schemes that can establish entanglement between adjacent quantum nodes by preparing and distributing entangled qubits.

\begin{figure}[t]
	\centering
	\includegraphics[width=0.96\linewidth]{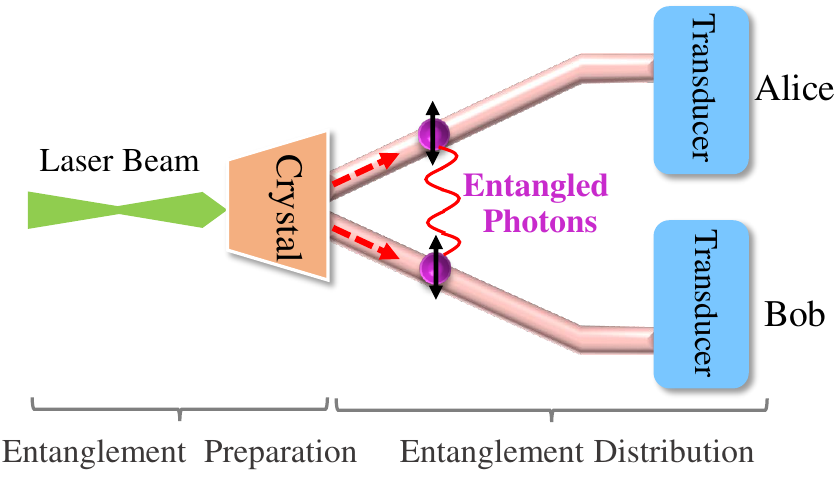}
	\caption{Entanglement distribution between Alice and Bob based on parametric down-conversion.}
	\label{distribution-polarization}
\end{figure}

\textbf{Parametric Down-conversion Scheme.} The first scheme is realized with the aid of the nonlinear-crystal-based  Spontaneous Parametric Down-Conversion (SPDC) process. Due to the availability of high-efficiency polarization-control elements and the relative insensitivity of most materials to birefringent thermally induced drifts, polarized photons are usually adopted in experiments to generate entangled qubits. The well-known method for generating polarization-entangled photon pairs is realized via the down-conversion process to share non-classical correlations \cite{hariharan199649Quantum}. In a nutshell, a laser beam is directed toward a nonlinear crystal, which occasionally splits photon beams into pairs of polarization-entangled photons. Based on this idea, Kwiat \textit{et al.} \cite{kwiat1995new} proposed a simple entanglement preparation scheme relying on noncollinear type-II phase matching, called type-II parametric down-conversion. In this way, true polarization-entangled photon pairs can be produced directly out of a single nonlinear crystal~\cite{kwiat1994proposal}. \textbf{Fig. \ref{distribution-polarization}} presents the implementation of the type-II parametric down-conversion scheme. Specifically, a laser beam is pointed toward a nonlinear crystal, and the down-converted photons are emitted into two cones with the help of type-II phase matching. At the two intersections of two cones, a pair of polarization-entangled photons is generated in one of the four Bell states. Experimentally, this scheme has been maturely employed to demonstrate quantum dense coding, teleportation, and a postselection-free test of Bell's inequality \cite{strekalov1996postselection}. Currently, the SPDC process based on non-linear optical materials is still a hot research topic in entanglement preparation. The future direction of the SPDC-based entanglement light source is to reduce the loss, increase the purity and degree of entanglement, and combine with micro and nanophotonic devices to improve the scalability and practicality of the entanglement light source.

\begin{figure}[htbp]
	\centering
	\includegraphics[width=1.0\linewidth]{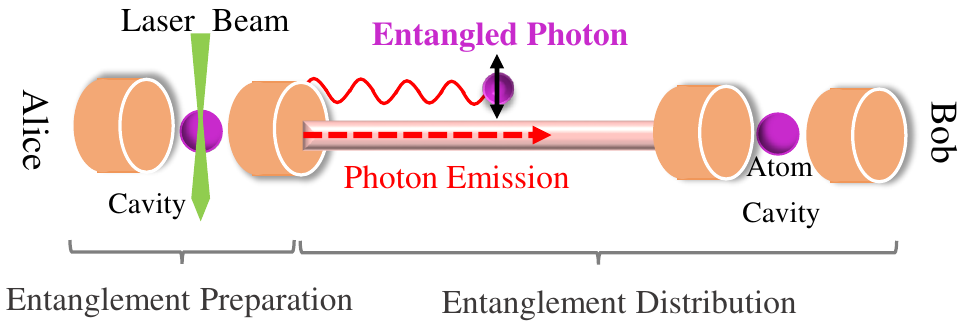}
	\caption{Entanglement distribution between Alice and Bob based on single atom excitation.}
	\label{distribution-single}
\end{figure}

\textbf{Single-Atom Excitation Scheme.} Another scheme conceived for preparing and distributing entangled qubits between two spatially isolated quantum nodes is implemented based on single-atom excitation \cite{matsukevich2006entanglement,ritter2012elementary,welte2018photon}. As shown in \textbf{Fig. \ref{distribution-single}}, this scheme utilizes atoms tightly coupled with optical cavities to build entanglement between two quantum nodes directly linked by a photonic channel. Specifically, the quantum node Alice transfers the internal state of an atom to the optical state of the cavity mode by employing a laser beam. In other words, an atom is first excited by a laser beam at Alice, and the emitted photon becomes entangled with the atom's internal state. Then, the atom-entangled photon is released from Alice's cavity, travels along the photonic channel, and enters another cavity at the quantum node Bob. In Bob's cavity, the photon is absorbed coherently, and its polarization is mapped onto the internal state of an atom. As a result, two atoms located at Alice and Bob are entangled remotely.

\begin{figure}[htbp]
	\centering
	\includegraphics[width=1.0\linewidth]{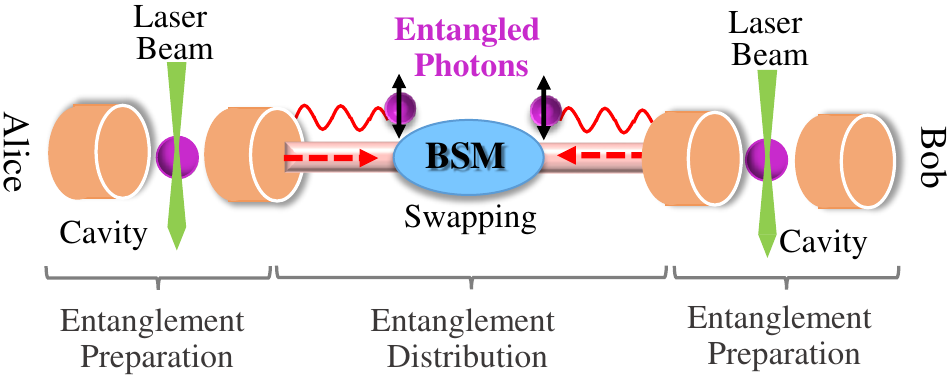}
	\caption{Entanglement distribution between Alice and Bob based on two atoms excitation simultaneously.}
	\label{distribution-two}
\end{figure}

\textbf{Two-Atom Excitation Scheme.} The third entanglement distribution scheme is shown in \textbf{Fig.~\ref{distribution-two}}. This scheme is implemented based on two simultaneous atoms' excitation~\cite{olmschenk2009quantum,ferrari2010entanglement}. Firstly, two atoms are excited by laser beams at Alice and Bob simultaneously, which leads each local cavity to emit a photon entangled with the corresponding atom. Then, both atom-entangled photons depart from the local cavity, propagate as a wave packet along the quantum channel, and reach a beam splitter where the BSM operation is performed to realize entanglement swapping discussed in Section \ref{Sec3.5}. After the BSM operation, the atom located at Alice establishes entanglement with Bob's atom, i.e., a pair of entangled atoms is distributed to Alice and Bob. The two-atom excitation scheme has also been proposed in the context of Nitrogen-Vacancy (NV) defect centers in diamonds \cite{bernien2013eralded}. Compared to the single-atom excitation scheme, the two-atom excitation scheme can effectively extend the distance of entanglement distribution with the assistance of a third party. However, the current implementation of the two-atom excitation scheme requires two quantum nodes to be symmetrically linked to a third party responsible for performing BSM operations and to distribute entangled qubits at the same time, which significantly hinders the application of this scheme in entanglement-assisted quantum networks.

Here, we comprehensively compare three schemes described above in terms of achievable distance, scalability, applicability, maturity, and field trials. \textit{(1) Achievable distance:} Although all three schemes utilize the transmitted entangled photons to establish entanglement between Alice and Bob, the specific “location” of the entanglement preparation varies among them. As a result, these schemes show different performances in terms of achievable distance. Concretely, with the assistance of an intermediate device, the two-atom excitation and parametric down-conversion schemes perform well and are better than the single-atom excitation scheme in terms of the achievable distance. \textit{(2) Scalability:} The parametric down-conversion scheme is mainly realized by easy-to-operate optical devices, so it exhibits better scalability than the atom-based excitation scheme. Besides, the single-atom excitation scheme is easier to scale than the two-atom excitation scheme because it does not require a third party to perform BSM operations. \textit{(3) Applicability:} Entangled systems generated by the atom-based excitation scheme have a longer lifetime than the parametric down-conversion scheme. Hence, the atom-based excitation scheme facilitate the storage and processing of entangled qubits. As a result, the atom-based excitation scheme are more applicable than the parametric down-conversion scheme. Moreover, the single-atom scheme performs better than the two-atom excitation scheme in applicability since it does not require two endpoints to perform operations simultaneously. \textit{(4) Maturity:} Because the parametric down-conversion scheme is easy to implement at the current quantum technical level, it is more mature than the atom-based excitation schemes. Besides, the single-atom excitation scheme performs better than the two-atom excitation scheme since it does not require a third party to perform BSM operations. \textit{(5) Field trial:} The implementation of the parametric down-conversion scheme has moved out of the laboratory. However, the atom-based excitation scheme is mainly demonstrated in the laboratory. In summary, each of three schemes has its advantages and disadvantages, and their practical application still needs to be explored.

So far, numerous experiments have been reported on entanglement distribution in the 1.5-$u$m wavelength band over optical fiber \cite{takesue2004generation,li2005storage,marcikic2004distribution,takesue2006long,hubel2007high,zhang2008distribution}. Besides, the record of entanglement distribution distance is constantly being broken. In 2009, \cite{fedrizzi2009high} realizes entanglement distribution between adjacent quantum nodes over 144km in free space. \cite{dynes2009efficient} achieves entanglement distribution over 200km of optical fiber. Inagaki \textit{et al.} \cite{inagaki2013entanglement} successfully conducted an experiment distributing entangled qubit pairs over 300km of fiber. Satellite-based entanglement distribution over 1200 km was implemented by Pan's group in 2017 \cite{yin2017satellite}. Furthermore, entanglement distribution between two quantum memories has been implemented~\cite{lipka2021entanglement,liu2021heralded}. In order to promote the application of entanglement distribution in a real-world quantum network, some ground-breaking works have been successfully experimented \cite{ciurana2014entanglement,wang2021experimental}. Although entanglement distribution between quantum nodes directly linked through a quantum channel has been demonstrated many times, efficient entanglement distribution in a realistic environment still requires more work to be done. Fortunately, the second quantum technology revolution promotes the development of quantum hardware, thus facilitating the implementation of efficient entanglement distribution in entanglement-assisted quantum networks.

\begin{figure}[t]
	\centering
	\includegraphics[width=1.0\linewidth]{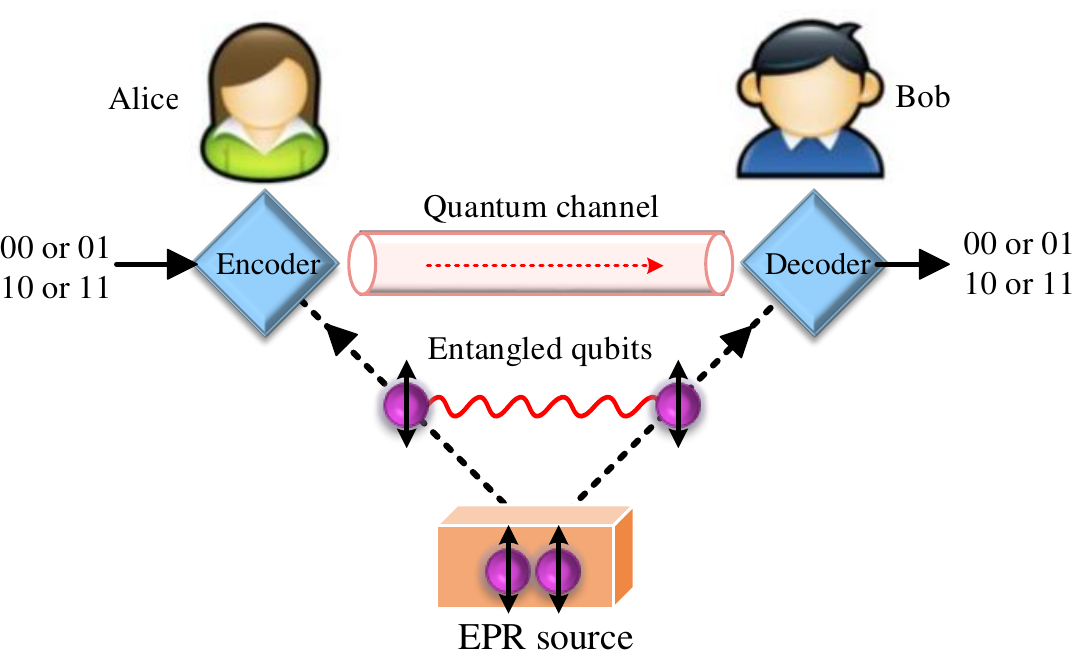}
	\caption{The implementation of quantum dense coding.}
	\label{dense-coding}
\end{figure}

\subsection{Quantum Dense Coding}\label{quantum-dense-coding}
In classical information theory, the upper limit of the amount of classical information a channel can transmit is called the channel's capacity \cite{kullback1997information}. However, with the assistance of quantum entanglement and the superposition principle, the amount of classical information transmitted through a quantum channel will far exceed this upper limit. To realize channel capacity improvement, Bennett first theoretically proposed a quantum dense coding scheme in 1992 \cite{bennett1992communication}, and its emergence breaks the Holevo Boundary~\cite{holevo1973bounds,guo2019advances}, i.e., a single qubit can carry the classical information of up to one classical bit. Bennett's scheme shows that two classical bits can be transported from Alice to Bob, who is entangled with Alice, only at the cost of transferring a single qubit. In this way, the channel capacity of the entanglement-assisted classical information transmission method is twice as much as the original. Formally, quantum dense coding is a quantum-based communication technology that uses entanglement characteristics to transport classical bits.

The implementation of quantum dense coding utilizing Bell-state entangled systems is shown in \textbf{Fig. \ref{dense-coding}}. Alice and Bob first share a pair of entangled qubits (i.e., Bell state) distributed by an EPR source through a quantum channel. Then, Alice encodes two classical bits (00, 01, 10, or 11) by applying corresponding quantum manipulation to the entangled qubit she owns locally and sends this entangled qubit to Bob through a quantum channel. Upon receiving the encoded entangled qubit, Bob performs a local BSM operation on the two entangled qubits to decode the transmitted classical bits. Consequently, Bob can obtain the two classical bits Alice wants to send according to the final state of the entangled system he owns. Unlike that QKD technology can only distribute random keys used to secure classical communication, quantum dense coding can directly transmit a specific string of binary bits between two communicating parties using entangled qubit pairs, further improving the security of classical communication.

The implementation of quantum dense coding involves four steps: entanglement distribution, encoding, sending, and decoding. The four steps are described in detail with the help of the quantum circuit -- a qubit processing collection composed of acyclically connected quantum gates -- depicted in \textbf{Fig.~\ref{dense-coding-circuit}} as follows:
\begin{enumerate}
\item [1)] Entanglement distribution aims to establish entanglement between two communicating parties. After the preparation of entangled state, the Bell state $\ket{\Phi^{+}}=\frac{1}{\sqrt{2}}(\ket{00}+\ket{11})$ is generated. Then, two entangled qubits are distributed to Alice and Bob through a quantum channel, respectively. As a result, Alice and Bob share the Bell state $\ket{\Phi^{+}}$.

\begin{figure}[htbp]
	\centering
	\includegraphics[width=1.0\linewidth]{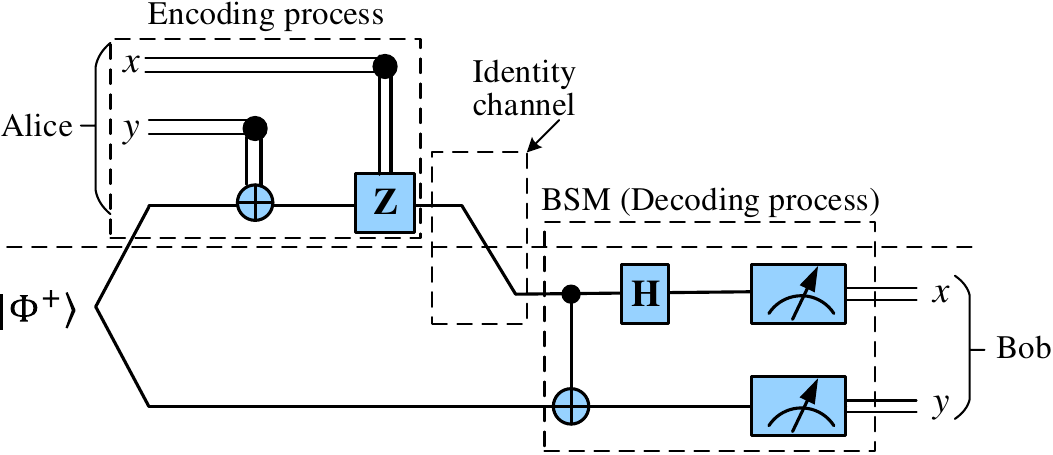}
	\caption{The quantum dense coding circuit.}
	\label{dense-coding-circuit}
\end{figure}
\item [2)] Alice encodes her two bits ($xy$) of classical information by applying quantum gate to her entangled qubit locally. With this, the entangled state $\ket{\Phi^{+}}$ can be transformed into any of four Bell states, i.e., $\ket{\Phi^{+}}$,$\ket{\Phi^{-}}$,$\ket{\Psi^{+}}$, and $\ket{\Psi^{-}}$. Formally, the corresponding relationships between the classical information and quantum gate can follow the following rules. If Alice wants to send $00$ to Bob, then she adopts an identity quantum gate to process her entangled qubit. In this way, there is no change in the Bell state $\ket{\Phi^{+}}$. If Alice wants to send classical two-bit string $01$ to Bob, then she uses Pauli-X gate to process her entangled qubit. In this way, the bit flip occurs on the Bell state $\ket{\Phi^{+}}$. The result is that Bell state $\ket{\Phi^{+}}$ transforms into
$\ket{\Phi^{-}}=\frac{1}{\sqrt{2}}(\ket{00}-\ket{11})$. If Alice wants to send $10$ to Bob, then she applies Pauli-Z gate to her entangled qubit. In this way, the Bell state undergoes phase flip. As a result, the Bell state becomes
$\ket{\Psi^{+}}=\frac{1}{\sqrt{2}}(\ket{01}+\ket{10})$. If Alice wants to send classical two-bit string $11$ to Bob, then she adopts $Z*X$ quantum gate to manipulate her entangled qubit. Consequently,  $\ket{\Phi^{+}}$ transforms into
$\ket{\Psi^{-}}=\frac{1}{\sqrt{2}}(\ket{01}-\ket{10})$.
	
\item [3)] After having performed one of four encoding processes mentioned above, Alice sends her entangled qubit to Bob through an identity quantum channel.
\begin{table}[hbtp]
	\centering
	\caption{The relationship between $xy$ and Bell state.}
	\label{xyvsBell}
	\begin{tabular}{m{2cm}<{\centering}m{4.5cm}<{\centering}}
		\toprule
		\textbf{\textit{xy}} &\textbf{Resultant Bell State} \\
		\midrule
		00 & $\frac{1}{\sqrt{2}}(\ket{00}+\ket{11})$ \\
		01 & $\frac{1}{\sqrt{2}}(\ket{00}-\ket{11})$   \\
		10 & $\frac{1}{\sqrt{2}}(\ket{01}+\ket{10})$ \\
	    11 & $\frac{1}{\sqrt{2}}(\ket{01}-\ket{10})$  \\
	    \bottomrule
	\end{tabular}
\end{table}

\item [4)] Upon receiving the entangled qubit sent by Alice, Bob performs the BSM operation to get the transmitted classical bits. The BSM operation consists of three steps. Firstly, Bob applies the CNOT unitary operation to his entangled qubit. Then, he applies the H gate to Alice's entangled qubit. Finally, the resultant Bell state will be one of the four entangled states, and Bob can obtain the state of the final entangled system by measuring two entangled qubits over Bell bases. According to the corresponding relationships (as shown in \textbf{Table \ref{xyvsBell}}) between $xy$ and the resultant Bell state, Bob can easily obtain the classical information sent by Alice.
\end{enumerate}

With the development of quantum information technology, quantum dense coding has made a great progress in theoretical study and experimental demonstration. So far, many quantum dense coding schemes have been proposed. Gorbachev~\cite{gorbachev2000teleportation} and Cereceda \cite{cereceda2001quantum} proposed a quantum dense coding scheme based on the GHZ state, respectively. Shimizu \textit{et al.} \cite{shimizu1999dense} proposed a scheme for enhancing the channel capacity to more than two classical bits in quantum dense coding that involves transmitting a polarization-entangled twin photon and a subsequent joint measurement with a Bell-state analyzer--an experimental apparatus performing projective measurement onto maximally entangled two-qubit state. Hao \textit{et al.} \cite{hao2001controlled} introduced controlled quantum dense coding to increase the information capacity by controlling the entanglement of a quantum channel, and Huang \textit{et al.} \cite{huang2009controlled} came up with a scheme for multi-party secure communication based on controlled dense coding. Liu \textit{et al.} \cite{liu2002general} proposed a general scheme for superdense coding among multiple parties. In the experiment, researchers adopted various experimental schemes to verify the superiority of quantum dense coding technology in improving information capacity \cite{mattle1996dense,fang2000experimental,yan2004scheme,lan2004quantum,ye2005scheme}. \cite{hu2018beating} realizes a channel capacity of $2.09$ using high-dimensional entanglement based quantum superdense coding.

\begin{table*}[hbtp]
	\centering
	\caption{ The comparison between different quantum dense coding schemes.}
	\label{dense-coding-schemes}
	\begin{tabular}{m{1.5cm}<{\centering}m{3.0cm}<{\centering}m{4.5cm}<{\centering}m{4.2cm}<{\centering}m{3.0cm}<{\centering}}
		\toprule
		\textbf{Topics} &\textbf{Schemes} & \textbf{Participants}& \textbf{Implementation cost}& \textbf{Channel capacity} \\
		\midrule
		\multirow{2}{*}{\textbf{Two-party}} & General schemes & Alice and Bob  & Maximally two-qubit entangled state &  2 bits \\
		
		&\multirow{2}{*}{Controlled schemes } & Alice, Bob, and Controller & GHZ state & $1+2\sin^{2}{\theta}$ bits \\
		& &Alice, Bob, and Controller & MS state& $1+\sin^{2}{(\theta-\gamma)}+\cos^{2}{\theta}$ bits \\	
		\multirow{3}{*}{\textbf{Multi-party}}  & Three-party schemes & Alice, Bob, and Charlie  & three-qubit entangled state & $2+2\alpha^2$ bits   \\
		& $(N+1)$-party schemes& $N$ senders and one receiver &$(N+1)$-qubit entangled state & $2N+2\alpha^2$ bits \\
		&\multirow{2}{*}{ Simultaneous dense coding }& Alice, Bob, and Charlie  &  two pair of Bell state &  total 4 bits  \\
		& & Alice, Bob, Charlie, and Controller & two pair of GHZ state & total 4 bits \\	
		\bottomrule
	\end{tabular}
\end{table*}

Based on the number of communicating parties, we can divide quantum dense coding schemes into two categories as shown in \textbf{Table \ref{dense-coding-schemes}}: two-party and multi-party quantum dense coding schemes. The two-party scheme contains general schemes represented by Bennett's scheme and controlled schemes. The general scheme can transmit 2 classical bits from Alice to Bob at the cost of a maximally two-qubit entangled state. The controlled scheme represented by Hao's scheme introduce a third party, i.e., controller, and use the GHZ state to implement quantum dense coding, and the average channel capacity of this scheme is $1+2\sin^{2}{\theta}$, where $\theta$ is a parameter used to determine the measurement base of the third party and $0\leq \theta \leq \frac{\pi}{4}$. Other controlled quantum dense coding schemes are realized by consuming certain partially entangled states called maximal slice (MS) states. The average channel capacity of dense coding is $1+\sin^{2}{(\theta-\gamma)}+\cos^{2}{\theta}$ using the MS-state-based controlled scheme, where $\theta$ is used to determine the measurement base of the third party, $\alpha$ is used to determine the MS state, $0\leq \theta \leq \frac{\pi}{4}$, and $0 \leq \gamma \leq \frac{\pi}{2}$~\cite{liu2016controlled}. The multi-party quantum dense coding scheme can be divided into three categories: three-parity schemes, $(N+1)$-party schemes, and simultaneous dense coding schemes. The three-party scheme represented by Cheng's scheme use a three-qubit entangled state to transmit a total of $2+2\alpha^2$ classical bits from senders to a receiver, where $\alpha$ is used to determine the three-qubit entangled system's state, and $\alpha \leq \frac{\sqrt{2}}{2}$ \cite{wei2007quantum}. Expand three-party schemes to $N+1$ parties, and channel capacity can be up to $2N+2\alpha^2$~\cite{bose1998multiparticle}. Notably, there is only one receiver in three-party schemes and $(N+1)$)-party schemes. If there are multiple receivers and one sender, the multiple-party scheme can also be called simultaneous dense coding schemes. The Bell-state-based three-party simultaneous dense coding scheme can transmit a total of 4 bits between participants~\cite{situ2010simultaneous}. Similar to controlled two-party schemes, a total of 4 bits can be transmitted by introducing a controller in GHZ-state-based controlled simultaneous schemes represented by Situ's scheme~\cite{situ2014controlled}. By comparing these schemes, we can find that multi-party schemes perform well in channel capacity, but these schemes consume more entangled qubits than two-party schemes and are more challenging to implement. In summary, quantum dense coding is becoming a mature quantum technology. However, the practical application of these quantum dense coding schemes need to be more thoroughly explored to provide secure communication in an entanglement-assisted quantum network.
	
\subsection{Quantum Teleportation}\label{quantum-teleportation}

In addition to QKD and quantum dense coding, \textit{quantum teleportation} is another vital technology to realize secure communication. The concept of quantum teleportation was first proposed by Bennett \textit{et al.} \cite{bennett1993teleporting} in 1993, which pioneered the exploration of using classical communication and entangled qubits to directly teleport unknown qubits from one node to another, and it was demonstrated experimentally in 1997 \cite{bouwmeester1997experimental,boschi1998experimental}. Formally, quantum teleportation is a quantum technology that utilizes classical communications and the properties of quantum entanglement to transmit quantum information between two communicating parties, even if they are not connected via quantum channels~\cite{pirandola2015advances}. In other words, Bob, who shares a Bell state with Alice, can make local perfect ``copy'' of the unknown qubit Alice wants to transmit based on classical information sent by Alice, thus realizing quantum inforamtion transmission between two communicating parties without suffering quantum channel noise.

\begin{figure}[t]
	\centering
	\includegraphics[width=1.0\linewidth]{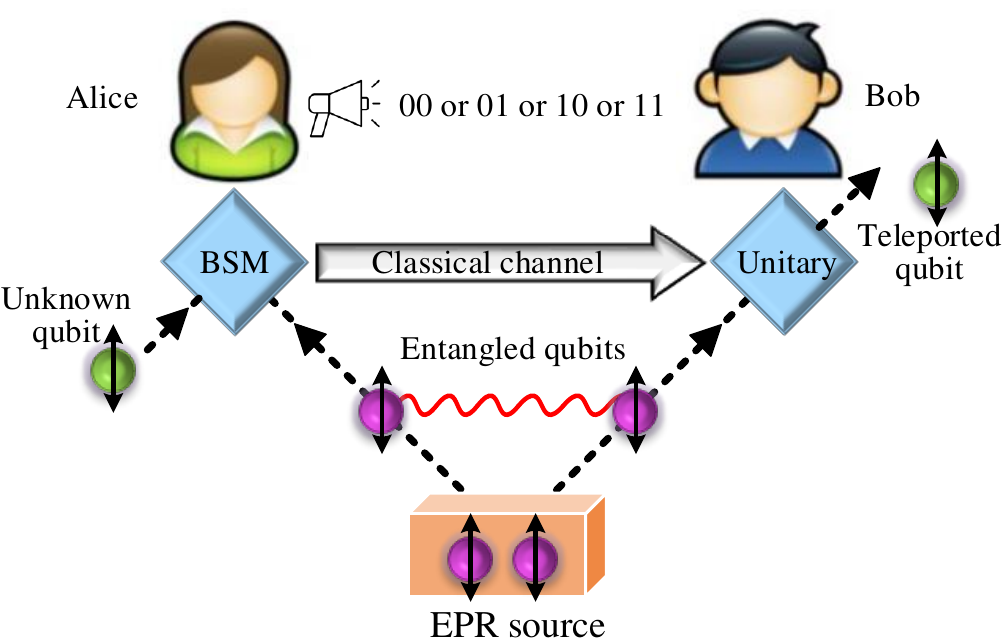}
	\caption{The implementation of quantum teleportation.}
	\label{teleportation}
\end{figure}
	
As described in Section~\ref{Introduction} and Section~\ref{quantum-dense-coding}, QKD and quantum dense coding can only realize the transmission of classical information (random keys and specific binary bits) using quantum mechanics' properties. In other words, QKD and quantum dense coding are usually used as auxiliary tools to protect classical communication between Alice and Bob. QKD is easy to implement under existing technical conditions and is the first quantum-based secure communication technology to be deployed and used to provide security services. Unlike QKD technology, which aiming at distributing random keys between two communicating parties, quantum dense coding can implement the transmission of specific classical binary bits. However, its practical application is limited by the development of physical devices. Similar to quantum dense coding, quantum teleportation requires two communicating parties to share entangled qubits. However, quantum teleportation fundamentally differs from quantum dense coding since it realizes the transmission of quantum information rather than classical information. Besides, the state of the teleported qubits is not subject to collapse-after-measurement during quantum teleportation. Hence, quantum teleportation play a more pivotal role in entanglement-assisted quantum networks than QKD and quantum dense coding.
	
In order to achieve the ``copy'' of qubits, all information of an unknown qubit must be divided into classical information and quantum information and sent to Bob via the classical channel and quantum channel, respectively. Notably, quantum information of an unknown qubit is not transmitted to Bob via a realistic physical channel. Due to the non-local correlation properties of Bell states \cite{bell1964physics,popescu1994quantum,tittel1998experimental}, the quantum information of an unknown qubit can be instantaneously ``transferred'' to Bob's entangled qubit after the BSM operation. Hence, the implementation of quantum teleportation only requires classical information to be transmitted through classical channels after two communicating parties share entangled states, i.e., the transmission of an unknown qubit free from the interference of quantum channel noise. \textbf{Fig. \ref{teleportation}} depicts a standard quantum teleportation system. After entanglement distribution, Alice and Bob share Bell state, i.e., each of them has an entangled qubit locally. Then, Alice performs a BSM operation on the unknown qubit she wants to send to Bob and the local entangled qubit together and sends the measurement result (two classical bits 00, 01, 10, or 11) to Bob via classical channels. Finally, according to the measurement outcomes, Bob applies the corresponding unitary operation to his entangled qubit to obtain a ``copy'' of the teleported qubit, i.e., the state information of the teleported qubit is mapped to Bob's local entangled qubit. Hence, with the help of quantum teleportation, the unknown qubit can be teleported from Alice to Bob, no matter how far apart they are. Most notably, although quantum teleportation allows quantum information to be teleported without a real physical channel, the unknown qubit does not travel faster than light because classical communication is indispensable for quantum teleportation. Besides, although the teleported qubit does not directly suffer from channel noise, channel noise will lead to a low-fidelity entangled system shared by two communicating parties, thus affecting the success probability of quantum teleportation. Hence, it is required to realize high-fidelity entanglement distribution between quantum nodes to efficiently realize qubits transformation.

Suppose that Alice and Bob share an entangled qubit pairs in Bell state, $\ket{\Phi^{+}}=\frac{1}{\sqrt{2}}(\ket{00}+\ket{11})$. Besides, Alice prepares a third qubit $\ket{\psi}=\alpha\ket{0}+\beta\ket{1}$. $\ket{\psi}$ is an unknown single qubit. In other words, Alice does not know the values of $\alpha$ and $\beta$. As shown in \textbf{Fig.~\ref{teleportation-circuit}}, the required steps for teleporting $\ket{\psi}$ from Alice to Bob by adopting quantum teleportation technology are discussed as follows with the aid of a quantum circuit:
\begin{figure}[t]
	\centering
	\includegraphics[width=1.0\linewidth]{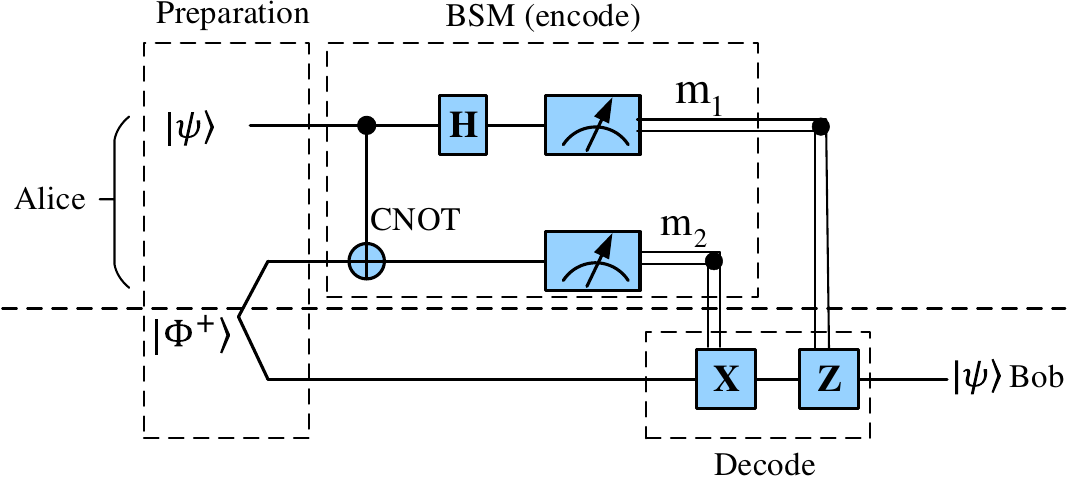}
	\caption{The quantum teleportation circuit.}
	\label{teleportation-circuit}
\end{figure}
\begin{enumerate}
\item [1)] In the quantum teleportation system, three qubits form a new Hilbert space, i.e., the tensor product of the unknown qubit and the Bell state. More concretely, Alice takes the local unknown qubit $\ket{\psi}$ together with the shared Bell state $\ket{\Phi^{+}}$ as the initial state of the quantum teleportation system, and the initial state is written as:
\begin{equation}
	\begin{aligned}
	\ket{\Psi_{0}}&=\ket{\psi}\otimes\ket{\Phi^{+}}\\
	&=\frac{1}{\sqrt{2}}(\alpha\ket{0}+\beta\ket{1})\otimes(\ket{00}+\ket{11}) \\
	&=\frac{1}{\sqrt{2}}[\alpha(\ket{000}+\ket{011})+\beta(\ket{100}+\ket{111})].
	\end{aligned}
\label{Initial-system}
\end{equation}
		
\item [2)] A CNOT gate is applied to Alice's two local qubits. In this local operation, the unknown qubit is used as the control qubit and the entangled qubit is used as the target qubit for the input of the CNOT gate. After performing CNOT process, the result of the quantum teleportation system is
\begin{equation}
	\begin{aligned}
		\ket{\Psi_{1}}&=CNOT\ket{\Psi_{0}}\\
		&=\frac{1}{\sqrt{2}}[\alpha(\ket{000}+\ket{011})+\beta(\ket{110}+\ket{101})].
		\end{aligned}
		\label{CNOT}
\end{equation}

\item [3)] Then, Alice applies the H gate to her local unknown qubit. In this local operation, $\ket{0}$ and $\ket{1}$ are mapped to $\frac{1}{\sqrt{2}}(\ket{0}+\ket{1})$ and $\frac{1}{\sqrt{2}}(\ket{0}-\ket{1})$, respectively. As a result, the transformation of the Eq.~\eqref{CNOT} is as follows:
\begin{equation}
	\begin{aligned}
	\ket{\Psi_{2}}&=H\ket{\Psi_{1}}\\
		&=\frac{1}{2}[\ket{00}(\alpha\ket{0}+\beta\ket{1})+\ket{01}(\alpha\ket{1}+\beta\ket{0})+ \\
		&\ket{10}(\alpha\ket{0}-\beta\ket{1})+\ket{11}{\alpha\ket{1}-\beta\ket{0}}].
	\end{aligned}
	\label{Hadamard}
\end{equation}
		
\item [4)]  After the above step, the measurement operation collapses the state of two qubits into one of four Bell states. In this step, Alice measures her two qubits jointly. The state of Bob's qubit is determined by Alice's measurement result $m_{1}m_{2}$. The corresponding relationship between $m_{1}m_{2}$ and the state of Bob's qubit is presented in \textbf{Table~\ref{m1m2vsqubit}}. If $m_{1}m_{2}=00$, Bob's qubit is in state $\alpha\ket{0}+\beta\ket{1}$. If Alice's measurement result is $01$, the qubit Bob possesses is in state $\alpha\ket{1}+\beta\ket{0}$. If $m_{1}m_{2}$ equals $10$, Bob's qubit is in state $\alpha\ket{0}-\beta\ket{1}$. Otherwise, Bob's state is in state $\alpha\ket{1}-\beta\ket{0}$. The measurement operation yields two classical bits, which are then sent to Bob through classical channels.
		
\begin{table}[hbtp]
\centering
\caption{The relationship between $m_{1}m_{2}$ and Bob's qubit.}
\label{m1m2vsqubit}
\begin{tabular}{m{2cm}<{\centering}m{4.5cm}<{\centering}}
	\toprule
	\textbf{$m_{1}m_{2}$} &  \textbf{Bob's Qubit} \\
	\midrule
    00  & $\alpha\ket{0}+\beta\ket{1}$ \\
    01  & $\alpha\ket{1}+\beta\ket{0}$  \\
    10  & $\alpha\ket{0}-\beta\ket{1}$  \\
    11  & $\alpha\ket{1}-\beta\ket{0}$  \\
    \bottomrule
\end{tabular}
\end{table}
		
\item [5)] According to the measurement result feedback by Alice, Bob applies the corresponding single-qubit unitary gate to his local qubit to obtain the initial unknown qubit $\ket{\psi}$. The corresponding operation applied by Bob can be represented as $Z^{m_{1}}X^{m_{2}}$, where $m_{1}$ and $m_{2}$ are the classical bits that result from Alice's measurement. For example, if Alice's measurement result is $10$, i.e., $m_{1}=1$ and $m_{2}=0$, Bob only needs to perform the $X$ transformation on his qubit. In a nutshell, the $X$ and $Z$ gates are applied conditionally to Bob's local qubit. As a result, the quantum formation of $\ket{\psi}$ can be completely transferred to Bob's qubit, thus recovering $\ket{\psi}$.
\end{enumerate}
	
Experimentally, quantum teleportation has achieved many breakthrough results and records. \cite{ursin2004quantum} increases the distance of quantum teleportation to 600 meters using optical fiber. Subsequently, the distance record of quantum teleportation has gradually increased from 16 km \cite{jin2010experimental} to 143 km~\cite{ma2012quantum}. The distance record using superconducting nanowire detectors for quantum teleportation reaches 102 km \cite{takesue2015quantum}. For material systems, the distance record is 21 meters \cite{nolleke2013efficient}. \cite{takeda2013deterministic} achieves the ``fully deterministic'' quantum teleportation in 2013. \cite{zhao2004experimental} demonstrates the ``open-destination'' quantum teleportation using five-photon entanglement. \cite{zhang2006experimental} realizes the teleportation of a composite quantum state of two qubits in 2004. \cite{feng2022quantum} proposes a scheme for transferring information by quantum teleportation. Pan's group~\cite{wang2015quantum,xia2017long} carried out the first experiment teleporting multiple degrees of freedom of a quantum particle. Recently, Valivarthi \textsl{et al.} \cite{valivarthi2020teleportation} achieved quantum teleportation over 44 km with a fidelity exceeding 90\% in 2020. Besides, \cite{ren2017ground} realizes ground-to-satellite quantum teleportation, which is an essential step towards a global-scale quantum internet. Moreover, qubit teleportation between non-neighboring quantum nodes~\cite{hermans2022qubit} and imperfect quantum dots~\cite{basso2021quantum} has been implemented. Currently, researchers have demonstrated quantum teleportation in the existing optical fiber network \cite{valivarthi2016quantum,sun2016quantum} and realized chip-to-chip quantum teleportation~\cite{llewellyn2020chip}. These encouraging breakthroughs make it convincing that quantum teleportation can be effectively implemented to provide services for secure communication and quantum computing in future entanglement-assisted quantum networks.

\subsection{Entanglement Purification}\label{entanglement-purification}

\textit{Entanglement purification} was originally introduced in the context of quantum communication as a solution to the problem of long-distance communication over noisy quantum channels \cite{dur2007entanglement}. Because qubits are extremely fragile, noise in quantum channels and the interaction with an uncontrollable environment, including quantum memory and measurement devices, have the effect that the expected entangled qubits are produced only with a certain non-unit fidelity. As a result, the pure entangled system will decay into a mixed entangled system after being distributed over noisy quantum channels, thus resulting in various errors in quantum information processing \cite{bennett1996mixed}. Notably, high-fidelity entangled systems shared by distant quantum nodes are crucial to achieving high-performance and reliable quantum applications. For example, fidelity directly affects the accuracy of measurements in quantum sensing. Higher fidelity can provide more accurate measurement results, thereby improving the reliability and accuracy of the application. An entangled system with low fidelity will result in increased inaccuracy or unpredictability of quantum sensing. Hence, a design that can protect quantum information is required in entanglement-assisted quantum networks. The most known method is entanglement purification, where quantum information is protected by improving the quality of the entangled system.

\begin{figure}[t]
	\centering
	\includegraphics[width=1.0\linewidth]{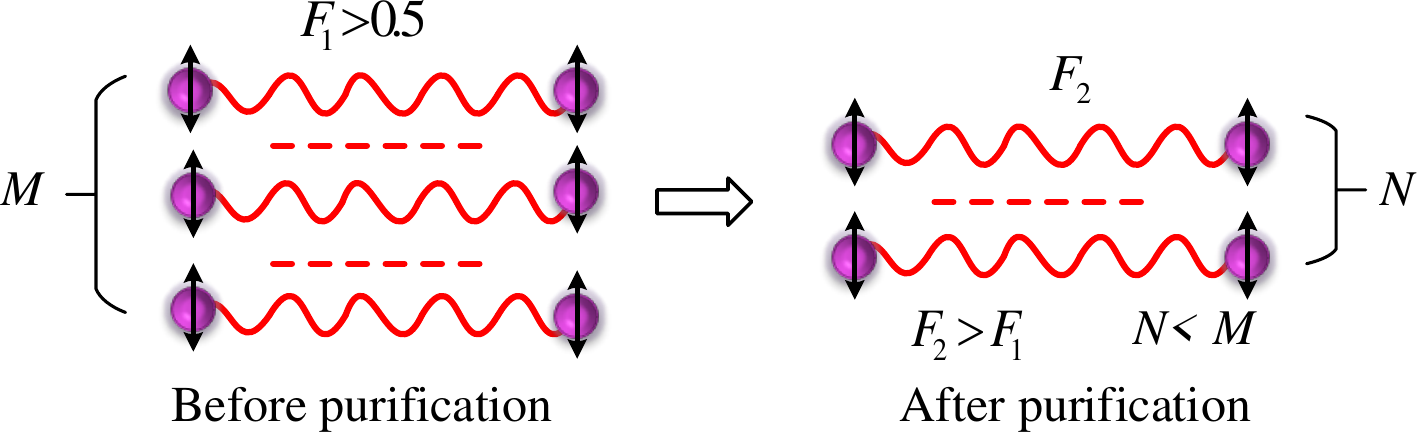}
	\caption{The function of entanglement purification.}
	\label{purification}
\end{figure}

Entanglement purification is a process of extracting the maximally entangled state from some partially entangled states using local measurement and classical communication (LOCC). That is, entanglement purification is a powerful tool to distill high-fidelity entanglement from low-quality entanglement ensembles, which plays a crucial role in long-distance quantum information transmission~\cite{Duan2001Long}. In a nutshell, entanglement purification can improve entangled systems' fidelities at the cost of a reduced number of entangled qubit pairs. As shown in \textbf{Fig.~\ref{purification}}, there are $M$ copies of non-maximally entangled states with fidelity $F_{1}$ before entanglement purification. During entanglement purification, these entangled qubit pairs are manipulated in such a way that a fewer number of copies with a reduced amount of noise are produced. As a result, $N$ pairs of entangled qubits with fidelity $F_{2}$ are generated from the original $M$ non-maximally entangled qubit pairs after entanglement purification, i.e., $N<M$ and $F_{1}<F_{2}$.

The most basic purification protocol is implemented based on two CNOT gates and is first proposed by Bennett \textit{et al.} in 1996 \cite{bennett1996purification}. As shown in \textbf{Fig. \ref{purification-circuit}}, the required steps for achieving entanglement purification using two CNOT gates are described as follows:
\begin{figure}[htbp]
	\centering
	\includegraphics[width=0.95\linewidth]{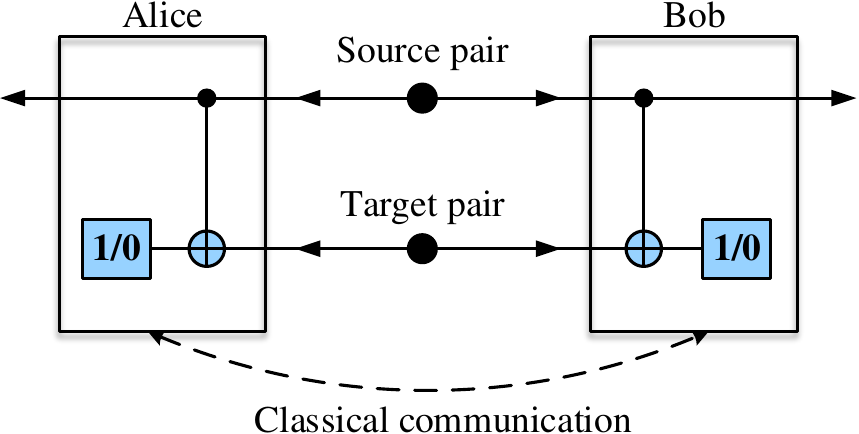}
	\caption{The implementation of entanglement purification based on two CNOT quantum gates.}
	\label{purification-circuit}
\end{figure}

\begin{enumerate}
\item [1)]  Preparing two entangled qubit pairs and performing local unitary operations. Both Bell states are in the state $\ket{\Phi^{+}}=\frac{1}{\sqrt{2}}(\ket{00}+\ket{11})$, and they constitute a pure entangled system. Then, distributing these two Bell states to two distant quantum nodes, Alice and Bob, through noisy quantum channels. The first operation in this purification protocol is to have Alice and Bob perform a random rotation on each shared pair, choosing a random rotation independently for each entangled qubit pair and applying it locally to both members of the entangled qubit pair. This transforms the initial general entangled state into a rotationally symmetric entangled system that can be presented by the Werner state of purity $F$:
\begin{equation}
	\begin{aligned}
		W_{F}&=F\ket{\Psi^{-}}\bra{\Psi^{-}}+\frac{1-F}{3}(\ket{\Psi^{+}}\bra{\Psi^{+}} \\
		&+\ket{\Phi^{+}}\bra{\Phi^{+}}+\ket{\Phi^{-}}\bra{\Phi^{-}}).
	\end{aligned}
\label{Werner}
\end{equation}

\item [2)] Alice and Bob apply a CNOT gate to their two locally entangled qubits, one of which acts as the control qubit and the other as the target qubit. Notably, using the CNOT gate to process qubits is equivalent to performing an XOR operation. Thus, after performing the CNOT gate, the target qubit will hold the parity of the control and target qubits. Then, Alice and Bob each calculate the parity of the two entangled qubits they hold. Traversing all 16 possible outcomes, we can get the truth table as shown in \textbf{Table~\ref{purificationtable}}.
Here, we assume that the states of two entangled qubit pairs shared by Alice and Bob are $\Phi^{+}$ and $\Phi^{-}$, respectively. The entangled system $\Phi^{+}$ consists of entangled qubits $e_1$ and $e_2$, and $\Phi^{-}$ consists of entangled qubits $e_3$ and $e_4$. We use $\Phi^{+}$ as the source pair and $\Phi^{-}$ as the target pair to illustrate the derivation of the truth table. In this hypothesis, two entangled qubits $e_1$ and $e_2$ act as control qubits, and $e_3$ and $e_4$ are target qubits. As discussed in Section II-D, the feature of the CNOT gate is that the target qubit is inverted as the output only when the control qubit is $\ket{1}$. Hence, the result of the operation that Alice and Bob perform CNOT gate simultaneously is presented as follows:
\begin{equation}
		\begin{aligned}
			&\ket{\Phi^{+}}\otimes\ket{\Phi^{-}} \\
			&=\frac{1}{\sqrt{2}}(\ket{00}+\ket{11})\otimes\frac{1}{\sqrt{2}}(\ket{00}-\ket{11})\\
			&\xrightarrow{}\frac{1}{2}\{\ket{00}(\ket{00}-\ket{10})+\ket{10}(\ket{00}-\ket{10})\}\\
			&\xrightarrow{}\frac{1}{2}\{\ket{00}\ket{00}-\ket{01}\ket{00}+\ket{10}\ket{00}-\ket{11}\ket{00}\}\\
			&\xrightarrow{CNOT}\frac{1}{2}\{\ket{00}\ket{00}-\ket{01}\ket{00}+\ket{11}\ket{00}-\ket{10}\ket{00}\}\\
			&=\frac{1}{2}\{\ket{00}\ket{00}-\ket{00}\ket{10}+\ket{10}\ket{10}-\ket{10}\ket{00}\}\\
			&=\frac{1}{2}\{\ket{00}(\ket{00}-\ket{10})-\ket{10}(\ket{00}-\ket{10})\}\\
			&=\frac{1}{2}(\ket{00}-\ket{10})(\ket{00}-\ket{10})\\
			&\xrightarrow{CNOT}\frac{1}{2}(\ket{00}-\ket{10})(\ket{00}-\ket{10})\\
			&=\frac{1}{\sqrt{2}}(\ket{00}-\ket{11})\otimes\frac{1}{\sqrt{2}}(\ket{00}-\ket{11})\\
			&=\ket{\Phi^{-}}\otimes\ket{\Phi^{-}}.
		\end{aligned}
\end{equation}
According to the results, we can know that target pair do no change and the source pair $\Phi^{+}$ becomes $\Phi^{-}$. Since there are four possible state for each of the target and source pairs, the truth table contains 16 outcomes.

\begin{table}[hbtp]
	\centering
	\caption{Truth value of entanglement purification.}
	\label{purificationtable}
	\begin{tabular}{m{1.6cm}<{\centering}m{1.6cm}<{\centering}m{1.6cm}<{\centering}m{1.7cm}<{\centering}}
		\toprule
		\multicolumn{2}{c}{\textbf{Before}} & \multicolumn{2}{c}{\textbf{After (n.c. = no change)}} \\
		Source & Target & Source & Target \\
		\midrule
		$\Phi^{\pm}$ & $\Phi^{+}$ & n.c. & n.c. \\
		$\Psi^{\pm}$ & $\Phi^{+}$ & n.c. & $\Psi^{+}$ \\
		$\Psi^{\pm}$ & $\Psi^{+}$ & n.c. & $\Phi^{+}$  \\
		$\Phi^{\pm}$ & $\Psi^{+}$ & n.c. & n.c. \\
		$\Phi^{\pm}$ & $\Phi^{-}$ & $\Phi^{\mp}$ &  n.c.\\
		$\Psi^{\pm}$ & $\Phi^{-}$ & $\Psi^{\mp}$ & $\Psi^{-}$ \\
		$\Psi^{\pm}$ & $\Psi^{-}$ & $\Psi^{\mp}$ &  $\Phi^{-}$ \\
		$\Phi^{\pm}$ & $\Psi^{-}$ & $\Phi^{\mp}$ &  n.c.\\
		\bottomrule
	\end{tabular}
\end{table}

\item [3)] Next, Alice and Bob measure the target qubit they hold and exchange the measurement outcomes with the aid of classical communication. If their measurement outcomes are the same, control qubits are retained. Otherwise, they discard the control qubits. By analysis, we can obtain the probability that the entangled state is $\ket{\Phi^{+}}$ after entanglement purification is $F^{2}+\frac{1}{9}(1-F)^{2}$. The probability that Alice and Bob have the same measurement outcomes is $F^{2}+\frac{2}{3}F(1-F)+\frac{5}{9}(1-F)^{2}$. Consequently, the fidelity of the entangled state is
\begin{equation}
		F^{'}=\frac{F^{2}+\frac{1}{9}(1-F)^{2}}{F^{2}+\frac{2}{3}F(1-F)+\frac{5}{9}(1-F)^{2}}.
		\label{fidelityafterpurification}
\end{equation}
According to \textbf{Eq. \eqref{fidelityafterpurification}}, we can easily draw the conclusion that the CNOT gate based entanglement purification scheme is valid only if the fidelity of the purified initial entangled states is larger than 0.5.
\end{enumerate}

We can regard the purification operation as a quantum information processing module with inputs and outputs. This module takes two low-fidelity entangled states as input and outputs a high-fidelity entangled state. Notably, the fidelities of two input entangled states affects the output fidelity. Let us consider a simple case where an entangled qubit pair has only suffered a bit flip error and the two entangled states input to the purification module are identical. In this case, the fidelity of the output entangled state after one round of entanglement purification can be written as $F^{'}=\frac{F^{2}}{F^{2}+(1-F)^{2}}$. Besides, the rounds of entanglement purification affect the final entangled state's fidelity. If an entangled state is purified by consuming $N$ entangled qubit pairs with fidelity $F$, the final fidelity $F^{n}$, can be iteratively calculated by 	
\begin{equation}
	F^{n}=\frac{FF^{(n-1)}}{FF^{(n-1)}+(1-F)(1-F^{(n-1)})}.
	\label{N-rounds-purification}
\end{equation}
where $F^{0}=F$. As described above, the fidelities of two input entangled states and the rounds of entanglement purification are closely related to the overhead of entanglement sources, significantly affecting networks' performance. Hence, there are two pivotal purification-related problems in entanglement-assisted quantum networks: scheduling purification operations and determining the rounds of entanglement purification. We will discuss these two issues in Section~\ref{Sec6}.

Until now, entanglement purification has made great progress both theoretically and experimentally. In 1996, Deutsch \cite{deutsch1996quantum} improved Bennet's scheme by using two additional unitary transformations to achieve entanglement purification. Murao \textit{et al.} \cite{murao1998multiparticle} proposed an improved CNOT-gate purification scheme for multi-particle systems in 1998. However, limited by the incompleteness of physical devices, the above two schemes are challenging to achieve the CNOT-gate operation. In this context, Pan \textit{et al.}~\cite{pan2001entanglement} proposed an entanglement purification scheme based on a simple optical element, the polarizing beamsplitter (PBS) \cite{sato1992laminated}, in 2001 and then demonstrated this scheme experimentally in 2003~\cite{pan2003experimental}. After that, many entanglement purification schemes based on linear optical elements were proposed~\cite{reichle2006experimental,dong2008experimental,hage2008preparation,takahashi2010entanglement}. Notably, the purification scheme mentioned above all have probabilistic constraints, i.e., they are  probabilistically implemented to generate an entangled state with fidelity of 1.0. Although some research groups have proposed improved purification schemes~\cite{sheng2008nonlocal,xue2001efficient,ballester2009entanglement}, these schemes, collectively referred to as progressive purification schemes, are still not free from probabilistic constraints. To overcome probabilistic constraints, Sheng \textit{et al.} \cite{sheng2010deterministic} proposed a two-step deterministic purification scheme in 2010. After that, the one-step deterministic purification schemes~\cite{sheng2010one,li2010deterministic,deng2011one} and the other deterministic purification schemes~\cite{morikoshi2000recovery,gu2006protocol,wang2009scheme} were successively proposed by this group.

\begin{table}[t]
	\centering
	\caption{The comparison between different entanglement purification schemes.}
	\label{purificationcomparision}
	\begin{tabular}{m{1.6cm}<{\centering}m{1.6cm}<{\centering}m{1.6cm}<{\centering}m{1.7cm}<{\centering}}
		\toprule
		\textbf{Topic} &\multicolumn{2}{c}{\textbf{Progressive Scheme}} & \textbf{Deterministic Scheme} \\
		& CNOT-based & PBS-based & \\
		\midrule
		\textbf{Efficiency}& Low & Low & High \\
		\textbf{Overhead} & High &High &Low \\	
		\textbf{Implementation} &Complex &Simple &Complex \\
		\textbf{Probabilistic} & Yes & Yes & No \\
 		\bottomrule
	\end{tabular}
\end{table}

Here, we comprehensively compare the existing entanglement purification schemes, i.e., progressive and deterministic schemes, in terms of efficiency, overhead, implementation, and probabilistic, as shown in \textbf{Table~\ref{purificationcomparision}}. There are mainly two progressive entanglement purification schemes: CNOT-based and PBS-based schemes. The CNOT-based scheme requires applying the CNOT gate to entangled qubit pairs. However, the low success probability of the CNOT-based operation is hardly sufficient for long-range quantum communication. Hence, the CNOT-based scheme is inefficient and requires more entanglement resources to be consumed for improving entanglement fidelity. Although the PBS-based scheme is easier to implement than the CNOT-based scheme, it needs to detect single photons during entanglement purification. As a result, the efficiency of the PBS-based scheme is very low (only half of the CNOT-based scheme) due to photon loss, thus resulting in higher entanglement resource overhead. Moreover, the PBS-based scheme is successfully implemented with probability. The probabilistic characteristic significantly limits the efficiency of entanglement purification. Unlike the two progressive entanglement purification schemes mentioned above, deterministic purification schemes can generate entangled states with a fidelity of 1.0 in a definite way. Moreover, these schemes perform well in terms of efficiency since they do not suffer from photon loss. Hence, deterministic purification schemes can improve entanglement fidelity at the cost of less entanglement resource overhead. Notably, the implementation of deterministic purification schemes usually requires high-precision optical devices, such as high-quality PBS, polarization rotators, phase shifters. However, these optical devices must have high stability and accuracy in operation, which is difficult to satisfy these requirements in a real-world environment. Therefore, deterministic purification schemes are experimentally difficult to implement.

Entanglement-assisted quantum networks require efficient and low-overhead entanglement purification schemes for high-quality entanglement distribution. On the one hand, it is challenging to successfully distribute entangled qubit pairs between adjacent quantum nodes due to the inherent loss and noise in the quantum channel. The success probability of the point-to-point entanglement distribution is usually negative exponential to the physical length of the quantum channel. As a result, entangled qubit pairs shared by adjacent quantum nodes are a scarce network resource. Entanglement purification schemes need to be as efficient as possible to improve the utilization of entanglement resources. On the other hand, the fidelity of the entangled state shared by two distant quantum end nodes is approximately equal to the product of the fidelity of the measured entangled states on the selected path. In order to generate high-quality end-to-end entanglement, single-hop entanglement usually needs to be purified to have high fidelity. Therefore, it is required to perform purification operations in multiple rounds between adjacent quantum nodes, contributing to more entanglement resources being consumed. Entanglement purification schemes with low overhead facilitate improving entanglement-assisted quantum networks' performance. As discussed above, although deterministic purification schemes are experimentally difficult to implement due to the imperfection of current physical devices, they perform better than CNOT-based and PBS-based progressive schemes in terms of efficiency and resource overhead. Hence, deterministic purification schemes have a broad application prospect, especially in view of the groundbreaking development in physical devices.

Currently, various entanglement purification schemes have been demonstrated in realistic systems~\cite{dur1999quantum,fujii2009entanglement,chen2017experimental,kalb2017entanglement}. Guo's group~\cite{hu2021long} implemented the efficient entanglement purification experiment over 11 km in 2021. Besides, the best fidelity that the entanglement purification experiment can achieve, implemented in 2022 \cite{yan2022entanglement}, is $94.09\pm0.98\%$. As a key role in high-performance qubit transmission, entanglement purification is expected to mature for practical application soon, thereby supporting high-quality qubit transmission in entanglement-assisted quantum networks.

\subsection{Quantum Error Correction}	
Quantum communication is the most sophisticated application of quantum information processing. However, the transmission of qubits over noisy physical channels inevitably leads to errors in quantum information, which considerably hinders the commercial application of quantum communication. Thus, how to transmit quantum information in a way such that its quantum feature can be sufficiently preserved is a vital problem. Referring to classical communication, the \textit{quantum error correction} (QEC) design \cite{knill1996concatenated,devitt2013quantum,roffe2019quantum} that can protect quantum information during transmission or recover quantum information after the transmission is required for quantum communication. In this context, QEC code has been introduced as an efficient error correction method to protect quantum information using a specific encoding.

Although the idea of designing QEC codes is similar to the idea of classical error-correction codes, i.e., introducing redundant information in a suitable way to improve the resistance of information to interference \cite{shannon1948mathematical,hamming1950error}, it is not a simple extension of classical error-correction codes due to the unique characteristics of quantum mechanics. There are three major challenges in designing QEC codes:
\begin{itemize}
	\item \textit{No-cloning theorem:} For classical error-correction codes, redundant information is introduced by preparing multiple copies of a single bit. However, as discussed in Section~\ref{Sec2.1}, qubits strictly follow the no-cloning theorem. Consequently, copying a qubit to introduce redundant quantum information is impossible.
	
	\item \textit{Errors are continuous:} In classical communication, the state of a single bit is deterministic. Thus, only bit-flip error needs to be considered in the error of classical information. However, due to the superposition principle, the degree of error in quantum information is greater than that of classical information errors. Qubits are susceptible to both bit-flips and phase-flips. Consequently, QEC codes must be designed with the ability to detect error types simultaneously.
	
	\item \textit{Collapse after measurement:} In a classical system, it is possible to measure arbitrary properties of the bit register without the risk of compromising the encoded information to obtain an error pattern. However, any measurement operation in the quantum world destroys the state of qubits and makes recovery impossible.	
\end{itemize}
These problems make it is quite challenging to design QEC codes compared to classical error correction codes.

Fortunately, none of the problems discussed above is fatal for QEC code designs. We can adopt some ingenious ways to overcome these challenges. Firstly, to break the limitation of the no-cloning theorem, a single qubit can be encoded as a complex entangled state. In this way, we can introduce redundant information into QEC codes without violating the fundamental principle of quantum mechanics. Secondly, although the variety of quantum errors is a continuum, it is a linear combination of three basic quantum errors (corresponding to three Pauli matrices) \cite{knill1997theory}. Thus, all quantum errors can be corrected as long as these three basic ones are corrected. Lastly, the quantum error pattern can be obtained by using a special type of projective measurement referred to as a stabilizer measurement \cite{gottesman1998heisenberg}, i.e., only measure some additional qubits but not all of them. In this way, quantum coherence is maintained, while the results of measurement operations can completely reflect the quantum error pattern. In summary, these novel ideas pave the road for designing QEC codes to correct quantum errors, thus accelerating the development of quantum information technology.

The first two QEC code schemes were proposed by Shor~\cite{shor1995scheme} and Steane \cite{steane1996error} in 1995 and 1996. In Shor code, nine qubits are used to encode a single qubit, analogous to classical repetition codes but with low error correction efficiency. Steane’s scheme introduces the concept of a complementary basis, and seven qubits are used to encode a single qubit. After that, Calderbank, Shor, and Steane proposed a landmark QEC code scheme, known as \textit{Calderbank-Shor-Steane} (CSS) codes \cite{calderbank1996good,steane1996multiple}, inspired by the basic idea of classical theory of linear codes. It is worth noting that one of the most powerful applications of QEC is not merely the protection of transmitted quantum information but also the protection of qubits as it dynamically undergoes computation. In this context, a fault-tolerant QEC code is introduced for high-quality quantum computing. A QEC code is said to be fault-tolerant if it can account for errors that occur at any location in the quantum circuit \cite{gottesman1998theory}. As the smallest code capable of protecting against a quantum error model, the $[[4,2,2]]$ code \cite{vaidman1996error,grassl1997codes} is a promising candidate to achieve fault-tolerance QEC code. Several implementations of the $[[4,2,2]]$ code have already been experimentally demonstrated to be effective in \cite{cory1998experimental,roffe2018protecting,linke2017fault,harper2019fault,vuillot2017error}. In addition to the QEC code designs mentioned above, in recent years, some other designs have been proposed in theory or implemented on qubit hardware  \cite{chiaverini2004realization,waldherr2014quantum,noh2020fault,darmawan2021practical,campagne2020quantum,wang2022quantum}. No matter which code method is used to implement error correction, consuming redundant qubits is inevitable. Improving the qubit transmission rate is pivotal in implementing quantum error correction in entanglement-assisted quantum networks. Therefore, although achieving efficient and perfect QEC codes is still a challenging task, efforts to design efficient network schemes to realize high-performance entanglement-assisted quantum networks have to continue, thus providing support for the application of QEC codes. Besides, the development of quantum information technology enables QEC technology to evolve at an astonishing speed. Hence, it is expected that the technology will soon mature enough to serve various quantum applications, especially quantum computing.

\begin{figure}[t]
	\centering
	\includegraphics[width=1.0\linewidth]{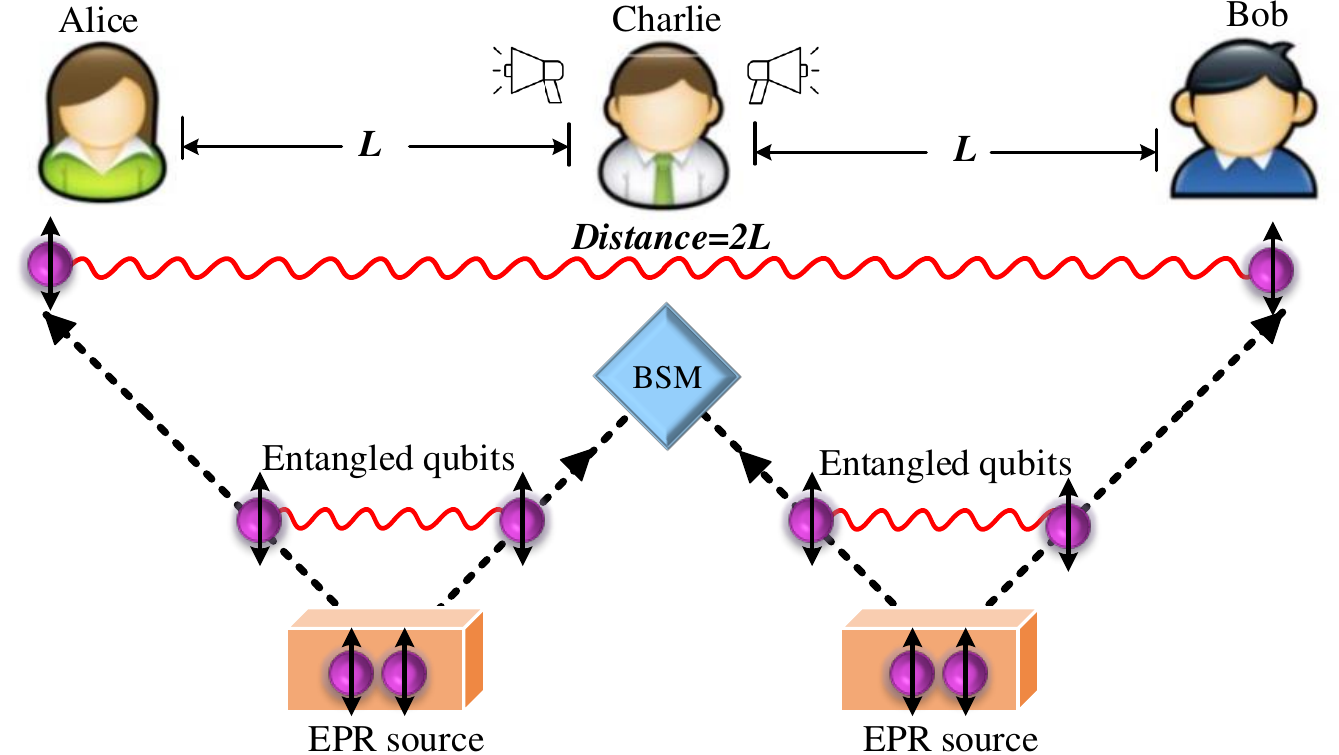}
	\caption{Establishing distant entanglement between Alice and Bob by performing entanglement swapping.}
	\label{swapping}
\end{figure}

\subsection{Entanglement Swapping}\label{Sec3.5}
Establishing entanglement between distant quantum nodes is one of the essential building blocks for achieving distributed quantum applications. However, the success possibility of direct entanglement distribution between two quantum nodes will decrease exponentially with the physical distance of a quantum channel~\cite{pirandola2009direct,pirandola2017fundamental}, i.e., it is extremely difficult to enable two distant quantum nodes to share entangled qubit pairs. Therefore, a quantum technology that can extend the distance of entanglement distribution from point-to-point over a short distance to remote end-to-end is required in entanglement-assisted quantum networks. However, quantum information technology strictly follows the no-cloning theorem. Consequently, the signal amplification and regeneration methods adopted in classical communication do not work in long-distance quantum communication. Inspired by quantum teleportation, an entangled qubit isolated from an entangled qubit pair can be teleported from one quantum node to another to establish entanglement between two distant quantum nodes, \textit{entanglement swapping} was proposed as an effective solution to generate long-haul entanglement between distant quantum nodes  \cite{zukowski1993event,sciarrino2002delayed,riedmatten2005long}.

Entanglement swapping is essentially a LOCC operation used to effectively extend the distance of entanglement distribution. \textbf{Fig. \ref{swapping}} depicts the principle of entanglement swapping. Initially, the distance between any two adjacent quantum nodes is $L$. After entanglement distribution between adjacent quantum nodes, the middle quantum node, Charlie, shares entangled qubit pairs with Alice and Bob, respectively. Then, Charlie performs BSM operation on his two non-entangled qubits separated from two pairs of entangled qubits and sends the measurement outcomes to Alice and Bob, respectively. Similar to quantum teleportation, the entangled qubit isolated from the entangled state shared by Charlie and Alice is teleported to Bob. Hence, the distance of entanglement distribution is extended from $L$ to $2L$, and Alice establishes a distant entanglement with Bob. With the assistance of entanglement swapping, any distant communicating parties can establish long-distance entanglement by ``coupling'' multiple single-hop entangled qubit pairs along a selected path. For example, assuming a $n$-hop path is selected to connect two communicating parties, and the physical length of the $i$-th hop is $l_{i}$, two parties can establish an entanglement of length $\sum_{i=1}^{i=n}l_{i}$ by performing entanglement swapping along this path.

Notably, there are some network problems caused by entanglement swapping during remote entanglement distribution. Firstly, entanglement swapping shows the probabilistic feature due to the imperfection of physical devices. As a result, selecting a path with a high success probability of remote entanglement distribution is pivotal in entanglement-assisted quantum networks. Moreover, unlike that packets are forwarded hop by hop from the source to destination nodes in classical network, swapping operations can be performed in parallel on a selected path to improve the entanglement distribution rate. However, it is required to track entanglement during remote entanglement distribution, considering the probabilistic feature of entanglement swapping. Due to the collapse-after-measurement phenomenon, it is also necessary to avoid competition for entanglement resources by swapping operations. Hence, entanglement-assisted quantum networks need to manage swapping operations performed in parallel. In Section~\ref{Sec6}, we will present network issues related to entanglement swapping in detail.
	
The quantum circuit of entanglement swapping in a quantum system consisting of three node is shown in \textbf{Fig. \ref{swapping-circuit}}. The required steps for implementing entanglement distribution between Alice and Bob by performing entanglement swapping are provided as follows:		
\begin{enumerate}
\item [1)] Entanglement preparation. Suppose Charlie shares Bell state
$\ket{\Psi^{+}}=\frac{1}{\sqrt{2}}(\ket{01}+\ket{10})$ and
$\ket{\Psi^{+}}=\frac{1}{\sqrt{2}}(\ket{01}+\ket{10})$ with Alice and Bob, respectively, after entanglement preparation. In other words, Charlie possesses entangled qubits 2 and 3, and the entangled qubits owned by Alice and Bob are 1 and 4, respectively. The tensor product of $\ket{\Psi^{+}}$ and $\ket{\Psi^{+}}$ is regarded as the initial state of the entanglement swapping system:
\begin{equation}
	\begin{aligned}
		\ket{\Psi_{0}}&=\ket{\Psi^{+}}\otimes\ket{\Psi^{+}}\\
		&=\frac{1}{2}[\ket{0101}+\ket{0110}+\ket{1001}+\ket{1010}].
	\end{aligned}
\label{Initial-system}
\end{equation}
\begin{figure}[t]
	\centering
	\includegraphics[width=1.0\linewidth]{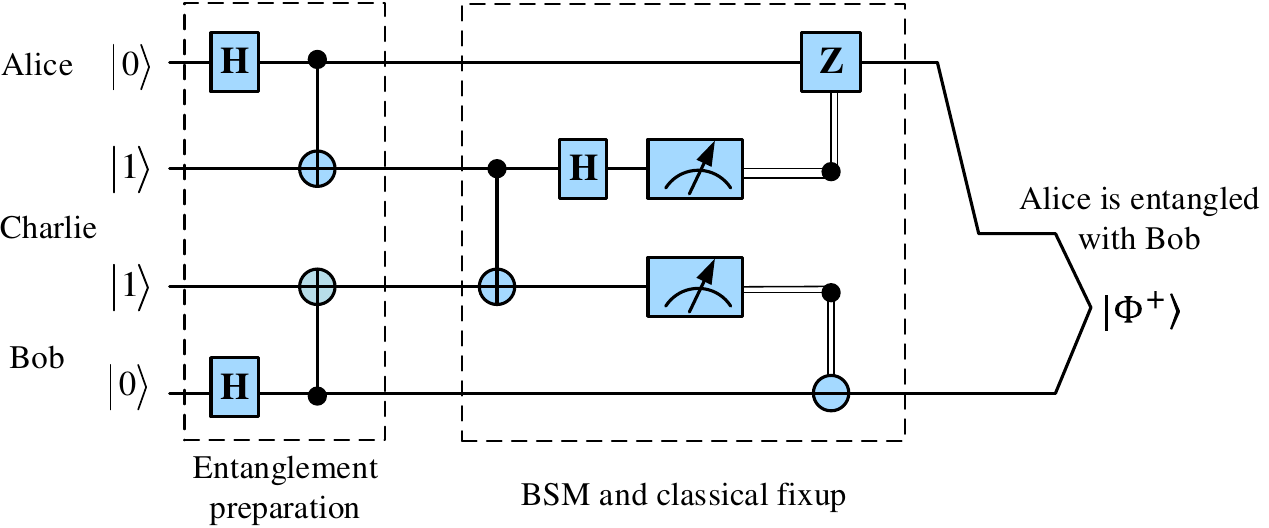}
	\caption{The quantum circuit of entanglement swapping.}
	\label{swapping-circuit}
\end{figure}
\item [2)] Then, Charlie takes the entangled qubit 2 as the control qubits and applies the CNOT gate to the entangled qubit 3. After that, the state of the quantum system goes from $\ket{\Psi_{0}}$ to $\ket{\Psi_{1}}$:
\begin{equation}
	\begin{aligned}
		\ket{\Psi_{1}}&=\frac{1}{2\sqrt{2}}\big[\ket{0001}-\ket{0111}+\ket{0010}-\ket{0100}\\
		&+\ket{1001}+\ket{1111}+\ket{1010}+\ket{1100}\big].
	\end{aligned}
	\label{1-system}
\end{equation}

\item [3)] Charlie applies the H gate to his entangled qubit 2. As a result, the state of the quantum system is
\begin{equation}
	\begin{aligned}
		\ket{\Psi_{2}}&=\frac{1}{2}\Big[\ket{\Psi^{+}}\otimes\ket{\Phi^{+}}+\ket{\Phi^{+}}\otimes\ket{\Phi^{-}}\\
		&+\ket{\Psi^{-}}\otimes\ket{\Psi^{-}}-\ket{\Phi^{-}}\otimes\ket{\Psi^{+}}\Big].
	\end{aligned}
	\label{2-system}
\end{equation}

\item [4)] Finally, Charlie performs the LOCC operation on his two entangled qubits. The system's state collapses into one of four Bell states, i.e., entanglement is built between entangled qubits 1 and 4, which are originally independent of each other. According to Eq.~\ref{2-system}, when the measurement result is $\ket{\Psi^{+}}$ or $\ket{\Psi^{-}}$ or $\ket{\Phi^{+}}$ or $\ket{\Phi^{-}}$, the Bell state shared by Alice Bob is $\ket{\Phi^{+}}$, $\ket{\Psi^{-}}$, $\ket{\Phi^{-}}$, and $\ket{\Psi^{+}}$, respectively, and the probability of which is one in four.
\end{enumerate}

\begin{figure}[htbp]
	\centering
	\subfigure[Sequence entanglement swapping.]{
		\includegraphics[width=0.99\linewidth]{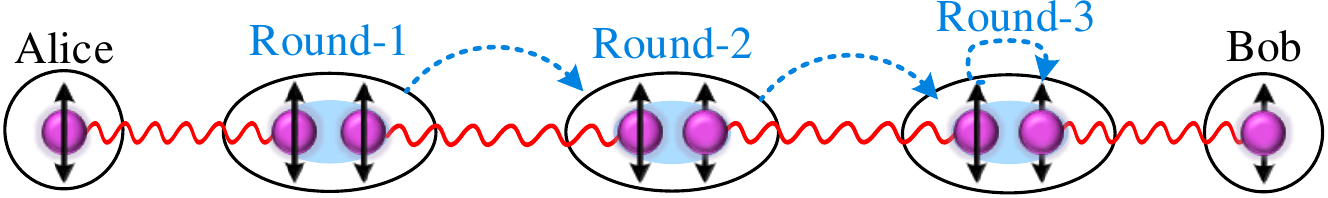}
		\label{sequence}}
	\quad
	\subfigure[Nested entanglement swapping.]{
		\includegraphics[width=0.99\linewidth]{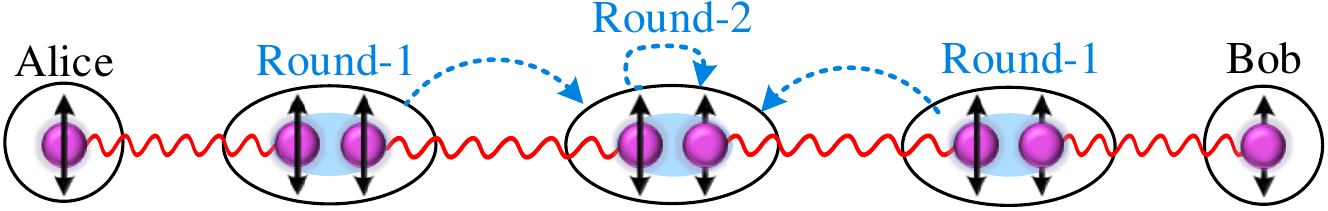}
		\label{nested}}
	\quad
	\subfigure[Parallel entanglement swapping.]{
		\includegraphics[width=0.99\linewidth]{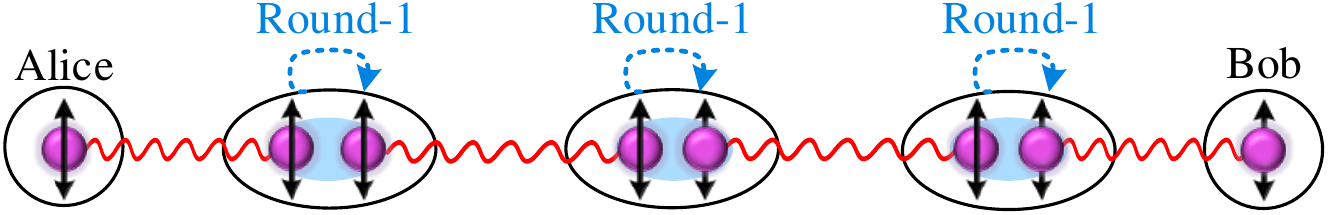}
		\label{parallel}}
	\caption{Three entanglement swapping methods for remote entanglement distribution.}
	\label{threemethod}
\end{figure}

As shown in \textbf{Fig.~\ref{threemethod}}, there are three methods to manage swapping operations on a selected path consisting of multiple intermediate quantum nodes. The first method is called the sequence method or hop-by-hop method. For the sequence method, swapping operations are performed hop-by-hop along the selected path. Hence, the entanglement swapping rounds equal the number of intermediate nodes on the selected path. As shown in \textbf{Fig.~\ref{sequence}}, establishing an end-to-end entanglement connection between Alice and Bob involves three rounds of entanglement swapping. The second method, i.e., the nested method, is shown in \textbf{Fig.~\ref{nested}}. For this method, non-adjacent intermediate nodes perform swapping operations simultaneously to create multi-hop entanglement in each swapping round, and these nodes that have swapped entanglement are removed from the selected path. End-to-end entanglement can be established after multiple rounds of nested entanglement swapping. The last method adopts a parallel manner in which all intermediate nodes perform BSM operations to swap entanglement simultaneously, and then Bob establishes entanglement with Alice based on convergent measurement results. Hence, only one round of entanglement swapping is required, as shown in \textbf{Fig.~\ref{parallel}}. The sequence method can effectively avoid entanglement resource competition between intermediate nodes and facilitate the tracking of entanglement correlations, but it results in complex interactions and a high delay time during remote entanglement distribution. The nested method can reduce remote entanglement distribution latency while avoiding resource competition. The parallel method significantly simplifies the management of swapping operations and reduces delay time. However, the parallel method contributes to inefficient remote entanglement distribution due to the probabilistic feature of entanglement swapping. Compared to sequence and parallel methods, the nested method is currently a good choice to support remote entanglement distribution in entanglement-assisted quantum networks. With the development of quantum physical devices, entanglement distribution between adjacent quantum nodes and entanglement swapping tend to be perfect, so the parallel method will be adopted to realize efficient remote entanglement distribution for future entanglement-assisted quantum networks.

There are two types of quantum systems in quantum information technology, i.e., discrete variable (DV) and continuous variable (CV) quantum systems~\cite{andersen2015hybrid}. In DV and CV quantum systems, quantum information is encoded in finite-dimensional and infinite-dimensional Hilbert spaces, respectively. DV and CV quantum systems have their pros and cons: in DV quantum systems, we can obtain maximally entangled states, but the generation of entanglement is usually probabilistic~\cite{van2011optical}. In contrast to DV quantum systems, CV quantum systems can realize the deterministic generation of entanglement, but it cannot generate perfect entangled states~\cite{braunstein2005quantum,weedbrook2012gaussian}. As the pivotal heart of entanglement-assisted quantum networks, entanglement swapping has been experimentally demonstrated in both DV and CV systems. In the DV regime, various EPR sources, e.g., polarization~\cite{pan1998experimental,ma2012experimental}, energy-time \cite{halder2007entangling}, orbital angular momentum \cite{zhang2017simultaneous}, and superconducting entangled qubits \cite{ning2019deterministic}, have been adopted to implement entanglement swapping. Entanglement swapping is also unconditionally demonstrated in the CV regime \cite{polkinghorne1999continuous,tan1999confirm,van1999uncondtional}, the realization of which is based on quadrature entanglement of optical fields from parametric down-conversion and feed-forward techniques \cite{jia2004experimental,takei2005high}. Besides, single-state entanglement swapping was successfully demonstrated by Zeilinger's group in 2001~\cite{jennewein2001experimental}, and multiple-state entanglement swapping of photonic Bell states was successfully demonstrated in 2008 by Pan's group~\cite{goebel2008multistage}. Recently, Adrien's group realized the interconnection between heterogeneous entanglement-assisted quantum networks by hybrid entanglement swapping \cite{guccione2020connecting}, which tremendously promotes the application of entanglement swapping in future entanglement-assisted quantum networks.

\subsection{Quantum Memories}\label{quantum-memory}
Quantum memories are important in many contexts, including the implementation of single-photon sources, quantum repeaters, loophole-free Bell inequality tests, communication complexity protocols, and precision measurements. There are several reasons for their significance. Firstly, as the smallest microscopic particle unit, quantum states are susceptible to noisy environments. Another reason is the probabilistic feature of both entanglement preparation and quantum manipulation during remote entanglement distribution. Hence, quantum memories are necessary for storing and synchronizing randomly generated entangled qubits in entanglement-assisted quantum networks \cite{simon2010quantum,heshami2016quantum}.

Unlike classical memories that can store copyable binary bit strings, quantum memories are used to store fragile qubits and are restricted by the fundamental laws of quantum mechanics. There are four pivotal requirements that need to be satisfied for quantum memories:
\begin{itemize}
	\item \textit{High fidelity}: Qubits stored in quantum memory should interact less with the noisy environment. In other words, the state of a single qubit extracted from a quantum memory should exhibit a high degree of coincidence with its pro-storage state.
	
	\item \textit{High efficiency}: Another important evaluation metric of quantum memories is store-and-retrieve efficiency, i.e., the ratio between the energies of stored and retrieved pulses. Generally, high efficiency implies a higher success probability of reading out a single qubit from quantum memories.
	
	\item \textit{Long lifetime}: The lifetime of quantum memories, also known as storage time, is crucial. Due to the inevitable decoherence of qubits during storage, quantum memories must be able to keep qubits coherent for as long as possible.
	
	\item \textit{Practicality at room temperature}: To build large-scale and wide-area entanglement-assisted quantum networks, quantum memories must be operable at room temperature while performing well in the three evaluation metrics discussed above.
\end{itemize}
These four requirements pose significant challenges for the implementation of quantum memories, especially for small-sized quantum memories.
		
Over the past few decades, quantum memories have been extensively studied in a variety of storage schemes, including the electromagnetically induced transparency (EIT) scheme \cite{Chaneliere2005Storage,Eisaman2005Electromagnetically,Zhang2011Preparation}, the Duan-Lukin-Cirac-Zoller (DLCZ) protocol \cite{manz2007collisional,Chrapkiewicz2017High}, the gradient echo memory (GEM) \cite{Moiseev2001Complete,Alexander2006Photon}, quantum memory based on Faraday interaction \cite{Julsgaard2001Experimental,Julsgaard2004Experimental}, far off-resonance Raman memory \cite{Reim2012Multipulse,Ding2018Raman}, and quantum memories based on off-resonant cascaded absorption (ORCA) \cite{Kaczmarek2018High}. Essentially, the various storage schemes listed above involve the interaction of light with matter, interference between internal states of matter, and the evolution and restoration of phase relations.
Quantum memories can be categorized into three types based on the media used: solid-state quantum memories, atomic quantum memories, and optical quantum memories. We will elaborate on these three categories of quantum memory as follows.

\textbf{Solid-state Quantum Memories.} There are two main categories of solid-state quantum memories: diamond-based memories and rare-earth ion-doped crystal memories. The lifetime of diamond-based quantum memories using optical photons as storage states is too short, only a few picoseconds~\cite{lee2011entangling,england2015storage}. Although the nuclear spin coherent times in a diamond with an NV center at room temperature have reached the order of seconds \cite{maurer2012room}, the optical photons in a diamond at room temperature can lead to a severe broadening of the leap frequency. Hence, the quantum memory scheme using the NV center in a diamond is not yet suitable for building quantum entanglement \cite{ghobadi2019progress}. The rare-earth ion-doped crystal scheme is one of the most promising quantum memory solutions for practical applications due to the excellent coherence properties at cryogenic temperatures~\cite{Hua2018Quantum}. In \cite{zhong2015optically}, the nuclear spins in a solid with a six-hour coherence time have been achieved. So far, various solid-state quantum memories have been demonstrated
\cite{morton2008solid,hedges2010efficient,saglamyurek2011broadband,zhong2017nanophotonic}, and specially \cite{zhu2022demand} demonstrated a spin-wave solid-state quantum memory with a fidelity of $(99.4\pm0.6)\%$.

\textbf{Atomic-ensemble Quantum Memories.} This category of quantum memories is primarily implemented based on the special properties of certain atoms. Generally, atomic-ensemble quantum memories are realized using alkali metal atoms such as rubidium or cesium atoms. Alkali metal vapour isotopes exhibit large optical depths at near-infrared wavelengths due to their relatively narrow spectral lines and high number densities at `warm' temperatures of 50–100$^{\circ}$C. These atoms typically have long coherence times, and the ensemble coherence lifetimes in warm alkali vapours are often limited by intrinsic coherence times. Alkali vapors have been utilized for some of the most significant memory advancements, thanks to their attributes of high optical depth, long coherence times, and easily accessible optical transitions in the near-infrared \cite{kuzmich2003generation, lukin2003colloquium}. In addition to alkali vapors, other implementations of atomic-ensemble quantum memories utilizing various storage schemes, such as EIT \cite{chen2013coherent}, DLCZ \cite{zhao2009millisecond, bashkansky2012quantum, yang2016efficient}, GEM \cite{hosseini2009coherent, hosseini2011high, cho2016highly}, and far off-resonance Raman \cite{reim2011single, saunders2016cavity}, have also been presented in recent years.

\textbf{Optical Quantum Memories.} This type of quantum memory represents the simplest approach to storing photons. To achieve time delay and storage functions, photons can travel in the optical circuit with low loss. Generally, an optical memory is an optical loop circuit that enables programmable control of the storage time. The circumference of the ring optical circuit determines the storage time~\cite{Pittman2002Cyclical}. Additionally, the storage and retrieval of photons are implemented by changing the photons' polarization~\cite{Kaneda2017Quantum,Pang2020Hybrid}. The low loss and minimal environmental noise in optical loop memory facilitate certain applications, such as multi-photon synchronization. However, the storage time of the memory is fixed by the length of the optical fiber. Alternatively, light can be stored in a cavity \cite{Leung2006Quantum,Maitre1997Quantum} to achieve variable and on-demand storage time, albeit at the cost of limited efficiency. In summary, optical quantum memory exhibits limitations in terms of efficiency and flexibility. Hence, this storage scheme is not suitable for quantum repeaters \cite{Lvovsky2009Optical}.

\begin{table}[htbp]
	\centering
	\caption{The comparison between different quantum memories.}
	\label{Comparison-memories}
	\begin{tabular}{m{2.0cm}<{\centering}m{0.8cm}<{\centering}m{1.0cm}<{\centering}m{1.0cm}<{\centering}m{2.0cm}<{\centering}}
		\toprule
		\textbf{Topic} &\textbf{Fidelity} &\textbf{Efficiency} &\textbf{Lifetime} &\textbf{Practicality at room temperature}\\
		\midrule
		\textbf{Solid-state}  &High &Low & High &Poor \\
		\textbf{Atomic-ensemble}  &Low  &High  &Low &Poor \\
		\textbf{Optical} &Low  & Low &Low & Well  \\
		\bottomrule
	\end{tabular}
\end{table}

Here, we provide a comprehensive comparison of three quantum memories in terms of four pivotal indicators, i.e., fidelity, efficiency, lifetime, and practicality at room temperature, as shown in \textbf{Table~\ref{Comparison-memories}}. Solid-state quantum memories perform well in fidelity and lifetime due to solid materials, especially rare-earth ion-doped solid materials, with long optical coherence time and wide optical absorption bandwidths. The fidelity of solid-state quantum memories can achieve 0.999 \cite{zhou2012realization}, and the lifetime can reach the order of seconds. However, currently implemented solid-state quantum memories have low efficiency (only $56\%$~\cite{sabooni2013efficient}) and perform poorly at room temperature. Since photons are not easily absorbed by atoms, atomic-ensemble quantum memories perform well in terms of efficiency. However, the fidelity and lifetime performances of atomic-ensemble quantum memories are poor because atoms move so violently that their collisions produce high noise. Optical quantum memories exhibit good practicality at room temperature but perform poorly in terms of fidelity, efficiency, and lifetime due to photon loss and channel noise. In summary, each type of quantum memory has its advantages and disadvantages, and it is still required to explore how to leverage their advantages in practical application scenarios.

\begin{figure*}[t]
	\centering
	\includegraphics[width=0.82\linewidth]{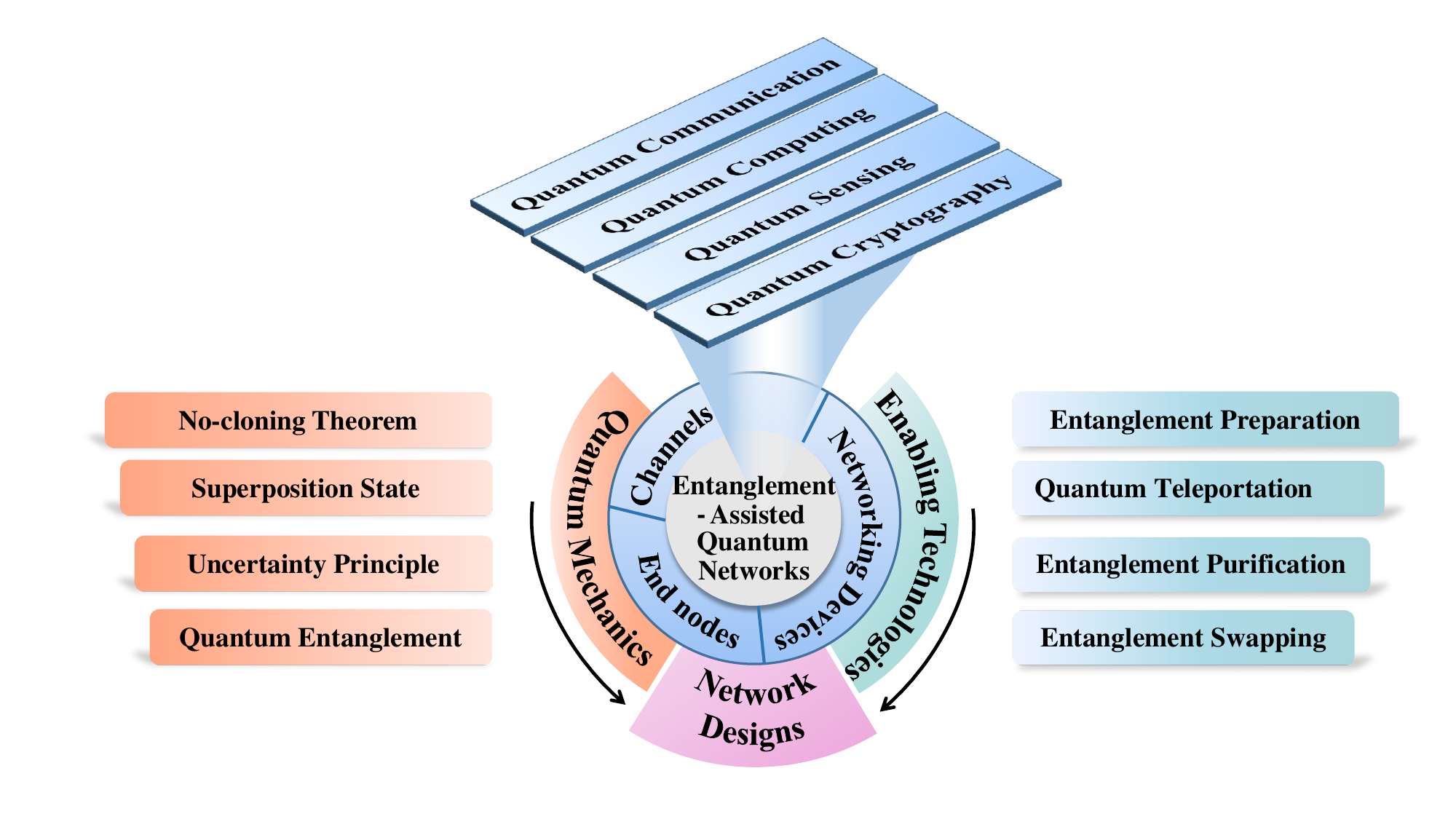}
	\caption{The knowledge map of entanglement-assisted quantum networks.}
	\label{ConnectionMap}
\end{figure*}

In the past decades, quantum memory has evolved from its original theoretical demonstrations to becoming close to practical nowadays, offering advantages in quantum information technology. In the latest research, the storage fidelity of quantum memory reaches up to $99\%$ under laboratory conditions \cite{liu2020demand}. Furthermore, the latest quantum memory can store a single photon for over one hour, as shown in \cite{Ma2021Elimination}. Analogous to classical memory, a quantum memory can be regarded as the combination of multiple independent memory cells \cite{Pu2017Experimental} with high fidelity, high efficiency, and long storage time. Quantum memories that can provide flexible storage services are expected to be realized in the near future, facilitating the development of quantum repeaters to further support entanglement-assisted quantum networks.
			
\section{Entanglement-assisted Quantum Networks}\label{Sec4}
This section mainly presents an introduction to entanglement-assisted quantum networks. More concretely, the development stages of entanglement-assisted quantum networks are discussed first. Then, we make a comprehensive comparison between classical communication and quantum communication, such as implementation steps, information resources, and security, to demonstrate that entanglement-assisted quantum networks are fundamentally different from classical networks. Based on this comparison, we further present the differences between classical networks and entanglement-assisted quantum networks, ranging from physical resources to protocol stacks. Moreover, network elements, including physical channels, EPR sources, quantum end nodes, quantum repeaters, and quantum routers, are described. Finally, we present a general structure of an entanglement-assisted quantum network and explain how quantum end nodes interact with each other to support various quantum applications.

\subsection{Definition}

Entanglement-assisted quantum networks are the network infrastructures formed by interconnecting numerous quantum nodes, which can realize quantum information transmission between arbitrary quantum nodes under the government of network designs and the fundamental laws of quantum mechanics, thus supporting various quantum applications. As shown in \textbf{Fig.~\ref{ConnectionMap}}, we elaborate on the connotation of entanglement-assisted quantum networks in terms of quantum mechanics, enabling technologies, network elements, network designs, and various quantum applications.

Firstly, entanglement-assisted quantum networks follow the fundamental laws of quantum mechanics, such as the no-cloning theorem, the superposition state, the uncertainty principle, and quantum entanglement. These unique properties, with no counterpart in classical mechanics, contribute to the enormous advantages of quantum information technology over classical information technology. The preparation, storage, transmission, and processing of quantum information are governed by the unique properties of quantum mechanics. As a result, entanglement-assisted quantum networks are essentially different from classical networks.

Secondly, enabling technologies described in Section~\ref{Sec3} are essential building blocks for entanglement-assisted quantum networks. Unlike classical networks, the interconnection and intercommunication between distant quantum nodes are realized based on entanglement. Hence, entanglement preparation, entanglement purification, and entanglement swapping play a pivotal role in establishing high-fidelity entanglement between distant quantum nodes, and these enabling technologies support quantum teleportation and thus realize quantum information transmission.

Thirdly, physical network elements, including physical channels, quantum end nodes, and networking devices, are the necessary physical devices to realize large-scale and wide-area entanglement-assisted quantum networks. Notably, quantum operations, including entanglement distribution, entanglement swapping, and quantum teleportation, are implemented with the help of classical communication. Hence, physical channels involve quantum channels and classical channels. In entanglement-assisted quantum networks, quantum end nodes and networking devices are connected via physical channels according to specific rules to form a mesh topology, thus building the underlying infrastructure. This infrastructure is a promising platform for implementing remote entanglement distribution between quantum end nodes, thus supporting various quantum applications. Although entanglement-assisted quantum networks are fundamentally different from classical networks, they are similar in the classes and functions of network elements. Therefore, the structural design principles of classical networks can provide guidance for the implementation of future entanglement-assisted quantum networks.

Fourthly, network designs are essential to implement effective and efficient remote entanglement distribution in entanglement-assisted quantum networks. Notably, an entanglement-assisted quantum network goes beyond a simple collection of multiple independent paths used to establish end-to-end entanglement connections. It is necessary to manage concurrent network tasks to satisfy applications' requirements in an orderly and efficient manner. Additionally, management solutions used in classical networks cannot be directly applied to entanglement-assisted quantum networks. Hence, network designs, such as routing algorithms, scheduling schemes, and resource allocation algorithms, need to be studied, acting as administrators to manage concurrent tasks to ensure the quality of service in entanglement-assisted quantum networks.

Lastly, quantum applications, including quantum communication, quantum computing, quantum sensing, and quantum cryptography, running on quantum end nodes can fully exploit the potential of quantum information technology. Quantum communication is one of the most interesting applications of applied quantum physics closely related to quantum teleportation. It enables the unconditional secure transmission of quantum information between communicating parties based on quantum properties. Quantum computing is a beautiful combination of quantum physics, computer science, and information theory. Due to the superposition state, quantum computing can provide exponential speed-up compared with classical computing. Hence, quantum computing has a wide application prospect in the era of big data. Quantum sensing is one of the most advanced quantum applications. It uses quantum resources to improve the sensitivity or precision of a measurement based on quantum properties beyond what is possible classically. Hence, quantum sensing can significantly boost the performance of a number of practical tasks, including gravitational wave detection, astronomical observations, microscopes, target detection, data readout, atomic clocks, biological probing, and so on. Quantum cryptography is the science of exploiting quantum properties to perform cryptographic tasks. The most well-known and developed application of quantum cryptography is QKD, which offers an information-theoretically secure solution to the key exchange problem in classical networks. Quantum cryptography also corresponds to a collection of other ideas broadly related to bit commitment, such as quantum secret sharing. Overall, quantum applications present significant advantages over classical applications based on the unique features of quantum mechanics. Notably, many quantum applications require quantum end nodes to establish entanglement. Hence, remote entanglement distribution between quantum end nodes is one of the building blocks of entanglement-assisted quantum networks.

In summary, entanglement-assisted quantum networks can be defined as a promising platform composed of quantum nodes and physical channels that follow the fundamental laws of quantum mechanics. These networks are built to support ground-breaking quantum applications. They work by realizing remote entanglement distribution and quantum information transmission between quantum end nodes, all under the control of network designs.
	
\subsection{Development Stages}\label{Sec4.1}
\begin{figure*}[t]
	\centering
	\includegraphics[width=1.0\linewidth]{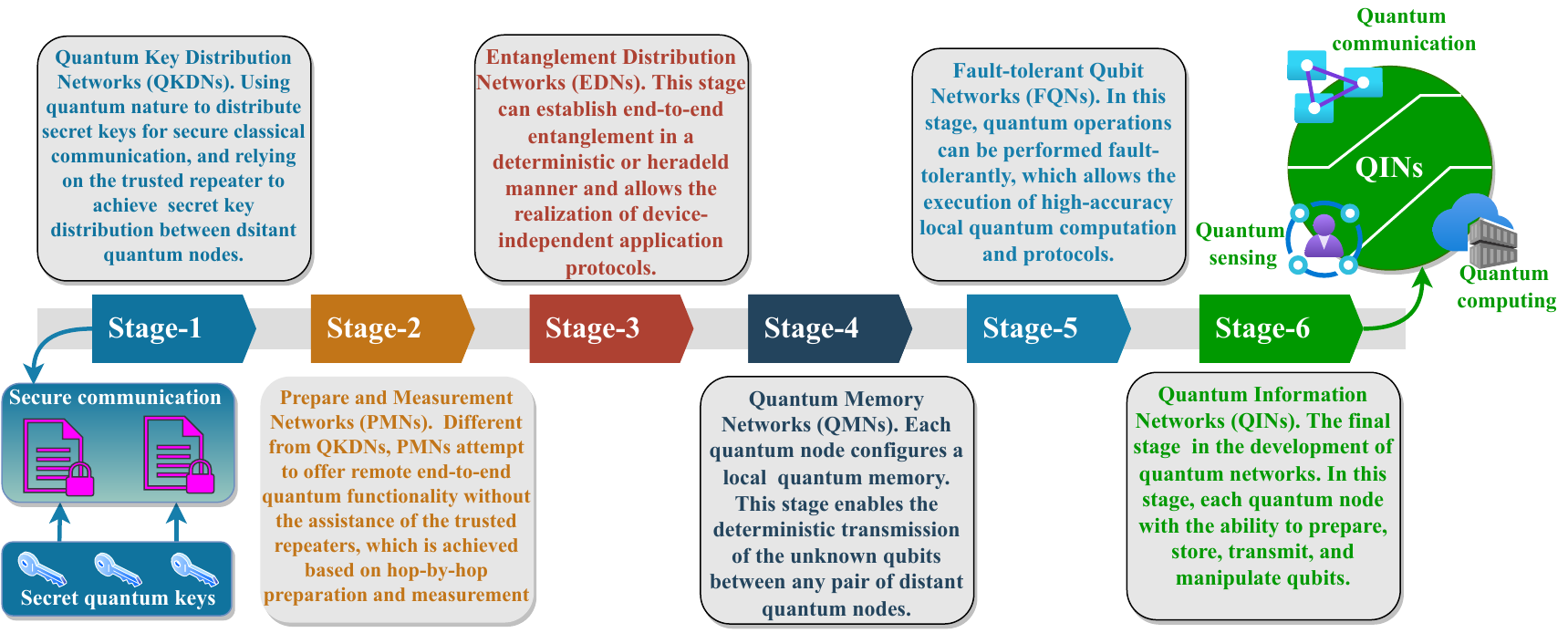}
	\caption{The development stages of entanglement-assisted quantum networks.}
	\label{development-stages}
\end{figure*}

Similar to the development trajectory of classical information technology, quantum information technology will evolve from point-to-point quantum communication to large-scale entanglement-assisted quantum networks for supporting various quantum applications. Entanglement-assisted quantum networks' application scenarios have been identified in \cite{wang2021application}. The functionality that an entanglement-assisted quantum network can achieve is driven by the development of quantum physical devices, so this development trajectory shows the vast diversity in functionality at different stages. Consequently, the development of entanglement-assisted quantum networks is not only reflected in network scale but also functionality. More concretely, the development stages of function-driven entanglement-assisted quantum networks are shown in \textbf{Fig.~\ref{development-stages}}.

\textbf{Quantum Key Distribution Networks (QKDNs).} A QKDN is a quantum network that can distribute random secret keys between QKD nodes based on the fundamental laws of quantum mechanics~\cite{elliott2002building}. This stage differs substantially from others in the sense that it mainly achieves unconditional secure key distribution in theory to enhance the security of classical communication rather than quantum information transmission. Optical devices and QKD protocols drive the development of QKDNs. The maturity of optical devices promotes QKD technology from laboratory to practical application. Currently, some small-scale QKDNs are available for commercial service, and a satellite-to-ground QKDN has also been demonstrated experimentally~\cite{liao2017satellite}. Notably, the inherent loss and noise of quantum channels significantly limit the rate of key distribution between adjacent QKD nodes. For example, the first QKD metropolitan network, DARPA, can only provide secret key distribution services with a maximum key rate of 10kbps. In order to effectively achieve long-distance key distribution, these QKDNs mainly adopt trusted repeaters to overcome distance limitations. Trusted repeaters, a class of quantum devices that contain multiple pairs of quantum transmitters and receivers, work by transmitting quantum keys using classical encryption operations, such as XOR operations. However, it is challenging to guarantee that all trusted repeaters are completely trusted in the real-world QKDN. Consequently, trusted-repeater-based remote key distribution is faced with severe security challenges. Fortunately, some innovative QKD protocols, e.g., measurement device independent QKD (MDI-QKD) \cite{rubenok2013real,da2013proof,tang2016measurement} and two-field QKD (TF-QKD) \cite{lucamarini2018overcoming}, are proposed to improve the rate of remote secret key distribution by introducing untrusted third parties to overcome distance limitations. However, large-scale QKDNs require the integration of trusted repeaters and untrusted third parties, which reduces the security level of QKDNs. One way to extend the key distribution distance with high-level security is to use quantum repeaters. Quantum repeaters work by performing entanglement swapping to establish long-distance entanglement. With the assistance of quantum repeaters, two distant QKD nodes can share entangled qubit pairs and realize unconditionally secure key distribution using entanglement-based QKD protocols, such as E91 protocol \cite{ekert1991quantum}. However, an idealized quantum repeater is still unavailable. Hence, building large-scale and wide-are QKDNs to distribute secret keys for supporting classical secure communication will remain a focus of quantum network research.

\textbf{Prepare-and-Measure Networks (PMNs).} Thanks to the dramatic development of quantum devices, including light sources and detectors, this stage attempts to offer end-to-end quantum functionality. In other words, encrypted information can be transmitted in the form of encoded qubits using specific coding rules. As a result, PMNs enable end-to-end information transmission by preparing and measuring qubits hop-by-hop, which is different from transmitting quantum keys hop-by-hop with the help of trusted repeaters in QKDNs. At this stage, any quantum node encodes qubits and transmits them to the next quantum node via quantum channels, i.e., hop-by-hop information transmission. Then, this node measures the received qubits and prepares qubits to be sent to the next hop based on the measurement outcomes. During information transmission between quantum end nodes, each quantum node on the communication path acts as both an encoder and decoder. The transmission of classical information based on the prepare-and-measure method is also known as quantum secure direct communication (QSDC), which consists of two implementations: one based on single photons and one based on entanglement. The first QSDC protocol was proposed by Long \textit{et al.} \cite{long2002theoretically} in 2000, and many QSDC protocols have been proposed in recent years \cite{deng2003two,wang2005quantum,zhu2006secure,hai2006quantum}. Besides, researchers demonstrated QSDC in a network of 15 nodes in 2021 \cite{qi202115}. Photon loss and quantum decoherence still are key obstacles for PMNs. It is challenging to keep the accuracy of classical information during long-distance communication. Moreover, PMNs strongly rely on quantum memory since the key to QSDC between adjacent quantum nodes is block (i.e., a sequence of qubits) transmission. Note that the prepare-and-measure functionality is not equivalent to transmitting arbitrary quantum information since the qubits being transmitted are not unknown, i.e., the essence of which is still the transmission of classical information.

\textbf{Entanglement Distribution Networks (EDNs).} An EDN can realize end-to-end entanglement distribution in a deterministic \cite{humphreys2018deterministic} or heralded \cite{liu2021heralded} manner. The development of entanglement distribution technology facilitates the implementation of EDNs. In this stage, end-to-end entanglement is established by repeatedly performing entanglement swapping along a repeater chain consisting of multiple quantum repeaters. As discussed above, the implementation of end-to-end secure communication in QKDNs and PMNs significantly relies on intermediate quantum devices. In QKDNs, trusted repeaters are required to perform classical encryption operations to expand the distance of key distribution. The premise that each trusted repeater will not maliciously disclose quantum keys ensures the security of end-to-end key distribution. In PMNs, end-to-end secure communication is implemented with the strong assumption that each quantum node is secure, i.e., encrypted information will not be leaked maliciously by quantum nodes. The major advance over the previous stages is that EDNs allow the realization of device-independent application protocols. More concretely, each quantum node in EDNs is quite untrustworthy, and each intermediate node between a pair of quantum end nodes is transparent to the end-to-end application protocol, i.e., the information leakage of intermediate nodes will not affect the end-to-end applications' implementation. Device-independent application protocols are realized based on entanglement characteristics. This stage does not strongly require quantum nodes to configure quantum memory. To reduce the negative effects of quantum decoherence, entanglement distribution is usually triggered in an on-demand manner. Hence, in addition to developing physical devices responsible for entanglement preparation, the entanglement distribution scheduling design is also pivotal to EDNs' performance. Notably, the generation of multi-party entanglement cannot be achieved at this stage.
	
\textbf{Quantum Memory Networks (QMNs).} The implementation of QMNs benefits from the fact that a quantum memory can operate like a classical memory in a room-temperature environment. This stage requires that each quantum node in the network have a local quantum memory. Different from entangled qubits being prepared on demand in EDNs, entangled qubits can be stored in a quantum memory for a period of time in QMNs, facilitating to alleviate the probabilistic feature of quantum operations such as entanglement distribution and entanglement swapping. As a result, QMNs allow the implementation of complex quantum applications, such as blind quantum computing \cite{broadbent2009universal,morimae2012blind}, secret sharing~\cite{hillery1999quantum}, and leader selection \cite{ambainis2004multiparty}, by exploiting the ability of local quantum memory. This stage is the turning point of networks' development. In other words, a quantum network enables the deterministic transmission of quantum information between any pair of quantum end nodes, thus achieving some less complicated distributed quantum tasks. As discussed in Section~\ref{quantum-memory}, a quantum memory needs to satisfy four pivotal requirements, and they directly affect quantum memories' performance. Therefore, the fidelity, lifetime, efficiency, and practicality of quantum memories will significantly affect QMNs' performance. Besides, the size of quantum memories is also pivotal to QMNs. Hence, developing large-size, high-fidelity, high-lifetime, efficient, and room-temperature-applicable quantum memories is essential to the implementation of QMNs.

\textbf{Fault-tolerant Qubit Networks (FQNs).} FQNs work by distributing entangled qubits to support some quantum applications. The fourth stage is distinguished by the feature that local quantum operations can be performed fault-tolerantly. Fault tolerance implies that all errors caused by noisy quantum channels, measurement devices, and quantum memories can be considered negligible by adding more network resources. Hence, high-performance quantum memories are strongly required in FQNs to store more network resources. Besides, QEC technology also plays a pivotal role in FQNs. At this stage, the available fault-tolerant local operations allow the execution of higher-accuracy local quantum computation and protocols with an arbitrary number of rounds of communication. In a nutshell, we can consume more network resources to alleviate quantum decoherence to support high-quality qubit manipulation. Hence, in addition to the deterministic transmission of quantum information that QMNs can realize, many quantum applications, e.g., clock synchronization and distributed quantum computation, can be realized in FQNs.
	
\textbf{Quantum Information Networks (QINs).} This stage is the final stage in the development of entanglement-assisted quantum networks. At this stage, numerous quantum nodes capable of preparing, storing, transmitting, and manipulating qubits are interconnected to form large-scale QINs. Thanks to the maturity of physical devices and quantum information technology, quantum decoherence can be effectively and efficiently overcome, and network resources are sufficient to satisfy concurrent network tasks at this stage. In this stage, high-fidelity entanglement can be established between quantum end nodes, no matter how far apart, to support the implementation of high-performance application protocols. QINs allow each quantum end node to arbitrarily exchange quantum information with others. Similarly to the current Internet, a global quantum internet can be realized to support various quantum applications. At this stage, physical devices and quantum information technology have been significantly improved. Therefore, the main challenge in QINs is to solve network issues, such as routing design, request scheduling, resource allocation, and qubit transport control, to improve entanglement-assisted quantum networks' performance, thus serving users with QoS.
	
\subsection{Classical VS Quantum Communications}
Entanglement-assisted quantum networks are governed by the fundamental laws of quantum mechanics, which are essentially different from classical networks. In other words, an entanglement-assisted quantum network is a revolutionary network rather than the development product of classical networks. There are many differences between entanglement-assisted quantum networks and classical networks. Here, we consider a future entanglement-assisted quantum network in the final development stage, i.e., QIN, and briefly conclude the differences from classical networks. To better understand how entanglement-assisted quantum networks vary from classical networks, we first comprehensively compare classical communication with quantum communication represented by quantum teleportation.

\begin{table*}[hbtp]
	\centering
	\caption{A comprehensive comparison between classical communication and quantum communication.}
	\label{communication-differences}
	\begin{tabular}{m{3.0cm}<{\centering}m{6.5cm}<{\centering}m{6.5cm}<{\centering}}
		\toprule
		\textbf{Topics} & \textbf{Classical Communication}& \textbf{Quantum Teleportation} \\
		
		\midrule
		\textbf{Steps} & Encoding, transmission, and decoding.  & Entanglement distribution and LOCC operation. \\
		
		\multirow{2}{*}{\textbf{Information Resources}} & Macroscopic objects, e.g., voltage and electric current.  &  Microscopic particles, e.g., photons and cold atoms. Macroscopic objects, e.g., voltage and electric current. \\
		& Features: it can be cloned, deterministic state. &  Features: the no-cloning theorem, superposition state. \\	
		
		\textbf{Third party}  & Absent. & EPR source: distributing entangled qubit pairs between adjacent quantum ndoes to establish entanglement links. \\
		
		\multirow{2}{*}{\textbf{Sender}} &  Classical transmitter: encoding information into classical signal suitable for transmission over a classical channel. & Classical transmitter: encoding the outcomes of the BSM operation into classical signal.   \\
		&Quantum Transmitter: absent.  &  Quantum transmitter: performing BSM operation on local qubits for conveying the entanglement at receiver. \\
		
		\multirow{2}{*}{ \textbf{Channel}}   & Classical channel: medium used to transmit classical signal. Bandwidth is determined by the difference between the maximum and the minimum frequency. &Classical channel: medium used to realize the transmission of the classical results of the BSM operation between two communicating parties.  \\
		&Quantum channel: Absent. & Quantum channel: medium used to transmit entangled qubit to quantum nodes. Bandwidth is determined the number of entanglement links.  \\
		
		\multirow{2}{*}{ \textbf{Receiver}}  & Classical receiver: decoding the classical information from the received classical signal.  & Classical receiver: decoding the classical input of the unitary operation from the received classical signal. \\	
		&Quantum receiver: Absent. & Quantum receiver: performing unitary operation to decode the entangled input of the BSM operation and obtain the teleported qubit.\\
		
		\textbf{Security}  &Security is guaranteed by classical encryption algorithms, and it is challenging to realize unconditional secure information transmission in information theory.  & It can achieve unconditional security based on the fundamental principles of quantum mechanics.  \\ 		
		\bottomrule
	\end{tabular}
\end{table*}

Quantum teleportation is a new communication method that uses quantum superposition and entanglement properties for quantum information transmission. Based on the three fundamental principles of quantum mechanics, i.e., the uncertainty principle, the collapse-after-measurement phenomenon, and the no-cloning theorem, quantum teleportation provides an absolute security guarantee that cannot be eavesdropped on and decrypted for information transmission. Here, we compare quantum teleportation between two directly linked nodes with classical communication. Detailed comparisons are shown in \textbf{Table~\ref{communication-differences}} and discussed as follows.

\textbf{Steps.} Classical communication between adjacent nodes is implemented based on three steps: encoding, transmission, and decoding, i.e., classical information is directly transmitted from the sender to the receiver via a classical channel as the specific encoded signal. However, it is unsuitable to directly transmit quantum information between adjacent quantum nodes due to the inherent photon loss and quantum decoherence of quantum channels. Quantum teleportation is an effective communication method to transmit quantum information based on entanglement properties, consisting of two steps: entanglement distribution between sender and receiver, and LOCC operation.

\textbf{Information Resource.} In classical communication, the basic unit of classical information is 0 or 1 bit. Binary bit strings are encoded into the deterministic signal characterized by macroscopic objects, such as voltage and electric current, for transmission over a classical channel. For quantum teleportation, the basic unit of quantum information is a single qubit. Qubits are represented by microscopic particles, e.g., photons and cold atoms. Unlike deterministic classical bits that can be cloned, qubits follow the fundamental principles of quantum mechanics, such as the no-cloning theorem and superposition state. Besides, the state of a single qubit is determined by its probability amplitudes. Most notably, the implementation of quantum teleportation requires classical communication to transmit measurement results. Hence, the information resource of quantum teleportation also contains macroscopic objects.

\textbf{Third party.} In classical communication, a third party is absent. The reason is that classical information is transmitted directly in the channel in the form of classical signals without the assistance of a third party. However, quantum teleportation requires two communicating parties to share entangled qubit pairs before qubit transmission since it is implemented based on entanglement properties. Hence, the third party, EPR sources, is necessary for quantum teleportation and aims at distributing entangled qubit pairs to adjacent quantum nodes.

\textbf{Sender.} In classical communication, the classical sender is responsible for encoding data into classical signals suitable for transmission over a classical channel, and the quantum transmitter is absent. As described in Section~\ref{quantum-teleportation}, quantum teleportation, essentially an LOCC operation, is implemented with the help of classical communication. Hence, each quantum node needs to deploy classical and quantum transmitters. The quantum transmitter performs BSM operations on the local entangled qubits and the data qubit for conveying the entanglement to the receiver. The classical transmitter in quantum teleportation is responsible for encoding the results of the BSM operation into classical signals and sending them to the receiver.

\textbf{Channel.} In classical communication, the classical channel is the transmission medium used to transmit classical signals and can be used to extend the communication range through classical amplify-and-forward or decode-and-forward technologies. The bandwidth of a classical channel is determined by the difference between the maximum and minimum frequency. The quantum channel is absent in classical communication. For quantum teleportation, the classical channel is used to transmit the classical results of the BSM operation, and the quantum channel is responsible for distributing entangled qubit pairs to establish entanglement links. A quantum channel's bandwidth equals the number of entanglement links established between adjacent quantum nodes. Notably, it is challenging to use the quantum channel to extend the communication distances due to the no-cloning theorem. Hence, long-distance quantum teleportation generally requires quantum repeaters to be deployed to realize remote entanglement distribution.

\textbf{Receiver.} In classical communication, the classical receiver works by decoding the classical information from the received classical signal, and the quantum receiver is absent. For quantum teleportation, the classical receiver is responsible for decoding the classical information required by unitary operation. According to classical information decoded by the classical receiver, the quantum receiver can perform the corresponding unitary operation to decode the entangled input of the BSM operation, thus obtaining the teleported qubits.

\textbf{Security.} In classical communication, the security of the transmitted information is guaranteed by classical encryption algorithms without quantum-resistant features. That is, the classical encryption algorithm can be broken in a short time by an efficient quantum computer running an efficient quantum algorithm. Hence, it is challenging to realize unconditional secure classical communication in information theory. However, quantum teleportation can realize unconditional secure information transmission based on the fundamental principles of quantum mechanics.

In summary, quantum communication essentially differs from classical communication. Unlike classical signals that are forwarded hop-by-hop along classical channels, the transmission of qubits is free from the noise of quantum channels. Besides, quantum teleportation can realize unconditionally secure communication due to the no-cloning theorem, which is quite challenging in classical communication. For quantum communication, two quantum nodes must share entangled qubit pairs. Qubits can be transmitted no matter how far away two communicating parties are as long as they are entangled. Most notably, long-distance qubit transmission should adopt quantum repeaters since the signal regeneration and amplification technologies adopted in classical communication cannot be used in quantum communication. Although quantum communication performs better in security than classical communication, it cannot completely replace classical communication since quantum teleportation is essentially an LOCC operation. Hence, classical communication and quantum communication cooperate in entanglement-assisted quantum networks to realize quantum information transmission.
		
\subsection{Differences from Classical Networks}	
As discussed above, quantum communication is fundamentally different from classical communication. Hence, as a promising platform that can realize quantum information transmission between arbitrary quantum nodes to support various quantum applications, entanglement-assisted quantum networks significantly differ from classical networks. Here, we comprehensively compare entanglement-assisted quantum networks and classical networks. The detailed comparisons, ranging from physical resource to protocol stack, are shown in \textbf{Table~\ref{differences}} and discussed as follows.

\textbf{Physical Resources.} The differences between classical networks and entanglement-assisted quantum networks begin with the physical resource. Classical information is encoded as a binary string of bits. Due to the fact that the state of each bit is determinate, either 0 or 1, thus the input of classical information processors is a deterministic signal characterized by macroscopic physical quantities such as voltage and electric current. Moreover, macroscopic physical quantities can be cloned and recovered during transmission through channels. However, the quantum state follows the principle of the superposition state, i.e., the state of a quantum system is determined by its probability amplitudes in the quantum world, as discussed in Section \ref{Sec2.1}. Hence, quantum information is encoded, transmitted, stored, and manipulated in the form of unknown qubits characterized by microscopic particles such as photons, cold atoms, and ions. Compared to classical physical resources, these physical resources are more susceptible to environmental noise and cannot be cloned. The vast difference in physical resources introduced by the unique features of quantum mechanics is why entanglement-assisted quantum networks are fundamentally different from classical networks.
	
\begin{table*}[hbtp]
	\centering
	\caption{A comprehensive comparison between entanglement-assisted quantum networks and classical networks.}
	\label{differences}
	\begin{tabular}{m{3.0cm}<{\centering}m{6.5cm}<{\centering}m{6.5cm}<{\centering}}
		\toprule
		\textbf{Topics} &\textbf{Classical Networks}& \textbf{Entanglement-Assisted Quantum Networks} \\
		\midrule
		\textbf{Physical Resources}  & Deterministic signal characterized by macroscopic physical quantities such as voltage and electric current. & Unknown quantum state characterized by microscopic particles such as photons, cold atoms.\\		
		\textbf{Transmission Method}   &  Three steps: signal modulation, transmission through channel, and signal demodulation.& Quantum teleportation based on entanglement distribution and LOCC operation.  \\
		\textbf{Transmission Resources}  & Time, frequency, and spatial resources. & Physical: Time, frequency, and spatial resources. Logical: Entanglement resources. \\			
		\textbf{Transmission Medium}   &  Fiber, cable, free space, etc.  &  Fiber and free space (loss and decoherence).  \\		
		\textbf{Relay Solution}  &Signal regeneration and amplification. &  Quantum repeaters based on quantum memory and entanglement swapping.  \\	
		\textbf{Networking Devices}  & Hubs, routers, repeaters, switches, etc.  &   EPR sources, quantum repeaters, routers, etc.   \\	
		\textbf{Protocol Stack}  &  TCP, UDP, HTTP. & Ongoing. \\	
		\bottomrule
	\end{tabular}
\end{table*}

\textbf{Transmission Method.} In classical networks, the transmission of a data packet involves three steps. Firstly, the sender modulates the data it wants to send into the deterministic signal. Then, the signal is transmitted to the receiver hop by hop through classical channels. After receiving the signal, the sender performs the demodulation operation to obtain the communication data. The transmission of classical information inevitably suffers channel noise. The difference in the transmission method is the change from store-and-forward style to teleportation style, which is realized with the aid of entanglement swapping. In entanglement-assisted quantum networks, quantum information is transmitted using quantum teleportation technology, a specific transmission method of quantum communications. For quantum teleportation, the data qubits that encode quantum information can be teleported from the sender to the receiver without suffering from quantum channel noise. In this way, two communicating parties only need to establish end-to-end entanglement by iteratively performing entanglement swapping. After the sender performs a LOCC operation, the receiver can achieve the ``copy'' of the data qubits locally owned by the sender using entanglement features. Hence, the transmission method in entanglement-assisted quantum networks can be intuitively regarded as a communication process that consists of only two steps: end-to-end entanglement distribution and LOCC operations. Most notably, although data qubits are not subject to channel noise in quantum teleportation since data qubits are not transmitted over quantum channels, entangled qubit pairs shared by two communicating parties inevitably suffer channel noise, which results in low-fidelity entanglement and thus directly affects the accuracy of the transmitted quantum information. Consequently, establishing high-fidelity end-to-end entanglement is pivotal to quantum teleportation.

\textbf{Transmission Medium.} As discussed above, classical information is modulated into the deterministic signal characterized by macroscopic physical quantities before being transmitted. Various media can be used in classical information systems to describe macroscopic physical quantities. Thus, there are many types of transmission media in classical networks. Classical communication can generally be categorized as wired and wireless communication, and the corresponding transmission media include cable, fiber, radio waves, and microwaves. Besides, these transmission media only introduce loss errors in classical communication. However, qubits can only be transmitted as photons. Consequently, only two transmission media, namely optical fiber and free space, can be used to transfer qubits. Due to the inherent loss and noise in photonic channels, quantum communication inevitably faces two problems, photon loss and quantum decoherence, resulting from the transmission media. Both of these problems significantly affect the performance of qubits transmission in entanglement-assisted quantum networks.

\textbf{Transmission Resources.} In classical networks, transmission resources are time, frequency, and spatial resources. Time resources are time intervals or time slots used to transmit signals. Frequency resources refer to the frequency range used to transmit signals over classical channels. Spatial resources refer to the physical space used to allocate transmission channels or paths. These transmission resources are crucial for the transmission rate of data packets. The more time, frequency, and spatial resources are allocated for signal transmission, the more classical information can be transmitted. In entanglement-assisted quantum networks, qubits are transmitted with the assistance of end-to-end entanglement, and each entangled qubit pair can only be used to transfer a single qubit. Before quantum teleportation, two communicating parties need to share entangled qubit pairs by performing entanglement swapping along the selected path to ``couple'' each hop's entanglement links. Hence, the more entanglement links per hop, the higher the expected number of end-to-end entanglement that can be established, i.e., more qubits can be teleported from the sender to the receiver. Intuitively, the transmission resources of entanglement-assisted quantum networks are entanglement links (also called logical transmission resources). Notably, entanglement links are essentially entangled qubit pairs shared by adjacent quantum nodes. Entangled qubits are transmitted through quantum channels in the form of photons. Hence, transmission resources in entanglement-assisted quantum networks also include time, frequency, and spatial resources. The more time, frequency, and spatial resources allocated to entangled qubit transmission, the more entanglement links can be established. Therefore, time, frequency, and spatial resources are also the transmission resources of entanglement-assisted quantum networks, and these resources are called physical transmission resources.

\textbf{Relay Solution.} The transmission media of both classical and quantum communications introduce loss error. Hence, relay solutions are necessary for long-distance communication in classical and entanglement-assisted quantum networks. Notably, the state of a classical bit is determinate and not affected by ``measurement'' operations. Hence, signal regeneration and amplification techniques can be used to alleviate loss errors, thus achieving high-quality long-distance communication. However, limited by the no-cloning theorem, signal regeneration and amplification techniques cannot be applied to extend the distance of entanglement distribution. Fortunately, entanglement properties provide a relay solution, called entanglement swapping, for long-distance quantum communication. Notably, entanglement swapping is a quantum operation with the probabilistic feature due to the imperfection of physical devices. In this way, the distance of entanglement distribution can be extended from adjacent quantum nodes directly linked by a short-distance channel to a pair of quantum nodes at any long distance by performing LOCC operations only. Quantum repeaters play a pivotal role in entanglement-assisted quantum networks' relay solution. In order to establish high-fidelity distant entanglement, a large number of entangled qubit pairs need to be consumed to correct errors caused by quantum decoherence. Consequently, a practical quantum repeater requires the assistance of quantum memory. In summary, the relay solution in entanglement-assisted quantum networks is realized based on two essential processes of quantum repeaters, generating high-fidelity entanglement links with the help of quantum memory and performing swapping operations locally to extend entanglement distribution distances.

\textbf{Networking Devices.} In terms of physical structure, both classical and entanglement-assisted quantum networks are the interconnection of various end nodes and support concurrent network tasks over a long distance. These two networks require relay and routing devices to extend communication distance and network scale. In classical networks, hubs (also called multi-port repeaters) and repeaters are used to mitigate channel loss during data transmission, and switches and routers are responsible for routing packets to the right destinations. Notably, although the structure of entanglement-assisted quantum networks is more or less the same as that of a classical network, their transmission methods are fundamentally different. An entanglement-assisted quantum network works by distributing entangled qubit pairs between distant quantum end nodes to achieve quantum teleportation. Thus, apart from basic network devices like repeaters and routers, entanglement-assisted quantum networks require a networking device that can build entanglement links between adajcent quantum nodes. This basic but essential networking device, named EPR sources, aims at generating and then distributing entangled qubit pairs to establish entanglement links. With the assistance of EPR sources, quantum end nodes can be networked and communicate with each other using entanglement features.

\begin{figure}[t]
	\centering
	\includegraphics[width=1.0\linewidth]{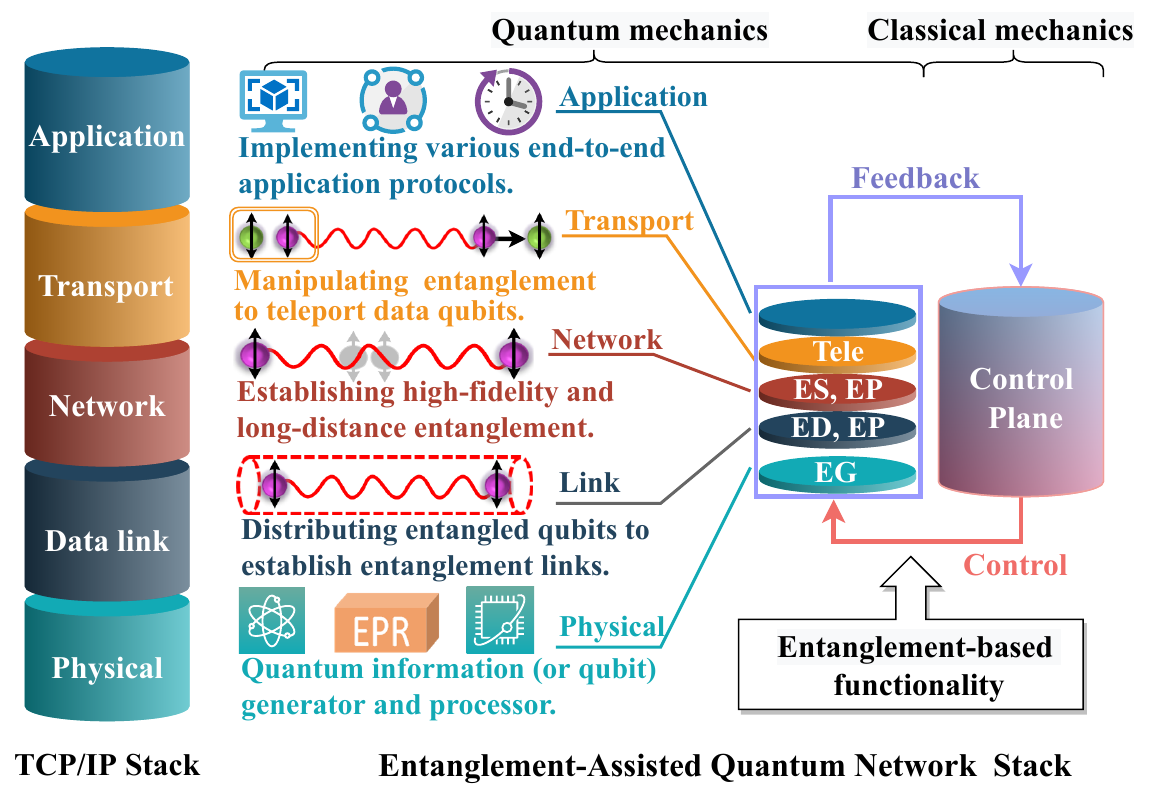}
	\caption{The comparison between TCP/IP stack and entanglement-assisted quantum network stack.}
	\label{stackcomparison}
\end{figure}

\textbf{Protocol Stack.} In order to simplify the complex process of end-to-end communication and support the network's iterative development, it is required to design a protocol stack for a network. Classical networks have been widely deployed, and some layered models have been proposed. There are two standard layered models, OSI (Open Systems Interconnection)~\cite{zimmermann1980osi} and TCP/IP \cite{cerf1974protocol}. Both models contain multiple layers, and each layer defines a set of specific functions or actions called protocols. For example, the application layer consists of HTTP protocol \cite{fielding1999hypertext}, the transport layer configures TCP or UDP protocol, and the network layer contains IP protocol. Each layer provides services to the layer above from the bottom to the top of the layered model. Besides, each layer only solves a part of the abstract problems and provides an overall solution through the cooperation of all layers.

Here, we take the five-layer TCP/IP stack and the implementation of end-to-end communication as examples to illustrate that the entanglement-assisted quantum network's protocol stack is significantly different from classical networks, as shown in \textbf{Fig.~\ref{stackcomparison}}. The stack design of entanglement-assisted quantum networks is driven by entanglement-based functionality. At the physical layer, both classical and entanglement-assisted quantum networks are responsible for processing information. In addition to processing data qubits, the entanglement-assisted quantum networks' physical layer also performs entanglement generation (EG) to prepare entangled qubits. Unlike the link layer in classical networks, quantum networks' link layer does not provide service for transmitting quantum information between adjacent quantum nodes. The link layer of entanglement-assisted quantum networks aims to provide high-quality entanglement resources for the network layer by entanglement distribution (ED) and entanglement purification (EP). Moreover, the classical network layer is responsible for routing and forwarding packets hop by hop. However, the network layer of entanglement-assisted quantum networks performs entanglement swapping (ES) to establish long-distance end-to-end entanglement. Notably, different from hop-by-hop information transmission in classical networks, swapping operations can be performed hop by hop (i.e., sequential), nested, or in parallel on a selected path. Classical and entanglement-assisted quantum networks provide end-to-end communication services at the transport layer, but end-to-end quantum communication is implemented by performing measurement operations, e.g., quantum teleportation. The application layer of classical and entanglement-assisted quantum networks enables applications to use the transmitted information. Most notably, all quantum operations require classical communication to control. Hence, quantum networks' protocol stack needs to design a control plane. In summary, each layer's function and implementation operation of entanglement-assisted quantum networks' protocol stack are different from those of classical networks, so the classical protocol stack cannot be directly applied to entanglement-assisted quantum networks.

In order to facilitate the development of quantum applications and their widespread use, it is also essential to develop methods that allow quantum protocols to make good use of the underlying quantum hardware and make fast and reactive decisions for efficiently generating entanglement in quantum networks to mitigate the limited qubits lifetime. However, it is hard to predict what the exact physical components and the upper user services will be, which means that the protocols in entanglement-assisted quantum networks have not been well defined. So far, some valuable studies have been done to pave the road for designing protocol stacks for entanglement-assisted quantum networks. \cite{illiano2022quantum} has  discussed three protocol stacks. Based on this work, we add additional work and present four proposals for the protocol stack in entanglement-assisted quantum networks, representing the most comprehensive state-of-the-art so far.

\begin{figure*}[hbtp]
	\centering
	\includegraphics[width=0.98\linewidth]{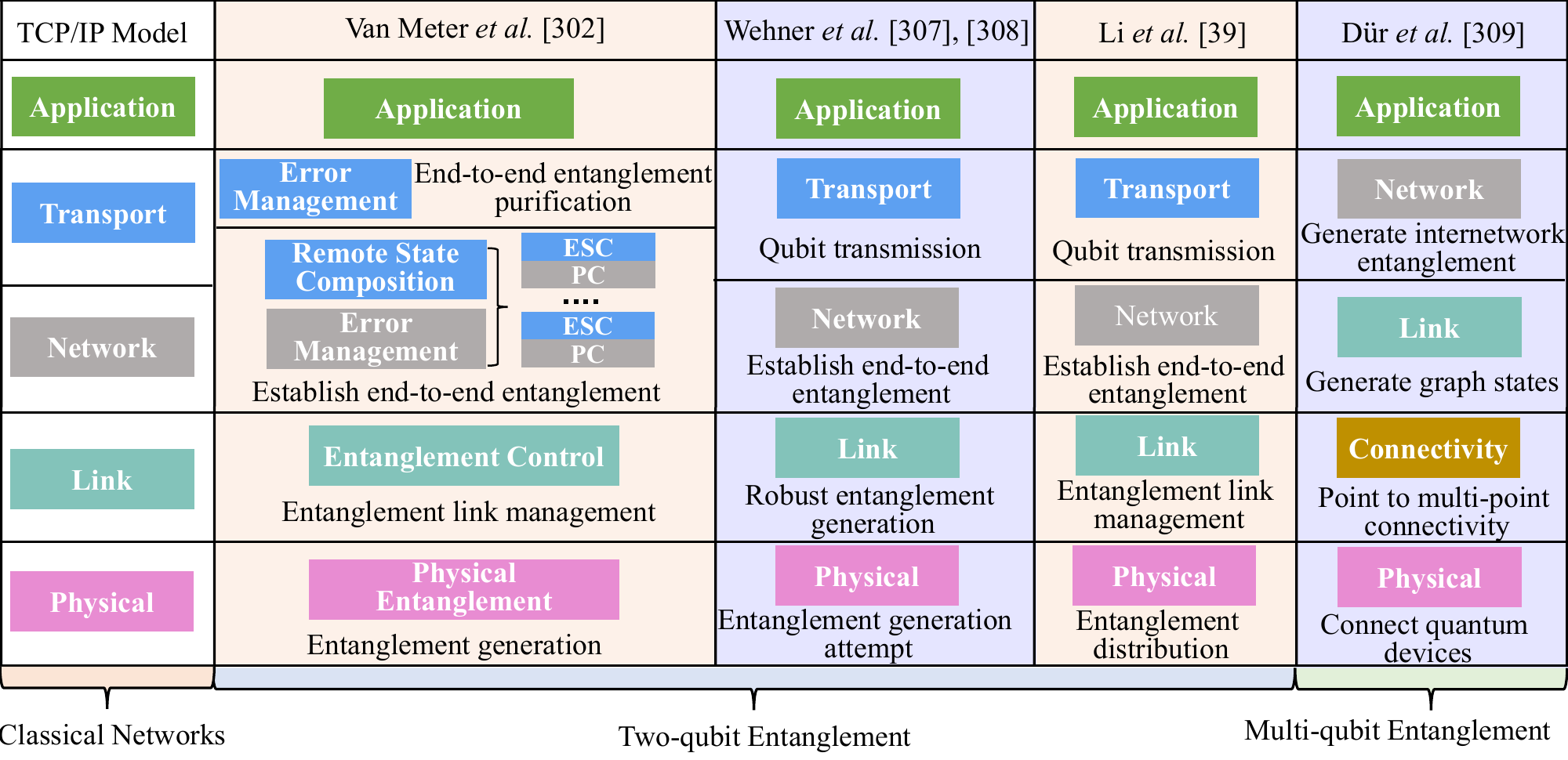}
	\caption{The representation of different protocol stacks designed for entanglement-assisted quantum networks.}
	\label{protocol-comparison}
\end{figure*}

\begin{enumerate}
	\item\textbf{Van Meter \textit{et al.}}: In \cite{van2013designing}, the authors provided the first description of a layered protocol stack for quantum repeater networks. This model emerged from the implementation processes of end-to-end entanglement distribution, including entanglement preparation, entanglement purification, and entanglement swapping. Specifically, the authors highlighted separate ``actions'' of entanglement, each action associated with a layer of the proposed protocol stack. The first action is the entanglement preparation attempt in the physical entanglement (PE) layer. The second layer, the entanglement control (EC) layer, is responsible for measuring some properties of the laser pulses implemented at the PE layer to determine the success of the entanglement preparation attempt. The error management layer then handles the entanglement purification with the assistance of a specific purification control (PC) protocol. The next layer is the remote state composition, which corresponds to the entanglement swapping action. This layer establishes end-to-end entanglement using an entanglement swapping control (ESC) protocol. The subsequent layer is the error management layer, which uses the PC protocol to establish high-fidelity end-to-end entanglement. Finally, the top layer is the application layer, where applications determine the necessity of end-to-end entanglement, or if our quantum states can be measured on a pay-as-you-go base. It is worth nothing that although the authors divided the implementation of end-to-end entanglement distribution into four types of quantum actions, in practical deployment, some quantum repeaters may deploy a protocol stack with more than six layers. For example, in addition to link-level entanglement purification, end-to-end entanglement purification is also required. Stemming from the aforementioned layered model, the authors introduced the quantum recursive network architecture (QRNA) \cite{lloyd2004infrastructure,van2011recursive}. Furthermore, the authors extended the above protocol stack by focusing on the issues of synchronization and signaling among quantum nodes~\cite{matsuo2019quantum}, as well as the higher layers from an entanglement-assisted quantum network services perspective \cite{van2022quantum}.
	
	\item \textbf{Wehner \textit{et al.}}: In \cite{dahlberg2019link,kozlowski2020designing}, the authors proposed a layered model of entanglement-assisted quantum networks based on bipartite entanglement. In this model, end-to-end entanglement is established on demand. The protocol stack is divided into five layers. Currently, the author are focusing on the physical layer, link layer, and network layer protocol design. The first layer, known as the physical layer, incorporates an auxiliary protocol called the midpoint herald protocol (MHP). Its purpose is to generate entanglement between two adjacent quantum nodes within a finite time slot. The MHP consults the upper link layer to determine if entanglement needs to be generated within a given time slot. The link layer is responsible for robust entanglement generation and utilizes the quantum entanglement generation protocol (QEGP). QEGP receives entanglement requests from high layers, which include parameters such as remote node ID, number of entanglement pairs, minimum fidelity, and request type. These instructions are then provided to the physical layer protocol. The entanglement preparation is triggered on demand to alleviate the negative effects of quantum decoherence. This means that entanglement is generated as needed, rather than continuously. The network layer is responsible for establishing entanglement over long distances by implementing entanglement swapping using link layer functions. Additionally, the network layer includes an entanglement manager that keeps track of each entangled qubit pair within the network to avoid confusion. Finally, the transport layer facilitates the transmission of qubits upon requests from the application layer.  Examples of such a request are quantum computing and quantum communication.
	
	\item \textbf{D{\"u}r \textit{et al.}}: Different from the two proposals mentioned above, the protocol stack of entanglement-assisted quantum networks proposed in~\cite{pirker2019quantum} is based on multi-particle entangled states. The layers' responsibilities range from establishing point-to-point connectivity, intra-network graph state generation, to inter-network routing of remote entanglement distribution. The authors assume that an entanglement-assisted quantum network operates in three phases: dynamic, static, and adaptive. In the first phase, quantum nodes prepare entangled qubits and utilize quantum channels to distribute them among each other. The number of entangled qubits shared by quantum nodes changes dynamically from 0 to some, hence, termed the dynamic phase. After the first phase, quantum nodes share entangled states. The number of shared entangled states between quantum nodes is fixed before providing service for remote entanglement distribution requests, resulting in the static phase. Finally, entangled qubits are manipulated to fulfill the nodes' requests in the adaptive phase. Stemming from the three phases, the authors organize the protocol stack into four layers: the physical layer, connectivity layer, link layer, and network layer. The physical layer corresponds to the quantum channels connecting adjacent quantum nodes, for example, optical fibers or free space channels. It is responsible for forwarding entangled qubits from one network to the other. The connectivity layer tackles errors caused by imperfect quantum channels and ensures point-to-point or point-to-multipoint connectivity in terms of high-fidelity entangled states between quantum nodes. The link layer defines the boundaries of an entanglement-assisted quantum network in terms of entangled, distributed, multi-particle network states. It aims to generate arbitrary graph-based entangled states on request in a network. The network layer is responsible for generating and manipulating inter-network entanglement to enable graph state requests spanning several different small-scale entanglement-assisted quantum networks through quantum routers. In this protocol stack design, each layer above the physical layer has access to auxiliary protocols for entanglement purification, entanglement swapping, and monitoring the internal state of small-scale entanglement-assisted quantum networks. Moreover, the authors proposed several auxiliary protocols operating on these layers. Within the above framework, the authors updated their proposal with a genuine network model \cite{miguel2021genuine}, capable of handling quantum superpositions of network tasks.	
	
	\item \textbf{Li \textit{et al.}}: Inspired by the OSI layering model of the current Internet, the authors described a five-layer protocol stack to support the iterative development of the quantum internet \cite{li2021building}. The authors assumed that entanglement links are continuously established and be regarded as the underlying network resource used to serve request. This work discusses the specific functions that need to be implemented at each layer to  build future entanglement-assisted quantum networks. More specifically, the physical layer of the quantum internet should shield the differences between physical devices and transmission media. The link layer aims to control entanglement links and process feedback from the network layer to maintain entanglement between two adjacent quantum nodes without breaking. The main work of the network layer is to pre-define a swapping path and perform entanglement swapping on the selected path to establish remote end-to-end entanglement. Additionally, entanglement purification must be performed at the link layer to correct loss errors and operation errors to generate high-fidelity entanglement. The responsibility of the transport layer is to ensure the reliable transmission of qubits by designing transmission control protocols. Finally, the application layer's function is to enable multiple quantum system application processes to communicate with each other and complete a series of services required for business processing. Based on the above hierarchical structure, the authors presented a cluster-based structure to describe how quantum nodes can be effectively interconnected, thereby achieving higher performance in the designed entanglement-assisted quantum network. Inspired by the hierarchical structure of classical networks, the concept of the quantum local area network (QLAN) is proposed as an essential component of the quantum internet. In this design, numerous quantum repeaters and routers are interconnected to form a quantum core network, and multiple QLANs are interconnected through the quantum core network.
\end{enumerate}

All four protocol stack models described above recognize quantum entanglement as the key resource of entanglement-assisted quantum networks. These models can be divided into two categories, i.e., the model designed for two-qubit entangled systems and the model designed for multi-qubit entangled systems, as shown in \textbf{Fig.~\ref{protocol-comparison}}. For the two-qubit-based model, the first protocol stack model proposed by Van Meter \textit{et al.} presents a clear distinction between layers based on the ``level'' of entanglement, including single-hop, multi-hop, and end-to-end entanglement. Besides, this model considers error correction during entanglement distribution and extension. In this model, the network layer runs multiple rounds of entanglement swapping and entanglement purification, which is essentially different from other models. The second protocol stack model proposed by Wehner \textit{et al.} focuses on on-demand entanglement distribution and mainly considers the probabilistic nature of entanglement preparation to design an auxiliary protocol for robust entanglement distribution between adjacent quantum nodes. The fourth protocol stack model proposed by Li \textit{et al.} adopts a five-layer model inspired by the OSI layering model, regarding swapping-based remote entanglement distribution as the establishment of communication connections in classical networks. The main difference between the second and fourth models is the function of the link layer. In the second model, the link layer is designed to realize robust entanglement generation for requests. However, in the fourth model, the authors adopted the continuous entanglement method, i.e., entanglement links are continuously established and stored in quantum memory as the underlying network resources. Hence, the link layer in the fourth model works by establishing and managing entanglement links, including entanglement distribution and entanglement link allocation. The fourth model relies more on the implementation of perfect quantum memories than the second model since the second model is conducive to the implementation of on-demand entanglement distribution. However, neither of them adequately considers entanglement purification in entanglement-assisted quantum networks. For the multi-qubit-based model, the third protocol stack model proposed by D{\"u}r \textit{et al.} envisions a physical layer focusing on single-hop entanglement. However, both the two upper layers provide services lying at the intersection between different classes. Similarly, the connectivity layer acts on the single hop as well as on intermediate nodes.

Although some valuable protocol stack models are proposed in recent years, the unique features of quantum mechanics still pose many challenges to designing protocol stack models for entanglement-assisted quantum networks. Firstly, considering that quantum manipulation cannot be achieved without classical communication and the ``collapse-after-measurement'' phenomenon, designing a mapping between classical information and the underlying qubits is required for the protocol stack. Besides, it is challenging to design an error correction protocol at the link layer. Error detection at the link layer needs to efficiently reduce the adverse effects of loss and operation errors. However, although quantum error-correcting codes can theoretically realize error detection, this method may be prohibitively expensive in practice. Moreover, qubits' lifetime is short, so the network layer is required to synchronize the randomly generated entanglement quickly to establish high-fidelity end-to-end entanglement. However, in a large-scale entanglement-assisted quantum network, it is extremely hard to achieve entanglement synchronization at the network layer due to the limited information on each quantum node. Overall, designing the protocol stack for an entanglement-assisted quantum network still requires researchers' persistent efforts.

\begin{figure*}[hbtp]
	\centering
	\includegraphics[width=0.76\linewidth]{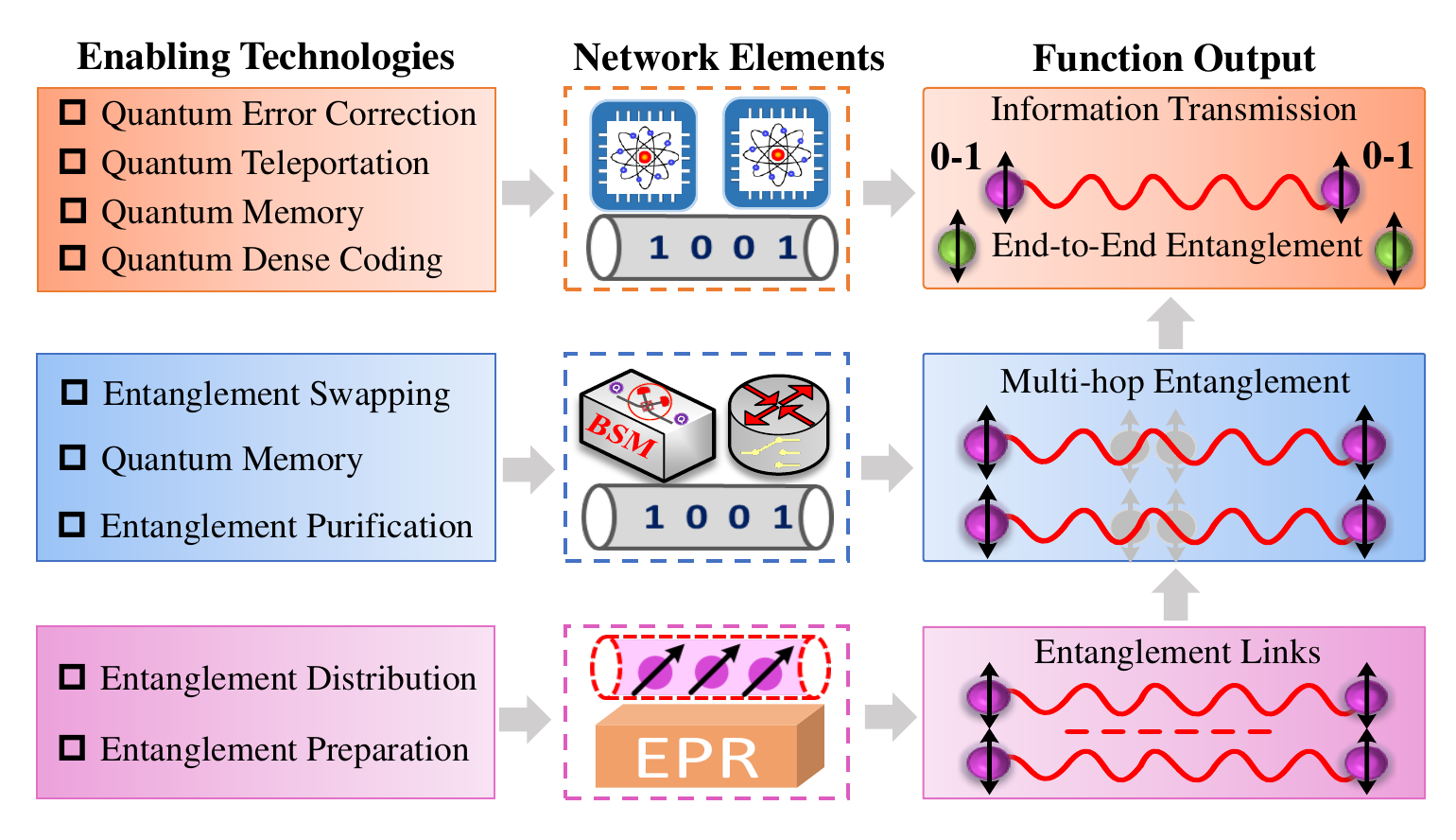}
	\caption{The connection map between enabling technologies and network elements in entanglement-assisted quantum networks.}
	\label{technologies-elements}
\end{figure*}

In summary, entanglement-assisted quantum networks essentially differ from classical networks. Since entanglement-assisted are governed by the fundamental laws of quantum mechanics with no counterpart in classical networks, it shows significant advantages compared with classical networks. Firstly, in terms of information transmission, entanglement-assisted quantum networks can realize unconditionally secure communication due to the no-cloning theorem, which is challenging to realize in classical networks. Besides, entanglement-assisted quantum networks perform better in computing power than classical networks. In the classical world, computing power is limited by Moore's Law. However, the superposition principle in quantum mechanics can effectively overcome the limitation of Moore's Law and thus achieve an exponential increase in computing power. Hence, entanglement-assisted quantum networks can further significantly improve computing power by interconnecting multiple quantum computers. Moreover, the non-local correlation feature of quantum entanglement makes quantum sensing more precise and sensitive than classical sensing. Although entanglement-assisted quantum networks have tremendous advantages over classical networks, their implementation also poses many challenges ranging from physical resources to network designs, such as entanglement preparation and routing design. Currently, researchers in academics and practitioners in industry are making efforts to overcome these challenges: academic researchers focus on designing efficient schemes to improve the performances of enabling technologies, e.g., entanglement purification and quantum memory, and those in the industry are exploring the application of enabling technologies and developing practical physical devices. A series of breakthroughs, such as the long-lifetime storage scheme and micro quantum chip, have been made, which pave the way for building high-performance entanglement-assisted quantum networks with the strong anti-noise ability and high qubit transmission rate.

\subsection{Network Elements}
In order to support concurrent qubit transmission tasks over long distances, numerous quantum end nodes are networked with the assistance of physical channels and networking devices. Hence, the elements of an entanglement-assisted quantum network mainly include physical channels, networking devices, and quantum end nodes. In entanglement-assisted quantum networks, network elements are the entities that enable quantum technologies to function. The connection map between enabling technologies and network elements is shown in \textbf{Fig.~\ref{technologies-elements}}. By adopting entanglement preparation and distribution technology, EPR sources can be used to establish entanglement links between adjacent quantum nodes with the assistance of quantum channels. Furthermore, long-distance entanglement can be generated by performing entanglement swapping and entanglement purification on the entangled qubit pairs stored in quantum memories. Quantum routers or quantum repeaters can cooperate with classical channels to realize this function. After end-to-end entanglement is established, quantum end nodes can adopt quantum teleportation or quantum dense coding to implement high-performance information transmission with the help of quantum error correction and quantum memories.

\begin{figure}[t]
	\centering
	\includegraphics[width=1.0\linewidth]{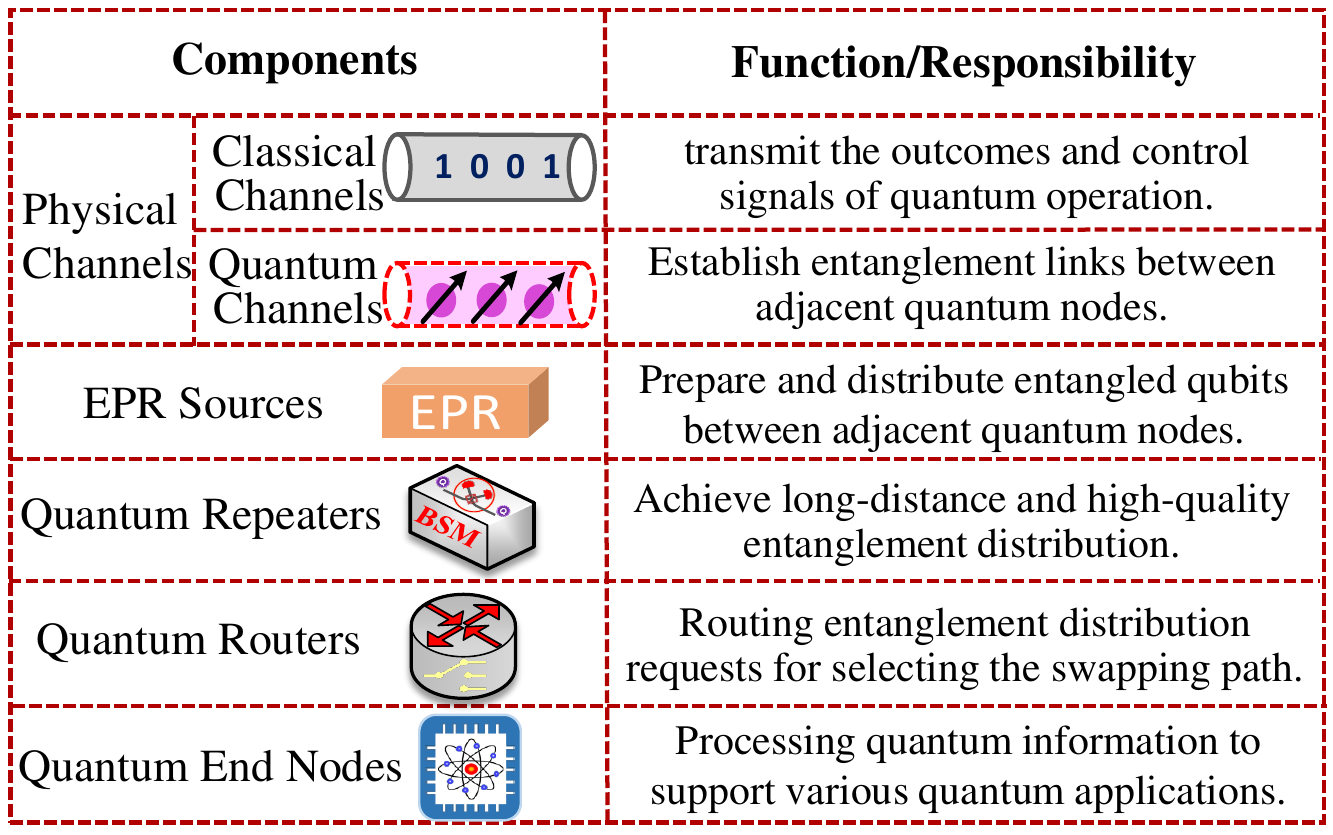}
	\caption{The components of entanglement-assisted quantum networks.}
	\label{components}
\end{figure}

As described above, the network elements and their functions can be summarized briefly as shown in \textbf{Fig.~\ref{components}}. The requirements and functions of these network elements are discussed in detail as follows.

\textbf{Physical Channels.} As the basic component of entanglement-assisted quantum networks, physical channels are communication links connecting adjacent quantum nodes. Most notably, entanglement-assisted quantum networks work with the assistance of classical communications. Consequently, there are two types of physical channels in entanglement-assisted quantum networks, i.e., classical and quantum. The classical channels are communication links adopted for transmitting classical messages, such as measurement outcomes and control signals, between communicating parties. Quantum channels serve as quantum links between adjacent quantum nodes. A ``good'' quantum channel possesses three features: low loss, low quantum decoherence, and high bandwidth. However, since qubits can only travel as photons over a channel, quantum channels can only be photonic channels. There are two types of photonic channels: free-space channels \cite{bloom2003understanding} and optical fiber channels~\cite{keiser2000optical}. Free space refers to the space in the atmosphere where optical waves can travel freely and the space beyond the atmosphere. Generally speaking, free space means air as opposed to material, transmission line, fiber-optic cable, etc. An optical fiber is a flexible, transparent fiber made by drawing glass (silica) or plastic. Optical fibers are used most often as a means to transmit photons, the transmission principle of which is ``total reflection of light''. Each quantum channel has its advantages and disadvantages. For example, free-space channels perform better than optical fiber channels in terms of the higher bandwidth of the channels, but optical fiber channels outperform free-space channels in terms of lower photon loss, e.g., the loss can be as low as 0.2dB in an optical fiber, but the photon loss of free-space is as high as 50dB even in a non-foggy environment. A future entanglement-assisted quantum network or the quantum internet may use a combination of these photonic channels.

\textbf{EPR Sources.} Most of the quantum applications require distant quantum end nodes to share entangled qubit pairs. Thus, EPR sources are required in entanglement-assisted quantum networks. EPR sources achieve the preparation and distribution of entangled qubits between adjacent quantum nodes, i.e., establish entanglement links. The preparation of entangled qubits can be achieved with NV centers in diamond, trapped ions, atoms, and superconducting circuits, and these schemes have been demonstrated experimentally. Besides, there are two fashions of entanglement distribution between two directly linked quantum nodes, i.e., deterministic \cite{turchette1998deterministic,riste2013deterministic,luo2017deterministic} and heralded \cite{bernien2013eralded,delteil2016generation,liu2021heralded}. The deterministic entanglement distribution fashion means that there is no inherent probabilistic nature to the EPR sources. Heralded fashion is a slightly weaker form of deterministic entanglement distribution in which the successful generation of entangled qubits can be denoted as an event independent of the measurement of entangled qubits themselves. Here, the generation of entanglement is deterministic, conditioned on such a successful heralded signal.  With the development of physical materials and devices, high-quality entangled qubits can be efficiently generated. Besides, entanglement preparation and distribution technologies discussed in Section \ref{entanglement-generation} are becoming mature. Hence, it is convinced that EPR sources can effectively work to serve entanglement-assisted quantum networks.

\textbf{Quantum End Nodes.} Quantum end nodes are the basic platform where various quantum applications can run. Similar to classical end nodes, a quantum end node is essentially a processor that can manipulate qubits. Quantum end nodes need to meet the following requirements. Firstly, quantum memory is an essential component for quantum end nodes. The reason is that some quantum applications require the simultaneous manipulation of multiple qubits, but only a single qubit can be teleported by performing quantum teleportation. Consequently, quantum end nodes must achieve the robust storage of qubits to synchronize random arrivals of qubits. Besides, quantum nodes need to achieve the high-fidelity processing of quantum information. Decoherence is inevitable in quantum information processing due to the environmental noise introduced by physical hardware. Thus, minimizing the fidelity attenuation of qubits during quantum information processing is necessary for quantum end nodes to support advanced tasks such as quantum computing and quantum sensing. Lastly, high compatibility with photonic communication hardware is required for quantum end nodes. This is because qubits can only be transmitted as photons over a channel, but quantum end nodes will adopt various physical resources to process quantum information. Thus, a high-efficiency converter is required to tackle the problem of incompatibility between different physical hardware. For this requirement, many research groups are currently working toward such a converter \cite{probst2013anisotropic,andrews2014bidirectional,bochmann2013nanomechanical}. It is expected that quantum end nodes will be developed into small-scale quantum computers like classical computers.

\begin{figure}[t]
	\centering
	\includegraphics[width=1.0\linewidth]{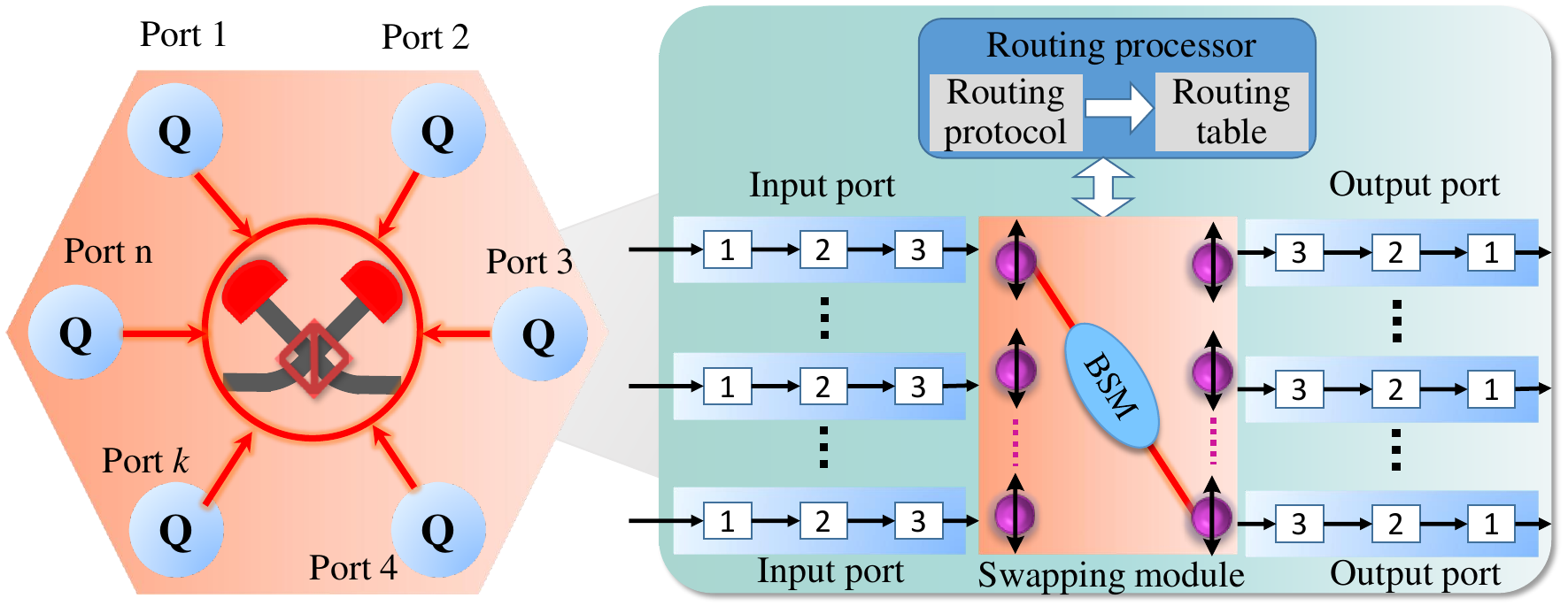}
	\caption{The abstract structure of a quantum router.}
	\label{router-architecure}
\end{figure}

\textbf{Quantum Repeaters.} Since the inherent loss and quantum decoherence in photonic channels, it is challenging to achieve the transmissions of entangled qubits between two distant quantum nodes. An intermediate device is required to overcome the limitation of distance. Fortunately, quantum repeaters can improve the transfer rate of entangled qubits between distant quantum nodes. Notably, the signal amplification technology adopted in classical repeaters cannot be applied to quantum repeaters because of the no-cloning theorem. Essentially, a quantum repeater works by performing entanglement swapping to solve the problem of low transmission rates over long distances. As discussed in Section~\ref{Sec3.5}, entanglement swapping can extend the distance of entanglement distribution without subjecting entangled qubits to channel noise. Thus, quantum repeaters can effectively overcome the distance limitation of entanglement distribution. Besides, to improve the robustness of establishing high-fidelity end-to-end entanglement, a quantum repeater needs to configure quantum memory to store entangled qubits. Lastly, in order to achieve high-quality entanglement distribution between distant quantum end nodes, quantum repeaters are required to enable erasing errors caused by photon loss and quantum decoherence during transmission. \cite{muralidharan2016optimal} classified all the proposed quantum repeater schemes into three generations depending on the methods used to correct errors. Basically, the first generation of quantum repeaters use entanglement purification to suppress operation errors~\cite{sangouard2011quantum}, the second generation of quantum repeaters use QEC to correct operation errors \cite{jiang2009quantum}, and the third generation of quantum repeaters rely on QEC to correct both loss and operation errors \cite{munro2012quantum,muralidharan2014ultrafast}. A future quantum repeater can correct loss and operation errors simultaneously, i.e., quantum repeaters support QEC and entanglement purification. Thanks to the development of quantum memory and entanglement-based technologies, efficient quantum repeaters are expected to be realized to effectively extend the communication distance.

\begin{figure*}[t]
	\centering
	\includegraphics[width=0.96\linewidth]{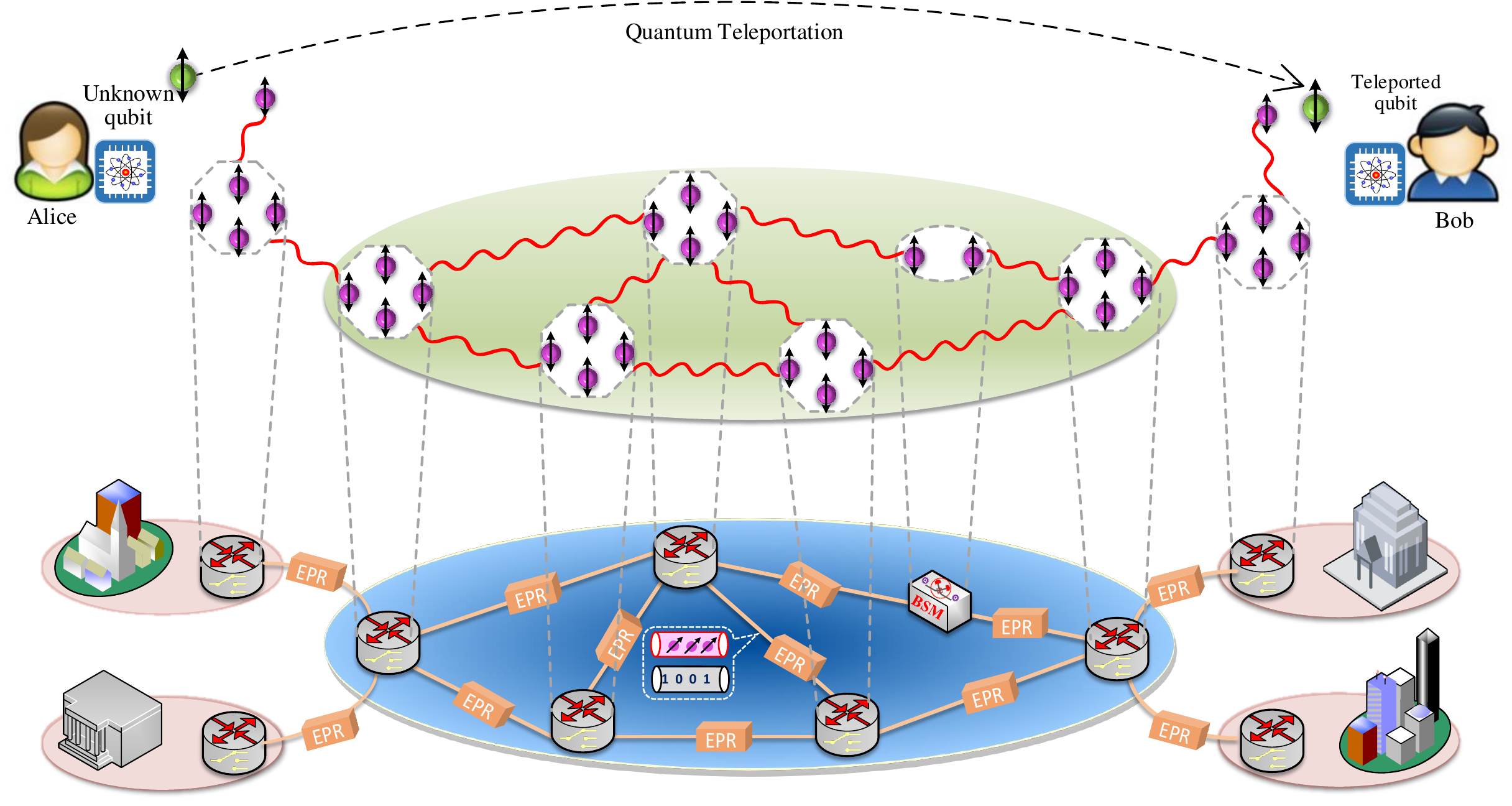}
	\caption{The structure of an entanglement-assisted quantum network.}
	\label{structure}
\end{figure*}

\textbf{Quantum Routers.} Entanglement distribution between any pair of distant quantum end nodes is achieved by performing entanglement swapping along quantum repeater chains to ``couple'' entanglement links into end-to-end entanglement. However, an entanglement-assisted quantum network is not a simple combination of multiple linear quantum repeater chains. Similar to the data packet routing in classical networks, quantum routing of entanglement from one source end node to one or many destination end nodes is an essential function for entanglement-assisted quantum networks. Besides, a quantum device is required to aggregate multiple quantum nodes together to extend the network scale. Therefore, quantum routers are requisite networking devices for routing entanglement distribution flows and extending network scale in entanglement-assisted quantum networks. The abstract structure of a quantum router is shown in \textbf{Fig.~\ref{router-architecure}}. A quantum router is a quantum physical device with multiple input and output ports \cite{yan2014single}, and it can route the input entanglement distribution flow to the correct output port \cite{lee2022quantum} with the assistance of a routing processor. In a quantum router, each port configures some quantum registers (or memory units) to store entangled qubits to establish entanglement links with adjacent quantum nodes. Besides, a quantum router needs to deploy a swapping module, i.e., a measurement device, which is similar to the packet switching module in a classical router. For each input entanglement distribution flow, the routing process involves three steps. Firstly, the routing processor selects an output port according to the routing table obtained by running the configured routing protocol. Then entangled qubits are retrieved from quantum registers at the input and output ports. Finally, the swapping module performs BSM operation on two entangled qubits to swap entanglement, thereby routing entanglement distribution flow to the next hop. In summary, the functions of quantum routers are similar to but more than the quantum repeaters. In addition to storing entangled qubits and extending the distance of entanglement distribution, quantum routers need to use local entanglement to link different neighbors' entangled qubits. In a nutshell, quantum routers are responsible for extending the entanglement distribution flow from the source node to the destination one via the path selected by the routing protocol. Currently, the basic research problem of quantum routing has been investigated in various systems \cite{aoki2009efficient,lemr2013linear,hoi2011demonstration,behera2019designing,ren2022nonreciprocal} but with numerous constraints. Driven by the development of physical devices, a future quantum router is expected to efficiently achieve the routing of entanglement distribution flows.

\subsection{Network Structure}
Although some valuable works \cite{van2011recursive,pirker2018modular} have explored the structure of entanglement-assisted quantum networks, these schemes are not conducive to network scaling and protocol stack design. The differences between entanglement-assisted quantum networks and classical networks are caused by the unique features of quantum mechanics. However, they have the same network components, except for EPR sources, and the networking devices present the same function. For example, repeaters are deployed to extend the distance of information transmission between end nodes, and quantum routers are adopted to scale network size and route requests. In entanglement-assisted quantum networks, EPR sources aim at distributing entangled qubits between adjacent quantum nodes to provide link resources for qubit transmission. An EPR source coupled with a quantum channel can be likened to a classical channel in classical networks' structure. Hence, entanglement-assisted quantum networks have a similar structure to classical networks.

\begin{figure*}[t]
	\centering
	\subfigure[Enc-to-end classical communication.]{
		\includegraphics[width=0.8\linewidth]{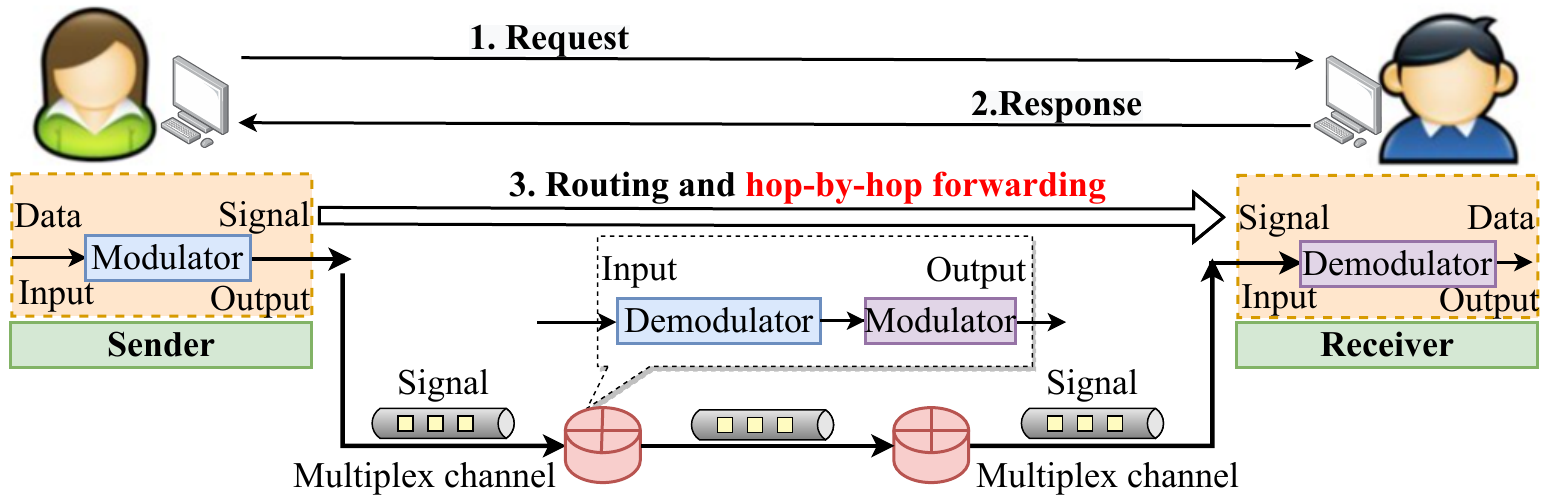}
		\label{classical}}
	\quad
	\subfigure[End-to-end quantum communication.]{
		\includegraphics[width=0.8\linewidth]{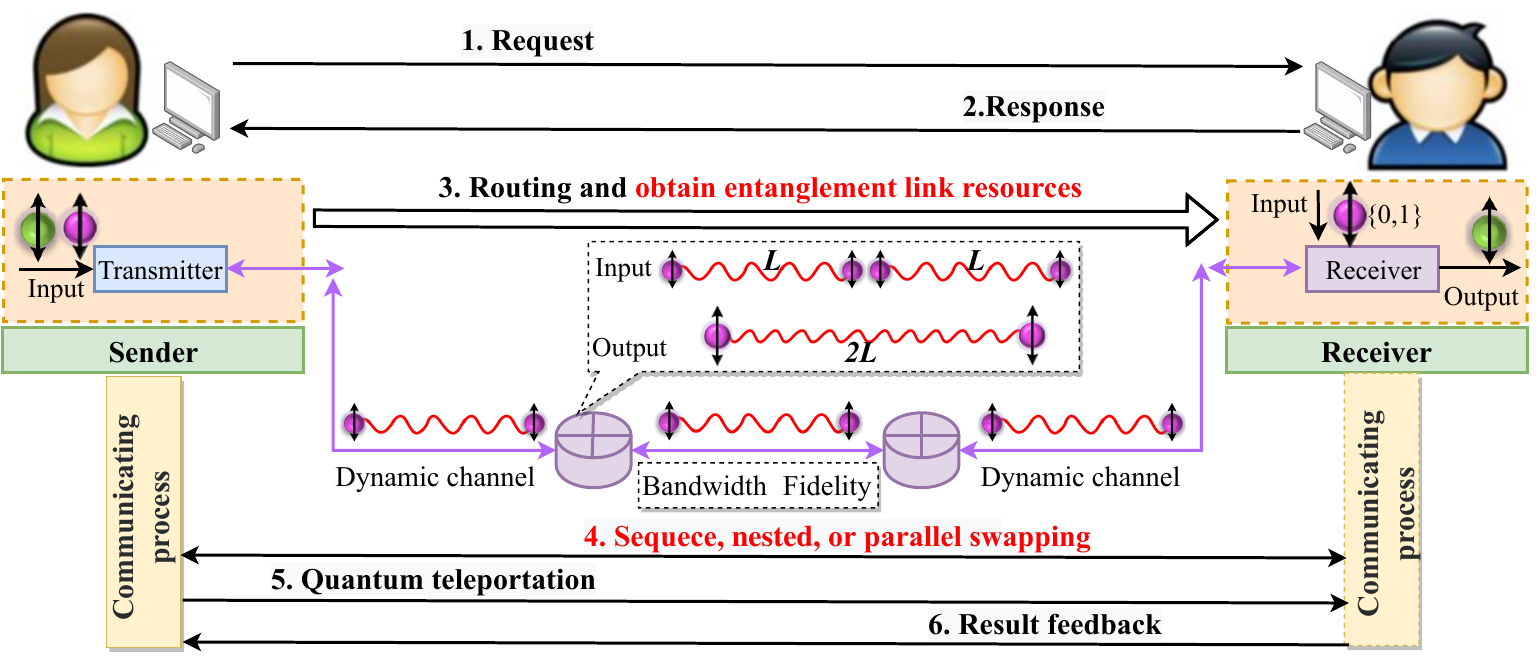}
		\label{quantum}}
	\caption{The implementation of end-to-end communication in classical networks and entanglement-assisted quantum networks.}
	\label{principlecompare}
\end{figure*}

Here, inspired by the hierarchical structure of classical networks, we present a general structure for developing wide-area entanglement-assisted quantum networks. As shown in \textbf{Fig. \ref{structure}}, a large-scale and wide-area entanglement-assisted quantum network consists of a backbone quantum network and many QLANs. The backbone network is the interconnection of many quantum routers, quantum repeaters, and EPR sources. More concretely, any pair of two adjacent quantum routers over short distances are directly linked via quantum channels in synergy with classical ones. Two adjacent quantum routers far apart are connected with the assistance of quantum repeaters. Besides, in the backbone network, EPR sources are deployed between any pair of adjacent quantum nodes to establish entanglement links. These pairs can be repeater to repeater, repeater to router, or router to router. The backbone network is designed to achieve entanglement routing and establish remote entanglement. Similar to classical LANs, QLANs are composed of three major parts: quantum information processing equipment, network connection equipment, and transmission medium. Each quantum information equipment is a quantum end node that can handle various quantum tasks. The network connection equipment consists of quantum routers and EPR sources. In QLANs, adjacent quantum nodes are linked via the transmission medium, i.e., quantum channels and classical channels. QLANs are responsible for quantum information processing, thus providing a promising platform for various quantum applications. In a wide-area entanglement-assisted quantum network, quantum information interaction between two QLANs is achieved via the backbone network.

\subsection{Working Principle}

A future entanglement-assisted quantum network works by distributing entangled qubit pairs between distant quantum end nodes and performing quantum teleportation to transmit quantum information, thus supporting various quantum applications. Here, we take the example of quantum communication between two quantum end nodes (Alice and Bob) belonging to different QLANs to describe the working principle of entanglement-assisted quantum networks. Because of the significant difference in enabling technologies, the implementation of end-to-end communication in entanglement-assisted quantum networks differs from that in classical networks. \textbf{Fig.~\ref{principlecompare}} presents the general implementation of end-to-end communication in classical networks and entanglement-assisted quantum networks.

As shown in \textbf{Fig.~\ref{classical}}, end-to-end classical communication is generally realized via three steps. Firstly, Alice sends a request for communication to Bob. Then Bob responds to Alice's communication request. Thirdly, Alice encodes the data into signals suitable for transmission over a classical channel, classical networks provide routing service to forward the data hop by hop along a selected path, and Bob decode the received signals to obtain the data transmitted by Alice. During end-to-end classical communication, a path connecting Alice and Bob is selected according to the specific rule (i.e., routing protocol), and each intermediate networking device in the selected path decodes the input signal and then encodes it again before being forwarded to the next hop. In summary, classical information is directly transmitted hop by hop through classical channels.

As shown in \textbf{Fig.~\ref{quantum}}, the implementation of end-to-end quantum communication generally follows six steps. Firstly, Alice sends a communication request to Bob. Then Bob responds to Alice's request. Thirdly, entanglement-assisted quantum networks route Alice's request to Bob along a selected path and allocate entanglement link resources for two communicating parties. Fourthly, swapping operations are performed using a sequence, nested or parallel manner on the allocated entangled qubit pairs to establish end-to-end entanglement and then Alice performs quantum teleportation to transmit quantum information. Finally, Bob feedback the results of quantum teleportation to Alice. In summary, end-to-end quantum communication requires establishing end-to-end entanglement first by ``coupling'' multiple entanglement links along a selected path, and quantum information is not directly transmitted hop by hop through quantum channels but through a LOCC operation.

As described above, although entanglement-assisted quantum networks' working principle differs from that of classical networks, the implementation of end-to-end classical communication can provide constructive guidance for end-to-end quantum communication. The reason is that end-to-end entanglement distribution is similar to end-to-end classical communication, especially when swapping operations are performed sequentially along the selected path. However, different from data flows that can share the transmission resources of classical channels, entanglement links cannot be shared by quantum communication requests due to the collapse-after-measurement phenomenon. Hence, the capacity of quantum channels, i.e., bandwidth, is dynamic. Besides, the quality of the entanglement link significantly affects end-to-end entanglement's fidelity and thus affects the efficiency of quantum teleportation. Hence, it is required to alleviate quantum decoherence by performing purification operations during end-to-end quantum communication. In summary, we can learn from end-to-end classical communication to design end-to-end quantum communication protocols, but the unique features of entanglement-assisted quantum networks need to be fully considered.

Inspired by end-to-end classical communication strategies, there are two typical strategies, namely connection-oriented and connectionless~\cite{li2022connection}, to achieve end-to-end quantum information transmission in entanglement-assisted quantum networks. The essential difference between the two quantum communication strategies is whether the entanglement link resources shared by adjacent quantum nodes are generated before path selection. The connection-oriented strategy selects a path and assigns the dedicated quantum memory unit for two communicating parties to establish end-to-end entanglement, then entanglement distribution is performed to fulfill the allocated memory units to generate the required entanglement links resources. The connectionless strategy selects a path and allocates the generated entanglement link resources to two communicating parties along the selected path. For two quantum communication strategies, swapping operations can be performed sequentially or in parallel to realize remote entanglement distribution. Here, we assume that entanglement swapping is performed hop-by-hop along a path to establish long-distance end-to-end entanglement and elaborate on the implementation of two strategies as follows.

\begin{figure}[t]
	\centering
	\includegraphics[width=1.0\linewidth]{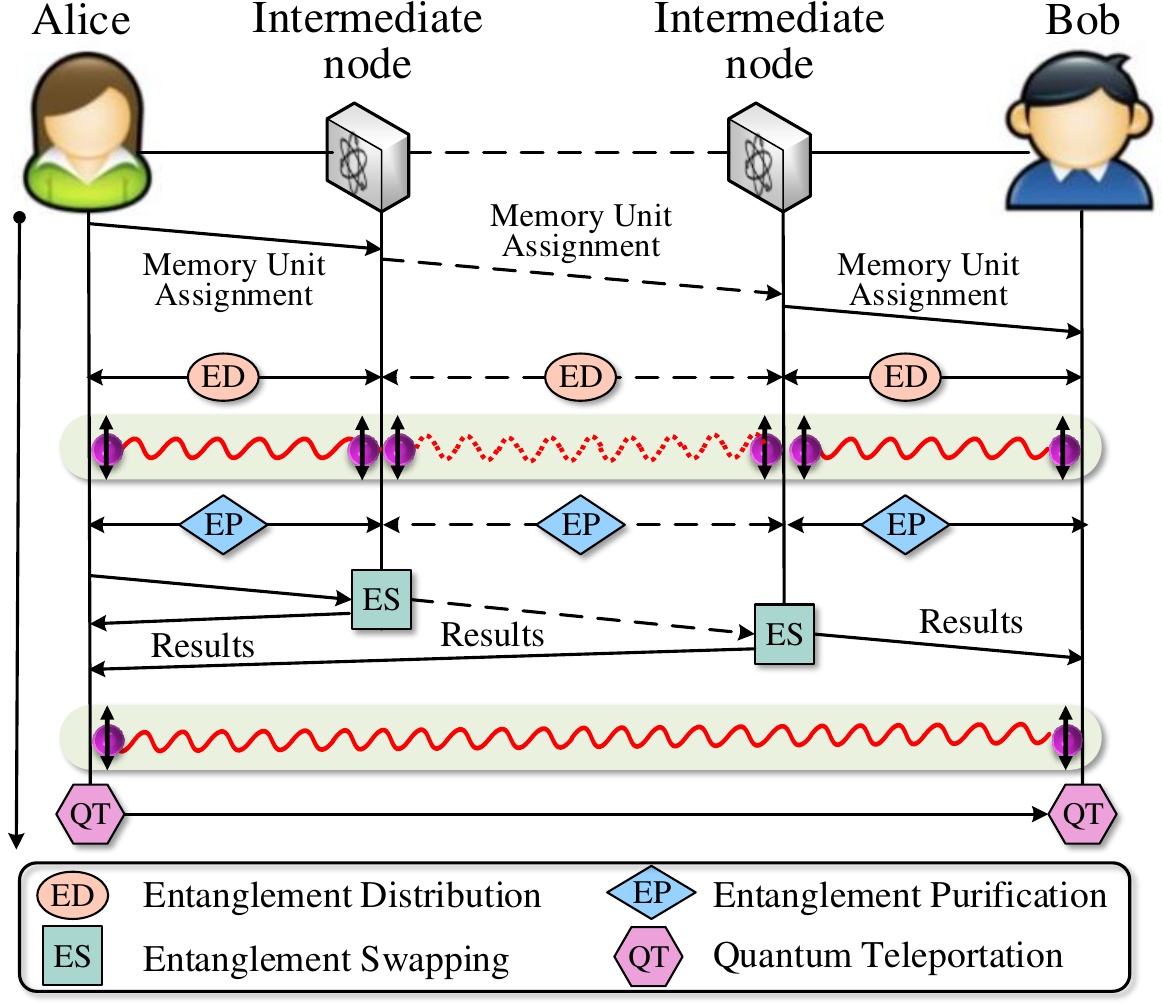}
	\caption{The connection-oriented qubit transmission between Alice and Bob.}
	\label{connection-oriented}
\end{figure}

For the connection-oriented strategy (as shown in \textbf{Fig.~\ref{connection-oriented}})~\cite{li2022connectiona}, each quantum node on the selected path allocates the dedicated quantum memory units for each request to establish the dedicated end-to-end entanglement connections, similar to the virtual circuit in classical communication. Concretely, a path connecting Alice and Bob is first selected, and each quantum node on the selected path allocates the dedicated memory units for this source-destination (SD) pair. Then, entanglement distribution is performed between adjacent quantum nodes to fill quantum memory units, i.e., entanglement links are established on demand. In order to realize high-quality quantum teleportation, entanglement purification is performed to improve the fidelity of entanglement links. After that, swapping operations are performed along the selected path to establish end-to-end entanglement connections. Finally, Alice can teleport qubits to Bob using quantum teleportation technology. For the connection-oriented strategy, qubits are teleported from Alice to Bob along the same path. After quantum communication, two communicating parties notice quantum nodes to free the dedicated quantum memory units.

\begin{figure}[t]
	\centering
	\includegraphics[width=1.0\linewidth]{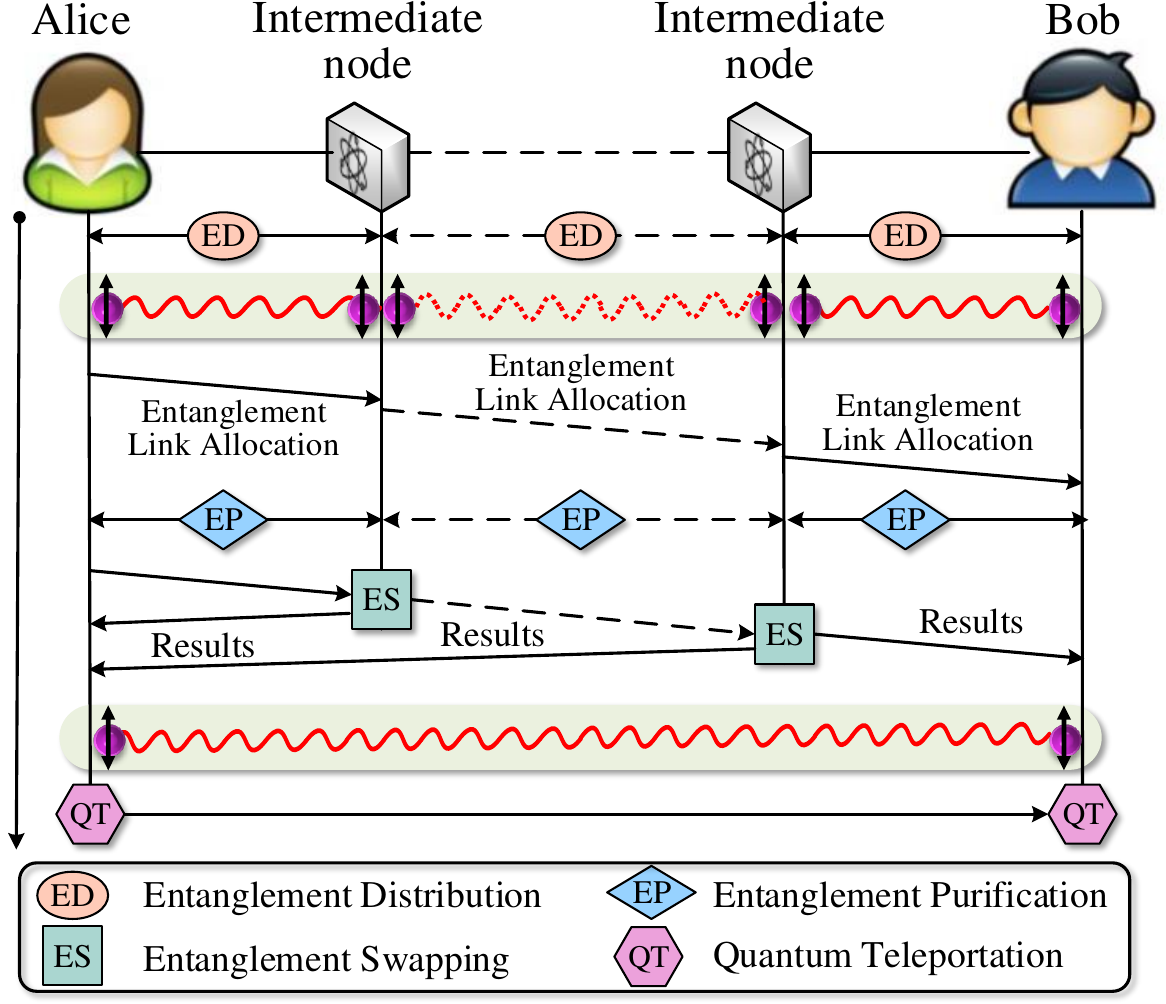}
	\caption{The connectionless qubit transmission between Alice and Bob.}
	\label{connectionless}
\end{figure}

For the connectionless strategy (as shown in \textbf{Fig.~\ref{connectionless}}), entanglement links are established between neighboring quantum nodes, and they can serve any quantum communication request before being allocated. Notably, in order to mitigate quantum decoherence caused by the channel noise during entanglement distribution, entangled qubit pairs generally are stored in quantum memory after being purified. Then, a path is selected with the help of entanglement routing algorithms, and each quantum node on the selected path allocates entanglement link resources for Alice and Bob. Similar to the connection-oriented strategy, entanglement purification is also required to improve each entanglement link's fidelity. After that, the allocated entangled qubit pairs are retrieved from quantum memory and swapping operations can be performed hop-by-hop along the selected path to ``couple'' multiple entanglement links into end-to-end entanglement connections. Finally, quantum teleportation can be performed to realize qubit transmission. For the connectionless strategy, all the end-to-end entanglement connections used to teleport qubits from Alice to Bob may be established through different paths, which is determined by routing protocols.

Both of the two strategies have their advantages and disadvantages. The connection-oriented strategy performs better than the connectionless one in terms of transmission reliability because the dedicated entanglement link resources are allocated, but it will lead to high network congestion in the scenario of high concurrent requests. Notably, quantum operations, such as entanglement distribution, entanglement swapping, and quantum teleportation, are usually implemented with a success probability. Hence, the connectionless strategy would provide a higher qubit transmission rate, using the pre-established entanglement links, than the connection-oriented strategy. Unlike classical communication, the qubit transmission rate in entanglement-assisted quantum networks is an expected value. Here, we take the quantum communication over a $k$-hop homogeneous linear path as an example to show the calculation of the qubit transmission rate. Assume that each hop provides $m$ homogeneous entanglement links to serve quantum teleportation per unit time after entanglement distribution (or entanglement link allocation) and entanglement purification, quantum repeaters with the same success probability of entanglement swapping, and each final end-to-end entanglement connection established between two communicating parties has the same effect on quantum teleportation. Let denote the success probability of entanglement swapping and quantum teleportation as $q$ and $p$, respectively. We can get the expected number of the shared entangled states per unit time: $mq^{k-1}$. Hence, the expected qubit transmission rate along the linear path is $mpq^{k-1}$. In summary, when a path is selected, the qubit transmission rate is mainly determined by the number of end-to-end entanglement connections established per unit time and their fidelity. In general, all network designs discussed in Section~\ref{Sec6} aim at maximizing the expected qubit transmission rate and thus realizing high-performance entanglement-assisted quantum networks.

\section{Challenges and Breakthroughs}\label{Sec5}
This section presents some challenges of building a future entanglement-assisted quantum network that supports distributed quantum applications from a networking perspective. We start with the problem of the imperfect quantum system mainly caused by various factors, including qubits, quantum channels, quantum memory, and quantum operation. Then, we describe the obstacle about the consolidation of different quantum technologies, and the challenge in the synergy between classical and entanglement-assisted quantum networks is discussed at the end. Finally, we show the great advances presented by the second quantum revolution, which makes it possible to build a large-scale and wide-area entanglement-assisted quantum network in the near future.

\begin{figure}[t]
	\centering
	\includegraphics[width=0.99\linewidth]{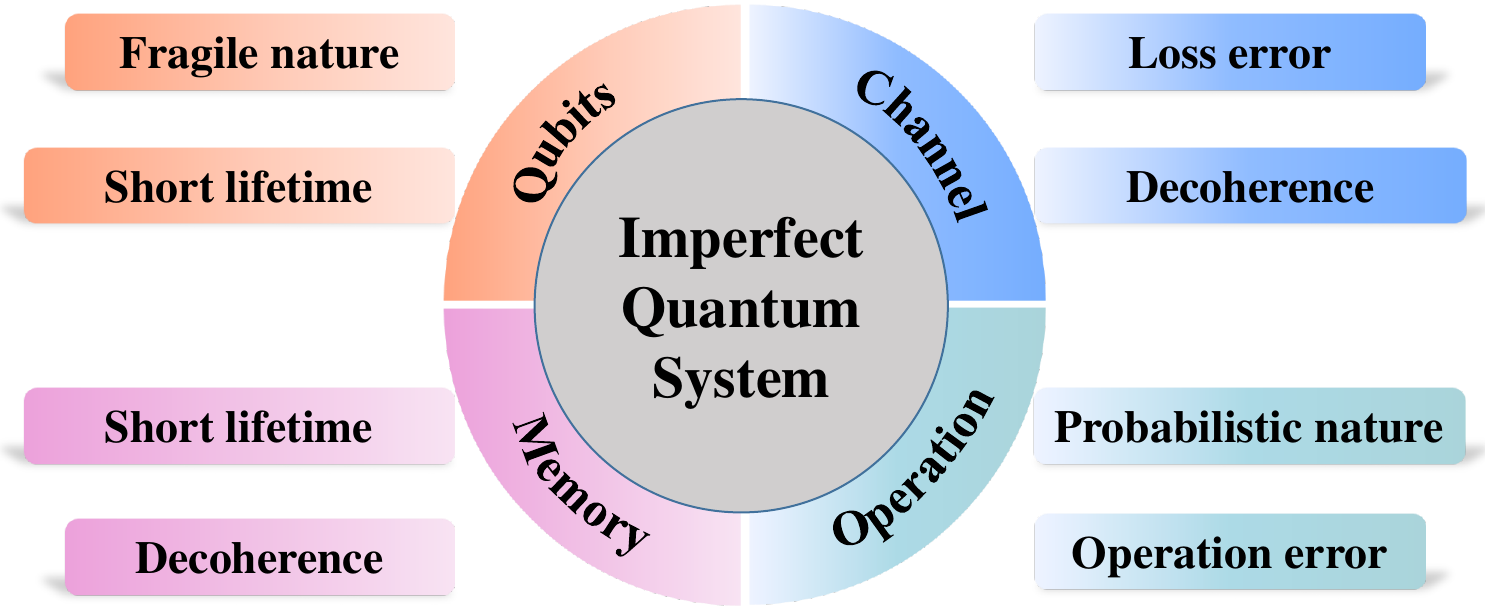}
	\caption{The imperfection of quantum system in entanglement-assisted quantum networks.}
	\label{imperfect}
\end{figure}

\subsection{Imperfect Quantum System}\label{Sec5.1}

The first and prime challenge for building future entanglement-assisted quantum networks is the inherent imperfection of quantum systems. As shown in \textbf{Fig. \ref{imperfect}}, a variety of factors contribute to the fact that quantum systems are imperfect \cite{chehimi2022physics}. Firstly, in an open system,  qubits are very fragile, thus being susceptible to noisy environments. As a result, qubits show a short lifetime, i.e., the state of a single qubit can only be maintained for a very short time after it is prepared. If the state of a single qubit changes, the quantum information it carries will be lost. Besides, qubits follow the no-cloning theorem. Preparing copies of a single qubit is impossible to reduce the effect of a short lifetime. Consequently, qubits need to be measured as soon as possible after they have been generated. Secondly, the inherent photon loss and noise of a quantum channel inevitably cause loss error and quantum decoherence during the transmissions of qubits, the result of which is that it is hard to establish a perfect entanglement link between adjacent quantum nodes. Thirdly, although quantum memory can enhance the lifetime of qubits, its inherent noise introduces redundant quantum decoherence for quantum systems. Besides, the capacity of quantum memory is limited by the incompleteness of the physical devices. Consequently, quantum memory hardly works as well as classical memory. Lastly, quantum operations in entanglement-assisted quantum networks show probabilistic feature. For example, entanglement preparation, entanglement swapping, and entanglement purification generally are successfully performed with a probability. Besides, quantum operations inevitably introduce operation errors due to the inherent noisy environment in quantum hardware. As discussed above, the inherent imperfection of quantum systems exists in the preparation, transmission, storage, and manipulation of qubits, severely increasing the difficulty of interconnecting various quantum nodes to form an entanglement-assisted quantum network with good operational performance.

\subsection{Consolidation of Various Physical Resources}\label{Sec5.2}
Similar to classical networks, a large-scale and wide-area entanglement-assisted quantum network usually is a combination of numerous heterogeneous and small-scale quantum networks. The difference between heterogeneous quantum networks is reflected in network structure, network scale, and especially physical resources. In quantum information technology, various physical resources can be used to support quantum networking and internetworking. Physical resources differ in the qubits' preparation, transport, and storage. For the preparation of qubits, a single qubit can be represented by ions, atoms, photons, spins, and superconductors. Each existing form of qubits shows unique superiority in different quantum applications. For example, optical-based quantum computing technology shows superiority in scalability and quantum coherence times, but it is impossible to program qubits and hard to miniaturize the size of computing devices \cite{obrien2007optical,kok2007linear}. Quantum computing based on superconductivity technology shows strong operability and integrability \cite{blais2004cavity,wu2021strong} compared to optical quantum computing technology. Besides, both NV centers \cite{hensen2015loophole}, trapped ions \cite{hucul2015modular}, neutral atoms~\cite{ritter2012elementary}, and superconducting circuits \cite{narla2016robust} can achieve the preparation of entangled qubits. However, qubits can only be transmitted as photons in quantum channels, including optical fiber and free space. Notably, photons show significant differences from matter qubits. Thus, a quantum signal converter is required to eliminate the difference between various physical resources. Summarily, entanglement-assisted quantum networks need to provide an suitable approach to abstract the underlying physical resources, thus making them scalable globally, including connecting physically and logically heterogeneous networks. However, considering the imperfect quantum systems and the no-cloning theorem, consolidating various physical resources in entanglement-assisted quantum networks is quite challenging.

\begin{figure*}[htbp]
	\centering
	\includegraphics[width=0.83\linewidth]{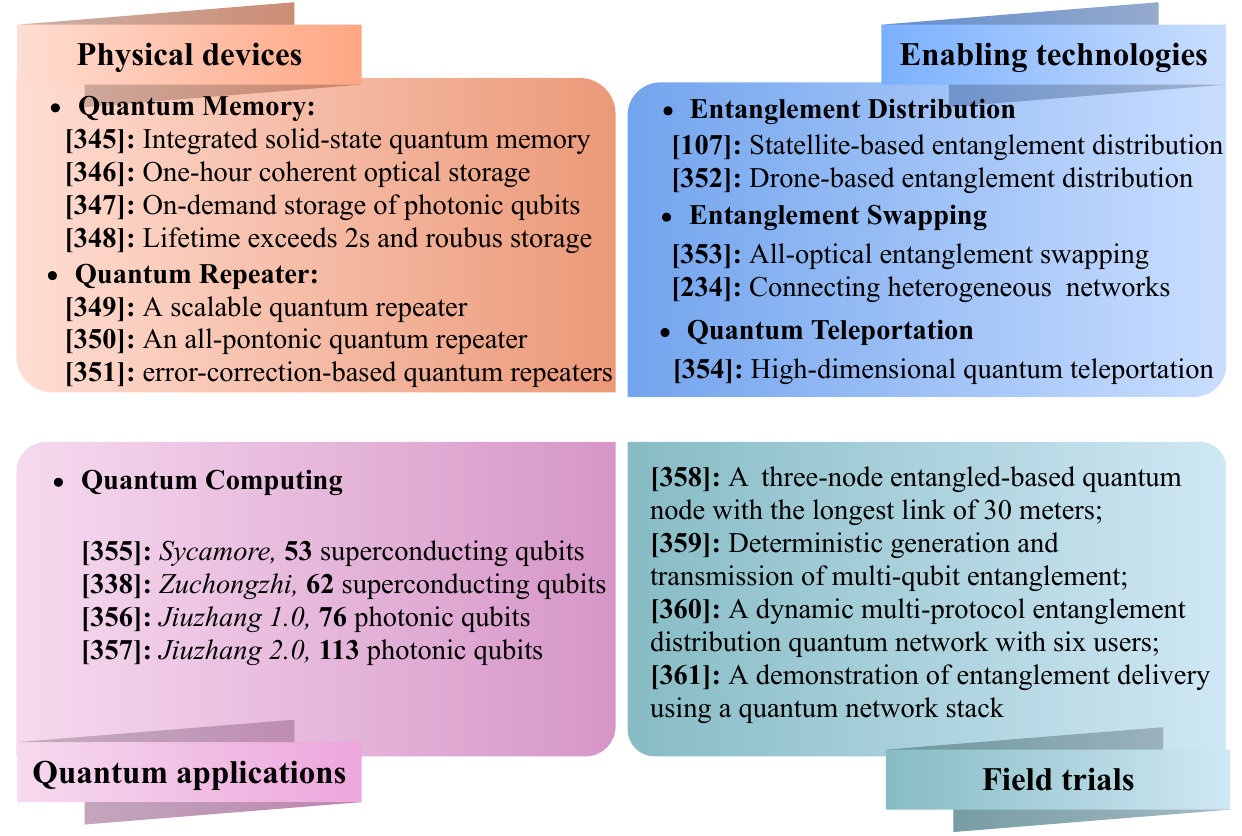}
	\caption{Breakthroughs related to entanglement-assisted quantum networks.}
	\label{current}
\end{figure*}

\subsection{Synergy between Classical and Quantum networks}\label{Sec5.3}
Entanglement-assisted quantum networks work by distributing entangled qubit pairs, i.e., establish entanglement, between quantum nodes to teleport qubits, thus supporting ground-breaking quantum applications. Notably, each quantum operation, such as entanglement swapping, entanglement purification, and quantum teleportation, requires the assistance of classical communication. Classical messages are usually used to control quantum operations and the feedback of operation outcomes. As a result, there is inevitably a cumbersome interaction of classical information between quantum nodes in entanglement-assisted quantum networks. The management of classical interactions directly determines whether quantum operations can be accurately performed. Besides, it is worth noting that realistic quantum systems are never completely isolated from their environment, and a quantum system undergoes quantum decoherence when interacting with its noisy environment. Thus, the additional latency introduced by classical interactions between quantum nodes will negatively affect the performance of entanglement-assisted quantum networks. In summary, there is a requirement to efficiently synergize entanglement-assisted quantum networks and classical networks. However, entanglement-assisted quantum networks are governed by the laws of quantum mechanics without counterparts in the classical world. Consequently, the inherent differences between classical and entanglement-assisted quantum networks still require further attention from researchers. Moreover, the security issue is also a vital obstacle hindering the implementation of entanglement-assisted quantum networks \cite{satoh2021attacking}. Although the laws of quantum mechanics ensure quantum superiority, especially the security of quantum communication, classical communication still faces security risks. Besides, quantum systems are incredibly fragile, and physical disruptions seriously affect the normal operation of entanglement-assisted quantum networks, such as attacks against quantum repeaters \cite{suzuki2015classification,satoh2018network}. Hence, the security system of classical networks must be improved to ensure the security of quantum operations in entanglement-assisted quantum networks.

\subsection{Breakthroughs}
Although it is challenging to build an entanglement-assisted quantum network, the second quantum revolution promotes the development of quantum information technology, significantly giving birth to entanglement-assisted quantum networks. As shown in \textbf{Fig.~\ref{current}}, we present some breakthroughs that are critical to building a future entanglement-assisted quantum networks, including physical devices, enabling technologies, quantum applications, and field trials.

In terms of physical devices, quantum memory and quantum repeater have made breakthrough progress. \cite{li2020hyperfine} presents an integrated solid-state quantum memory and \cite{ma2021one} proposes a one-hour coherent storage. A quantum memory that can store photonic qubits on demand was developed \cite{liu2022demand}. \cite{stas2022robust} develops a robust quantum memory, the lifetime of which exceeds 2 seconds. A scalable quantum repeater was developed in 2015 \cite{wang2015towards}. Besides, all-photonic and error-correction-based quantum repeaters were proposed successively \cite{li2019experimental,rozpkedek2021quantum}. It is expected that high-lifetime and scalable quantum memory can be realized and flexibly deployed in entanglement-assisted quantum networks in the near future.

In terms of enabling technologies, satellite-based and drone-based entanglement distributions were demonstrated~\cite{yin2017satellite,liu2020drone}, respectively. \cite{liu2022all} realizes all-optical entanglement swapping and \cite{guccione2020connecting} presents an experiment to connect two heterogeneous quantum networks by entanglement swapping. High-dimensional quantum teleportation was experimentally demonstrated in 2020 \cite{hu2020experimental}. We can foresee that entanglement distribution can be efficiently implemented even in harsh environments, and swapping operations can be effectively performed to establish end-to-end entanglement to support quantum teleportation.

Here, we use quantum computing as an example to present the dramatic development of quantum applications. Google developed a quantum computer, \textit{Sycamore}, and first demonstrated quantum supremacy using 53 superconducting qubits in 2019~\cite{arute2019quantum}. \cite{wu2021strong} realizes a quantum computer, \textit{Zuchongzhi}, to strengthen quantum computational advantage using 62 superconducting qubits in 2021. The quantum computer \textit{Jiuzhang} uses 76 and 113 photonic qubits to demonstrate quantum computational advantage successively~\cite{zhong2020quantum,zhong2021phase}. Obviously, quantum applications have been rapidly maturing in recent years.

In terms of filed trials of entanglement-assisted quantum networks, \cite{pompili2021realization} reports on the experimental realization of a three-node network with the longest link of 30 meters, and the authors achieved multi-party entanglement distribution across the three nodes and any-to-any connectivity through entanglement swapping. \cite{zhong2021deterministic} implements the deterministic generation and transmission of multi-qubit entanglement between two quantum nodes with three interconnected qubits. A dynamic multi-protocol entanglement distribution network with six quantum nodes is implemented in 2022~\cite{wang2022dynamic}, Additionally, \cite{pompili2022experimental} realizes an experimental demonstration of entanglement delivery using a quantum stack. In summary, the second quantum revolution significantly promoted the development of quantum devices, enabling technologies, and quantum applications. These breakthroughs make it possible to build entanglement-assisted quantum networks in the near future. Besides, developing quantum hardware for building the network infrastructure, network design is also a crucial aspect in realizing effective entanglement-assisted quantum networks that can support various quantum applications.

\section{Research Directions}\label{Sec6}
In this section, we first provide some future research directions for the pivotal problems that must be overcome to realize a wide-area entanglement-assisted quantum network and then suggest some solutions. These research directions focus on the architecture design of future entanglement-assisted quantum networks, entanglement-based network problems, and standardization. Specifically, architecture design and standardization aim at solving the problems caused by the challenges discussed in Section~\ref{Sec5}, namely, the consolidation of various physical resources and the synergy between classical and entanglement-assisted quantum networks. The proposal for studying entanglement-based network designs aims to solve the problems caused by imperfect quantum systems to improve network performance.

\begin{figure*}[htbp]
	\centering
	\includegraphics[width=0.99\linewidth]{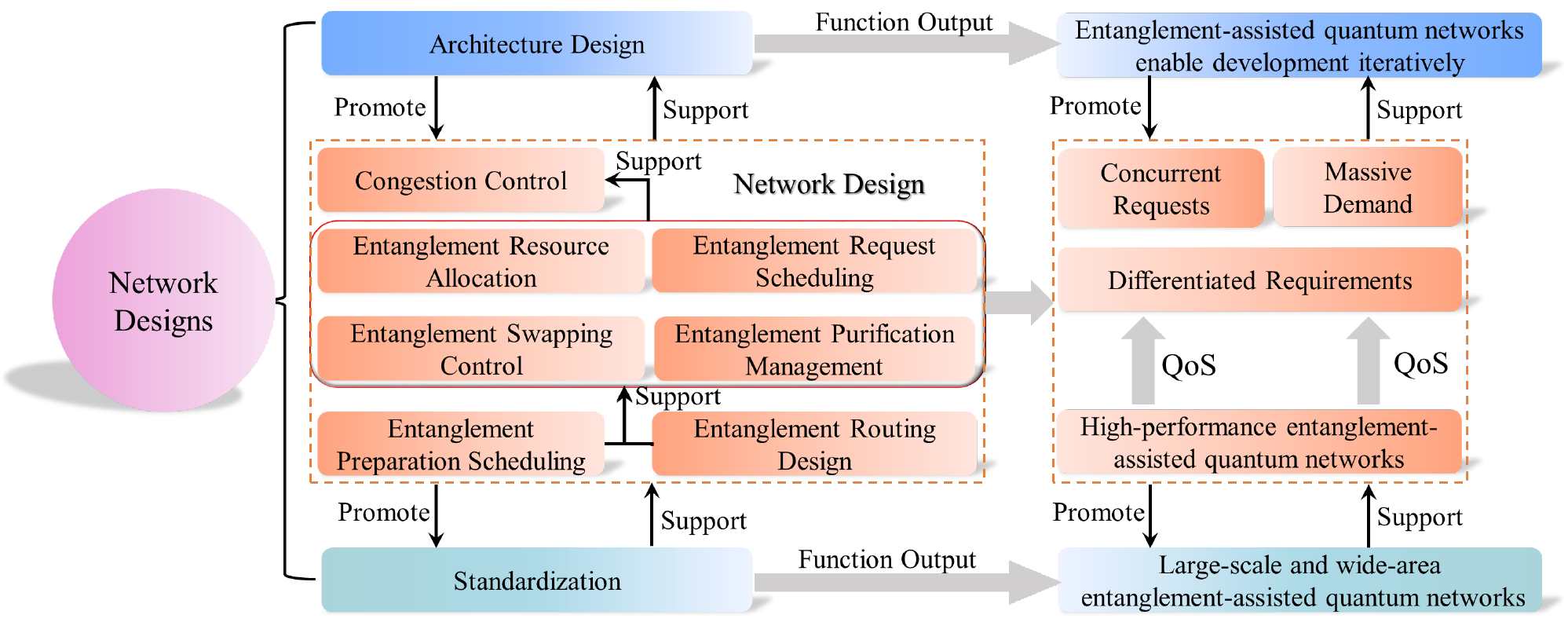}
	\caption{The illustration of the relationship between research directions.}
	\label{relation}
\end{figure*}

\subsection{Overview}
In this section, we describe the research direction from three aspects: architecture design, network design, and standardization. The architecture design aims to decompose the complex end-to-end qubit transmission problem into multiple sub-problems that are easier to solve. It proposes a layered protocol stack and abstracts the differences between different physical resources,enabling the iterative development of an entanglement-assisted quantum network. The goal of network design is to implement high-performance entanglement-assisted quantum networks and provide QoS for users. Network design focuses on the efficient generation and application of entanglement resources. Specifically, entanglement preparation scheduling aims to efficiently establish entanglement links between adjacent nodes to reduce remote entanglement distribution latency. Entanglement routing design is responsible for selecting a ``good'' path that can realize a high end-to-end entanglement establishment rate, thus improving the qubit transmission rate. Entanglement purification management is designed to effectively improve entanglement fidelity with a small entanglement resource overhead. Entanglement swapping control is conducted to efficiently establish end-to-end entanglement. Entanglement request scheduling aims to improve the utilization of entanglement link resources Entanglement resource allocation is responsible for the efficient and fair allocation of entanglement links. Congestion control aims to reduce network congestion, thus improving the service quality of entanglement-assisted quantum networks. The study of standardization promotes the development of the entanglement-assisted quantum network industry and facilitates the construction of large-scale and wide-area quantum internet. In summary, network design supports architecture design and promotes the study of standardization. Furthermore, architecture design and the study of standardization drive network design to become more efficient and reliable. These research directions jointly promote the development of entanglement-assisted quantum networks, and we will provide detailed descriptions of them in the following sections. The relationship among the three research aspects is shown in \textbf{Fig.~\ref{relation}}.

\subsection{Architecture Design}
The development of entanglement-assisted quantum networks is a process of iteratively updating technologies and increasingly extending network scale. In this process, it is required to eliminate the variability among heterogeneous quantum networks to improve networks' scalability . Although the OSI model and TCP/IP model have been proven to be hugely successful in the current Internet, it is doomed to fail by extending only some classical protocols to their quantum counterpart without global modifications. The reason is that entanglement-assisted quantum networks are governed by the laws of quantum mechanics, which inherently differs from classical network design principles. Hence, there is a need to have a novel design of networks' architecture from different points of view when building a large-scale and wide-area entanglement-assisted quantum network. Architecture design aims at providing feasible and effective solutions to solve the problem of integrating heterogeneous small-scale entanglement-assisted quantum networks to form a larger-scale and wide-area entanglement-assisted quantum network and simplifying the process of remote entanglement distribution. We can divide the architecture design into two parts: the design of the network structure and the protocol stack design. A ``good'' design of network structure should be one that can simply and efficiently interconnect heterogeneous quantum networks and reduce the cost of network deployment. Van Meter \textit{et al.} proposed a recursive structure for entanglement-assisted quantum networks \cite{van2011recursive,van2022quantum}, which is useful for building arbitrary distributed entangled states, such as Bell pairs and GHZ, W, and cluster states. However, network scaling will bring about large deployment overhead in such a recursive design. The protocol stack design aims at dividing the complex process of remote entanglement distribution into multiple easy-to-implement sub-processes. In this way, each layer is an abstract function or an independent system, contributing to the scalability and flexibility of entanglement-assisted quantum networks. Currently, the service requirements for quantum applications are unclear, and designing a high-performance entanglement-assisted quantum network architecture remains an open issue.

\subsection{Entanglement Distribution Scheduling}
Entanglement distribution aims to distribute entangled qubit pairs between adjacent quantum nodes to establish entanglement links. However, it is hard to establish an entanglement link, and each quantum node may make numerous attempts to establish entanglement with adjacent quantum nodes. As a result, the time spent on entanglement distribution cannot be ignored. There are mainly two fundamental models to schedule entanglement distribution, i.e., on-demand and continuous~\cite{chakraborty2019distributed}. For the on-demand model, SD pairs' requests trigger entanglement distribution, and entanglement distribution is only performed hop-by-hop along swapping paths selected for achieving remote entanglement distribution between SD pairs. In this way, the on-demand method will result in high latency and a waste of entanglement resources, especially considering the imperfection of entanglement swapping. For the continuous model, any pair of adjacent quantum nodes attempt to remain entanglement as much as possible, i.e., entanglement distribution is performed spontaneously in a consume-then-replenish manner. In this way, entangled qubit pairs are stored in quantum memory and wait to be measured, introducing unnecessary quantum decoherence. Summarily, these two models negatively affect the performance of remote entanglement distribution in latency and entanglement fidelity, respectively. Hence, an efficient design is required to schedule entanglement distribution to establish low-latency and high-fidelity entanglement in entanglement-assisted quantum networks.

In general, there is a trade-off between latency and the attenuation of fidelity during remote entanglement distribution. Since entanglement purification can mitigate the attenuation of entanglement fidelity, the entanglement distribution scheduling can be regarded as the problem of managing preparation operations on the selected swapping path to minimize the latency of end-to-end entanglement distribution \cite{chen2022heuristic}. A novel design is to perform entanglement distribution along the selected path in parallel. \cite{Dai2020Optimal} proposed an optimal entanglement distribution scheme on a quantum repeater chain, but the authors ignored the probabilistic feature of entanglement swapping. Notably, the failed entanglement swapping will trigger re-distribution operations on a swapping path, thus introducing redundant latency. To alleviate the negative impact of the imperfect swapping operation on remote entanglement distribution latency, we can split a long swapping path into multiple segments and make the pre-distribution operations be performed within the segments as much as possible. Besides, considering that the superiority of machine learning (ML) techniques have been demonstrated for network optimization in classical networks, we can also adopt machine learning in entanglement-assisted quantum networks to optimize entanglement distribution scheduling. For example,~\cite{zhang2022hybrid} reported a try to use ML to resolve the incompatibility of entanglement distribution rate between the DV and CV quantum systems. Moreover, we can utilize ML to predict end-to-end entanglement distribution requests based on the historical data and pre-establish entanglement links based on the predicted results. In this way, the latency caused by entanglement distribution can be effectively reduced, though prediction accuracy will affect the cost.

\subsection{Entanglement Routing Design}
The routing problem is fundamental but critical in achieving entanglement distribution between any pair of quantum end nodes in entanglement-assisted quantum networks. Similar to classical networks, the function of entanglement routing is to select one or multiple ``good'' swapping paths connecting SD pairs. The selected path is used to establish end-to-end entanglement by performing entanglement swapping rather than forwarding qubits hop-by-hop. As discussed above, entanglement-assisted quantum networks are fundamentally different from classical networks, and thus the routing algorithms adopted in classical networks cannot be directly applied to entanglement-assisted quantum networks. We need to redesign the routing algorithm for remote entanglement-distribution. In recent years, some routing algorithms have been proposed for swapping path selection. For example, \cite{van2013path} fully considers the nature that the success probability of establishing an entanglement link is negatively exponential related to the physical distance of a quantum channel and then proposed a Dijkstra-based algorithm for entanglement-assisted quantum networks. However, this routing algorithm ignores the effect of swapping operations' probabilistic feature on the performance of remote entanglement distribution. \cite{caleffi2017optimal} proposes a routing algorithm with a more complex routing metric, such as quantum decoherence time and the success probability of entanglement swapping, to find an optimal path between quantum end nodes. However, this algorithm relies on a single node to have the global information of the whole network topology, so the efficiency of path selection is not good. Pirandola \textit{et al.} proposed a path selection design for entanglement distribution in a diamond topology \cite{Pirandola2019End}. However, this routing design relies on the assumption that entanglement swapping can be perfectly performed, which obviously doesn't apply to real quantum systems. Schoute \textit{et al.} studied the path selection of remote entanglement distribution in the ring and spherical topologies~\cite{schoute2016shortcuts}, and Das \textit{et al.} evaluated routing designs under different specific network topologies~\cite{Das2018Robust}. However, these routing designs often perform poor in randomly generated network topologies and are not suitable for deployment in real wide-area quantum information networks. Pant \textit{et al.} adopted the greedy strategy to select a path with minimum hops from the source node to the destination node \cite{pant2019routing}. However, this algorithm only considers the special scenario where adjacent quantum nodes share a pair of entangled qubits and only studies the routing design under the special network model of a lattice network. \cite{li2023swapping} proposes a entanglement routing algorithm, taking into account the success probability of entanglement swapping and hop counts. However, this routing algorithm can only select the approximate optimal swapping path for an SD pair. In summary, these algorithms either do not sufficiently take into account the properties of entanglement-assisted quantum networks or perform poorly in scalability. Considering quantum decoherence and imperfect quantum operations, it is challenging to design an efficient entanglement routing algorithm for selecting swapping paths in entanglement-assisted quantum networks.

Three problems need to be solved before designing an entanglement routing algorithm. \textit{(1) What is the goal of a routing algorithm in entanglement-assisted quantum networks?} Generally, a routing algorithm in classical networks will reduce end-to-end communication latency as much as possible to increase network throughput with avoiding the routing loop. In addition to considering the entanglement distribution rate (EDR) in the routing algorithm design, entanglement fidelity is also a non-negligible objective. Hence, there are basically three requirements for entanglement routing designs in entanglement-assisted quantum networks: maximum EDR, the highest entanglement fidelity, and a trade-off between EDR and entanglement fidelity. \textit{(2) How to determine the metrics based on the properties of entanglement links?} Different from classical networks, quantum operations are imperfect, and the quality of the entanglement link is measured by fidelity. Hence, it is challenging to select the characteristics of an entanglement link as routing metrics. \textit{(3) How to define the cost function for a routing algorithm in entanglement-assisted quantum networks?} Generally, more than one metric is selected to achieve the goal of the routing algorithm. However, there will be a trade-off between multiple routing metrics \cite{leone2021qunet}. Consequently, we need to balance the weights of the different metrics in designing the cost function to realize the optimal routing goal in entanglement-assisted quantum networks.
	
The choice of routing metrics plays a vital role in designing entanglement routing algorithms. Here, we present some routing metrics~\cite{schoute2016shortcuts} that might need to be considered in routing algorithm design. \textit{(1) The success probability of entanglement preparation.} Notably, it is difficult to build an entanglement link between adjacent quantum nodes. Hence, the success probability of entanglement preparation is critical to end-to-end EDR. \textit{(2) The success probability of entanglement swapping.} Remote entanglement distribution is realized by performing entanglement swapping on a quantum repeater chain, and thus the success probability of entanglement swapping also affects the end-to-end EDR. \textit{(3) Hop count.} The higher the number of hops, the lower the probability that an end-to-end entanglement will be established. Besides, more hops results in more delay and thus fidelity degradation. \textit{(4) Entanglement fidelity.} Generally, the higher the fidelity, the closer the quantum manipulation is to perfection. Furthermore, high-fidelity entanglement will reduce the resource and time overheads associated with the entanglement purification. \textit{(5) Bandwidth.} Bandwidth represents the number of available entanglement links. The high-bandwidth path can reduce the risk of network congestion on the one hand and compensate for the imperfect quantum operations on the other. It, however, brings out the cost or poses the requirements for entanglement preparations and quantum memories. In summary, there are several routing metrics that can be selected to design an entanglement routing algorithm. But, we need to consider whether there is a trade-off between different metrics for a special routing target.

\subsection{Entanglement Swapping Control}
Entanglement swapping plays a vital role in networking numerous quantum information processors. However, the inherent features of swapping operations significantly affect remote entanglement distribution, thus affecting the performance of entanglement-assisted quantum networks. Firstly, entanglement swapping is an imperfect operation due to the inherent limitation of quantum hardware. If a failed swapping operation cannot be precisely detected, this error will be passed along a swapping path to end-to-end entanglement. In this case, remote entanglement distribution needs to be re-performed or consumes valuable entanglement link resources to correct errors. As a result, the precious entanglement link resource on the swapping path will be wasted. Hence, entanglement-assisted quantum networks need to provide a swapping operation management design that can effectively detect the failure of entanglement swapping. Secondly, a pair of entangled qubits shared by two quantum nodes can serve two  quantum nodes to perform swapping operations. However, the case that two quantum nodes attempt to swap entanglement simultaneously will make entanglement relationships hard to track, especially considering the imperfect swapping operation. Conversely, there will be competition for the shared entangled qubits between two quantum nodes, thus impeding remote entanglement distribution~\cite{wang2022asynchronous}. Hence, entanglement swapping control is required to avoid entanglement resource competition. Summarily, these two problems caused by the unique features of entanglement swapping impair networks' performance. Hence, entanglement-assisted quantum networks need to provide an effective entanglement swapping control design to tackle these two problems and improve the performance of remote entanglement distribution.

Entanglement swapping control aims at managing swapping operations on a swapping path to precisely detect the failure of entanglement swapping and avoid entanglement resource competition between quantum nodes as much as possible. As discussed in Section~\ref{Sec3.5}, although the parallel entanglement swapping method can reduce remote entanglement distribution latency and simplify complex information interactions, it inevitably contributes to the two problems mentioned above. Hence, the parallel entanglement swapping method is not a good choice for remote entanglement distribution, especially considering the scarcity of entanglement link resources. Besides, although the sequence entanglement swapping method can effectively solve the entanglement resource competition problem and track entanglement, it contributes to high entanglement distribution latency. For these two methods, the failed entanglement swapping contributes to the result that swapping operations must be re-performed from the source node. In this case, an SD pair will spend more time establishing an available end-to-end entanglement and lead to a waste of valuable entanglement link resources, especially in the scenario where only the quantum repeater directly linked by the destination node fails to swap entanglement in a swapping path.

A novel design is to split the swapping path into multiple segments, i.e., using the nested-based method to manage swapping operations on a selected path. In this way, swapping operations can be performed in parallel in different segments, and entanglement swapping just needs to be re-performed within a segment if a repeater fails to swap entanglement. However, determining the optimal segmentation points in the swapping path poses a greater challenge. The more quantum repeaters a segment contains, the fewer the number of entanglement swapping rounds. However, the success probability of establishing multiple-hop entanglement in a segment is inversely proportional to the length of the segment. There are two situations during remote entanglement distribution. Firstly, the multiple-hop entanglement established in some segments will undergo a long decoherence time, thus building end-to-end entanglement with low fidelity. Secondly, most segments will establish multiple-hop entanglement using the re-generated link-level entanglement with high fidelity. In this case, the SD pair establishes a high-fidelity entanglement at the cost of increased latency and wasted entanglement resources. In summary, there is a trade-off between the latency of remote entanglement distribution and end-to-end entanglement fidelity in the segment-based control. For the problem of entanglement swapping control, we can design a segmentation method according to SD pairs' requirements so as to realize remote entanglement distribution with QoS.

\subsection{Entanglement Purification Management}

End-to-end entanglement distribution requires purification operations to correct errors caused by channel loss and quantum decoherence. However, entanglement purification improves fidelity at the expense of reducing the number of entangled qubit pairs, which negatively impacts the performance of entanglement-assisted quantum networks in end-to-end EDR. Notably, the output fidelity and success probability of entanglement purification are closely related to the input fidelity of entangled qubit pairs. This implies that different purification designs used in end-to-end entanglement distribution will consume varying numbers of entangled qubit pairs. Furthermore, different application scenarios have different requirements for entanglement fidelity. For instance, quantum sensing demands high-fidelity entanglement to ensure accurate and reliable measurement results, often necessitating entanglement fidelity of 99\% or higher. Conversely, certain applications may tolerate lower-fidelity entangled systems, such as quantum simulations aimed at studying complex systems' behavior rather than replicating every detail precisely. Consequently, the fidelity requirements of different application scenarios influence the number of entangled qubit pairs consumed in entanglement purification. In this context, managing purification operations along the swapping path can reduce the number of entangled qubit pairs consumed for error correction, thereby enhancing the performance of entanglement-assisted quantum networks.

Entanglement purification management can be defined as the problem of controlling purification operations on a swapping path to minimize the consumption of entangled qubit pairs while meeting end-to-end entanglement fidelity requirements. Several insightful works have proposed solutions to address this problem. Briegel \textit{et al.} \cite{Briegel1998Quantum} introduced the concept of connecting a series of imperfect entangled qubit pairs using a novel nested purification protocol, enabling the creation of highly-fidelitous distant entanglement. Li \textit{et al.}  \cite{li2022fidelity} proposed a scheme to purify entanglement links by utilizing an ``expected cost value'' on a selected path, ensuring the fidelity of end-to-end entanglement. Similarly, Li \textit{et al.} \cite{li2021effective}  presented a purification scheme to improve the fidelity of entanglement links up to a feasible threshold. Zhao \textit{et al.} \cite{zhao2022e2e} introduced an entanglement purification scheme that determines the rounds of purification performed by each hop, optimizing network throughput in scenarios involving concurrent requests. However, these entanglement purification solutions primarily focus on managing purification operations performed on link-level entanglement resources. It is important to note that decoherence occurs during the transmission, storage, and measurement of entangled qubits, making it challenging to meet end-to-end entanglement fidelity requirements solely with link-level purification operations. Therefore, in order to establish high-fidelity end-to-end entanglement, entanglement purification must be performed after each entanglement swapping round in entanglement-assisted quantum networks.

For a selected swapping path, once the entanglement swapping control is determined, entanglement purification management can be divided into two problems. \textit{(1) How many rounds of purification operations should be performed?} The number of purification rounds is positively correlated with end-to-end entanglement fidelity. However, performing more rounds of purification consumes more entanglement resources. Considering the limited availability of entanglement resources, it is crucial to minimize the number of purification rounds performed on entangled qubit pairs while still meeting the fidelity requirements of SD pairs. \textit{(2) How should entangled qubit pairs be selected for each round of purification?} Entanglement purification can be seen as a processing unit with input and output. The fidelity of the input entanglement directly impacts the output fidelity and the success probability of the purification operation. Some purification scheduling designs, discussed in \cite{van2008system}, show that the purification gain achieved with different combinations of input entanglement fidelity can vary significantly. Therefore, selecting different entangled qubit pairs for purification can reduce the state transition of entangled qubit pairs from useful to useless caused by decoherence. This approach preserves more entanglement resources, ultimately improving network performance. In summary, entanglement purification management significantly affects the consumption of entanglement resources, making it a crucial area of research in the context of entanglement-assisted quantum networks.

\subsection{Entanglement Resource Allocation}

Most quantum applications require distant quantum nodes to establish entanglement, but entanglement is a scarce resource in entanglement-assisted quantum networks. This scarcity arises due to the following reasons: \textit{(1) Difficulty in establishing entanglement links:} Establishing an entanglement link is challenging due to inherent quantum channel losses and limitations of quantum hardware. The success probability of establishing an entanglement link decreases exponentially with the physical length of the quantum channel. As a result, adjacent quantum nodes in a network can only share a finite amount of entanglement resources. \textit{(2) Quantum decoherence mitigation through entanglement purification:} Entanglement purification operations can alleviate the negative impact of decoherence on remote entanglement distribution by improving entanglement fidelity. However, enhancing fidelity comes at the cost of reducing the number of available entanglement links. \textit{(3) Imperfections in entanglement swapping:} Each SD pair requires multiple attempts of remote entanglement distribution to establish end-to-end entanglement. Consequently, a significant number of entanglement links are consumed for each request. In summary, the inherent features of entanglement-related quantum operations contribute to the limited availability of entanglement resources in entanglement-assisted quantum networks. Therefore, efficiently allocating entanglement resources to support concurrent entanglement distribution requests between multiple SD pairs is a crucial issue that needs to be addressed.

Entanglement resource allocation can be defined as the problem of effectively distributing the limited entanglement resources on a shared quantum link to accommodate multiple remote entanglement distribution requests. The allocation of entanglement links to each SD pair directly impacts the rate of entanglement distribution. Fairness is an important criterion for resource allocation, but there is typically a trade-off between fairness and the overall rate of entanglement distribution in entanglement-assisted quantum networks. An ideal entanglement resource allocation design should ensure fairness while maximizing the total rate of remote entanglement distribution. \cite{li2021effective} proposes three methods for entanglement resource allocation. However, this work did not adequately address the design of entanglement purification management. Notably, the consumption of entanglement resources by purification operations significantly affects resource allocation. Therefore, the design of entanglement resource allocation methods must consider purification strategies and entanglement fidelity.

\subsection{Entanglement Request Scheduling}
Efficient and effective scheduling of remote entanglement distribution requests at a quantum node is essential. While some scheduling designs have proven successful in classical networks \cite{lu2013dynamic,peng2016engery,xue2018forward}, these designs are not suitable for entanglement-assisted quantum networks. The fundamental difference lies in the fact that an entanglement link can only serve a single entanglement flow, unlike classical networks where multiple data flows can share bandwidth. Additionally, each pair of adjacent quantum nodes on the selected path must allocate the same number of entanglement links for the source-destination (SD) pair. Consequently, a suboptimal scheduling scheme can lead to blocked requests at certain hops \cite{cicconetti2021request}, reducing the overall request service rate of entanglement-assisted quantum networks. Furthermore, quantum decoherence in quantum memory can cause the decay or even breakdown of the entangled system shared by adjacent quantum nodes or nodes spanning multiple hops, resulting in wastage of entanglement resources. Therefore, the development of effective request scheduling algorithms is crucial. Currently, research on request scheduling in entanglement-assisted quantum networks is still in its early stages due to unclear service requirements for future networks. \cite{cicconetti2021request} presents a solution to the scheduling problem through a general framework of heuristic algorithms, proposing three illustrative instances with the objective of minimizing application delay while achieving high system utilization in terms of entanglement rate and fidelity of remotely entangled qubits. However, the network model assumed by the authors overlooks the dynamic nature of link capacity (or entanglement link resources), which is unrealistic in actual deployments. As research on entanglement-assisted quantum networks progresses, the development of effective request scheduling mechanisms will become increasingly important for network performance and will require more effort from researchers.

\subsection{Congestion Control}

Entanglement links play a crucial role in quantum internetworking. However, establishing entanglement links between adjacent quantum nodes is challenging, and purification operations are required to increase fidelity at the expense of reducing the number of entanglement links. This creates a contradiction between concurrent entanglement distribution requests and the limited entanglement link resources in entanglement-assisted quantum networks. As a result, bottleneck links inevitably arise, leading to network congestion, degraded performance, and a poor user experience. Therefore, congestion control mechanisms that can regulate the rate of end-to-end entanglement distribution to mitigate network congestion are necessary in entanglement-assisted quantum networks.

Although various congestion control strategies have been proposed and successfully deployed in classical networks \cite{nagle1984congestion,low2002internet,afanasyev2010host}, these strategies cannot be directly applied in entanglement-assisted quantum networks due to the unique nature of entanglement. Hence, the development of effective congestion control strategies is essential. Before designing a congestion control strategy to ensure network performance, two issues need to be addressed. \textit{(1) How is congestion defined in entanglement-assisted quantum networks?} In classical TCP protocol, congestion is typically determined if the destination node fails to receive a packet within a fixed period of time or receives three consecutive repeated ACKs \cite{allman2009tcp}. However, in entanglement-assisted quantum networks, latency alone cannot serve as the sole criterion for detecting network congestion due to the imperfect nature of entanglement swapping. \textit{(2) How can entanglement flows be controlled?} This is the core challenge in designing congestion control strategies. Congestion control strategies can be broadly categorized as coarse-grained or fine-grained. Coarse-grained strategies effectively alleviate network congestion but come at the cost of reduced utilization of entanglement resources. For instance, \cite{zhao2023quantum} proposes a transport protocol named DTPs for entanglement-assisted quantum networks, which employs a ``Sending Window Control'' mechanism for coarse-grained congestion control. Specifically, the sending window size is halved when network congestion is detected, and it doubles when congestion is absent. On the other hand, fine-grained strategies can reduce network jitter in entanglement distribution rates to enhance resource utilization but may not effectively alleviate network congestion. There is a clear trade-off between relieving network congestion and maximizing entanglement resource utilization. Regardless of the strategy employed, implementing congestion control in entanglement-assisted quantum networks is challenging due to the probabilistic feature of quantum operations.

\subsection{Standardization}
Today, quantum key distribution (QKD) stands as a representative of quantum secure communication technology and has gradually transitioned from laboratories to commercial services. In the coming decades, future entanglement-assisted quantum networks and the quantum internet are expected to emerge as promising platforms supporting a range of quantum applications, including blind computing, enhanced sensing, and secret sharing. However, as discussed in Section~\ref{Sec5.2}, a major obstacle impeding the development of entanglement-assisted quantum networks is the challenge of consolidating diverse quantum hardware technologies to enable quantum internetworking. Therefore, to construct large-scale and wide-area entanglement-assisted quantum networks, the industry needs to actively engage in standardization efforts. Similar to classical networks, the main focus of standardizing entanglement-assisted quantum networks is to define abstractions and interfaces that decouple the underlying quantum hardware from upper-layer software. Currently, there are several international groups and standardization initiatives (such as those in ITU, IEEE, IETF, ETSI) working towards defining architectures, interfaces, and protocols that ensure interoperability between entanglement-assisted quantum networks (including QKDNs) and their seamless integration with existing telecommunications infrastructures~\cite{etsi2022official,IEEE2022,ITU-T2022Security,ITU-T2022Quantum,ITU-T2022QIRG,GSMA2022}.

\section{Conclusion}\label{Sec7}
In this survey, we discussed the fundamental principles of quantum mechanics to showcase the advantages of quantum information technology, specifically in terms of secure communication and immense computing power. Entanglement-assisted quantum networks stand as promising platforms for supporting groundbreaking quantum applications by enabling the distribution of entangled qubit pairs and the teleportation of qubits between distant quantum end nodes. Subsequently, we introduced essential enabling technologies for constructing future entanglement-assisted quantum networks, including entanglement preparation, quantum dense coding, entanglement swapping, quantum teleportation, quantum memories, and more. We also presented the current research progress in these technologies. Furthermore, we emphasized that the development of entanglement-assisted quantum networks is propelled by enabling technologies and quantum physical devices. We outlined six developmental stages of entanglement-assisted quantum networks, highlighting the capabilities of each stage. Viewing an entanglement-assisted quantum network as a mesh structure comprising numerous networking devices capable of qubit preparation, transmission, storage, and processing, we discussed the challenges associated with building future networks. These challenges encompass imperfect quantum systems, the integration of diverse physical resources, and the synergy between classical and entanglement-assisted quantum networks. Drawing inspiration from classical networks, we summarized research directions pertaining to inter-networking problems. We firmly believe that entanglement-assisted quantum networks will garner increasing attention from researchers and practitioners alike. It is expected that these networks will be realized in the near future, paving the way for the widespread application of quantum information technology.

\section*{Acknowledgments}
We are grateful for the support of Shengkai Liao, Chenzhi Peng and Jianwei Pan. This work is supported in part by Anhui Initiative in Quantum Information Technologies under grant No. AHY150300, National Scientific and Technological Innovation 2030 Major Project of Quantum Communications and Quantum Computers under grant No. 2021ZD0301301, Youth Innovation Promotion Association of CAS under grant No. Y202093, and Japan Society for the Promotion of Science (JSPS) KAKENHI under grant No. 23H03380.
	
\bibliographystyle{IEEEtran}
\bibliography{ref.bib}

\begin{IEEEbiography}[{\includegraphics[width=1in,height=1.25in,clip,keepaspectratio]{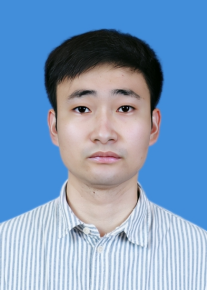}}]{Zhonghui Li} received the B.E. degree in Software Engineering University of Electronic Science and Technology of China, in 2018, and received his Ph.D degree in information security from the School of Cyber Science and Technology, University of Science and Technology of China (USTC), in 2023. He is currently a Post-Doctoral fellow with the School of Cyber Science and Technology, USTC. His current research interests include Quantum Internet architecture and Quantum networking.
\end{IEEEbiography}

\begin{IEEEbiography}[{\includegraphics[width=1in,height=1.25in,clip,keepaspectratio]{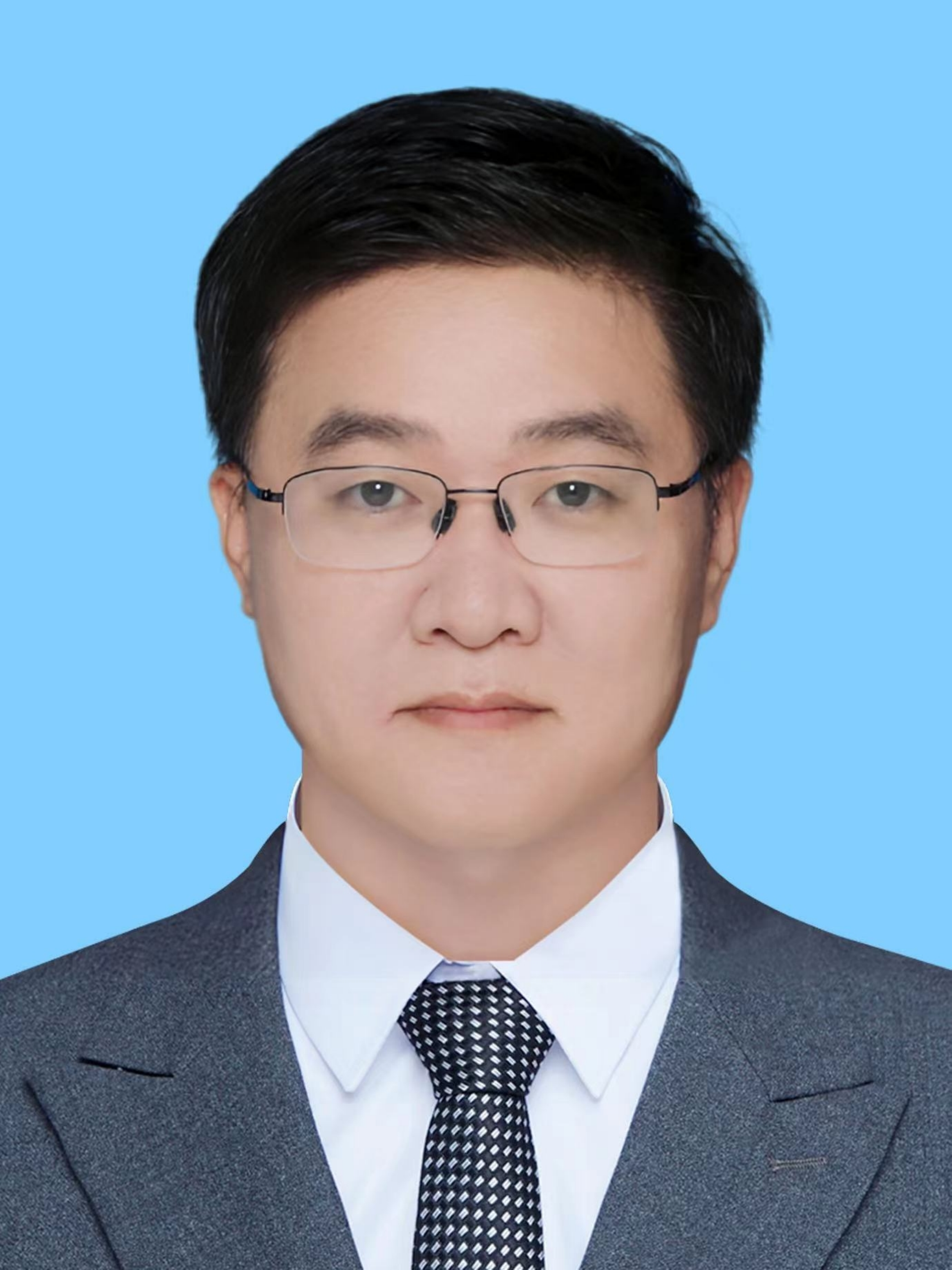}}]{Kaiping Xue (M'09-SM'15)} received his bachelor's degree from the Department of Information Security, University of Science and Technology of China (USTC), in 2003 and received his doctor's degree from the Department of Electronic Engineering and Information Science (EEIS), USTC, in 2007. From May 2012 to May 2013, he was a postdoctoral researcher with the Department of Electrical and Computer Engineering, University of Florida. Currently, he is a Professor in the School of Cyber Science and Technology, USTC. He is also the director of Network and Information Center, USTC. His research interests include next-generation Internet architecture design, transmission optimization and network security. His work won best paper awards in IEEE MSN 2017 and IEEE HotICN 2019, the Best Paper Honorable Mention in ACM CCS 2022, the Best Paper Runner-Up Award in IEEE MASS 2018, and the best track paper in MSN 2020. He serves on the Editorial Board of several journals, including the IEEE Transactions on Dependable and Secure Computing (TDSC), the IEEE Transactions on Wireless Communications (TWC), and the IEEE Transactions on Network and Service Management (TNSM). He has also served as a (Lead) Guest Editor for many reputed journals/magazines, including IEEE Journal on Selected Areas in Communications (JSAC), IEEE Communications Magazine, and IEEE Network. He is an IET Fellow and an IEEE Senior Member.
\end{IEEEbiography}

\begin{IEEEbiography}[{\includegraphics[width=1in,height=1.25in,clip,keepaspectratio]{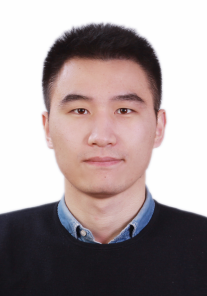}}]{Jian Li (M'20)}  received his bachelor's degree from the Department of Electronics and Information Engineering, Anhui University, in 2015, and received doctor's degree from the Department of Electronic Engineering and Information Science (EEIS), University of Science and Technology of China (USTC), in 2020. From Nov. 2019 to Nov. 2020, he was a visiting scholar with the Department of Electronic and Computer Engineering, University of Florida. From Dec. 2020 to Dec. 2022, he was a Post-Doctoral researcher with the School of Cyber Science and Technology, USTC. He is currently a research associate with the School of Cyber Science and Technology, USTC. He also serves as an Editor of China Communications. His research interests include wireless networks, next-generation Internet, and quantum networks.
\end{IEEEbiography}

\begin{IEEEbiography}[{\includegraphics[width=1in,height=1.25in,clip,keepaspectratio]{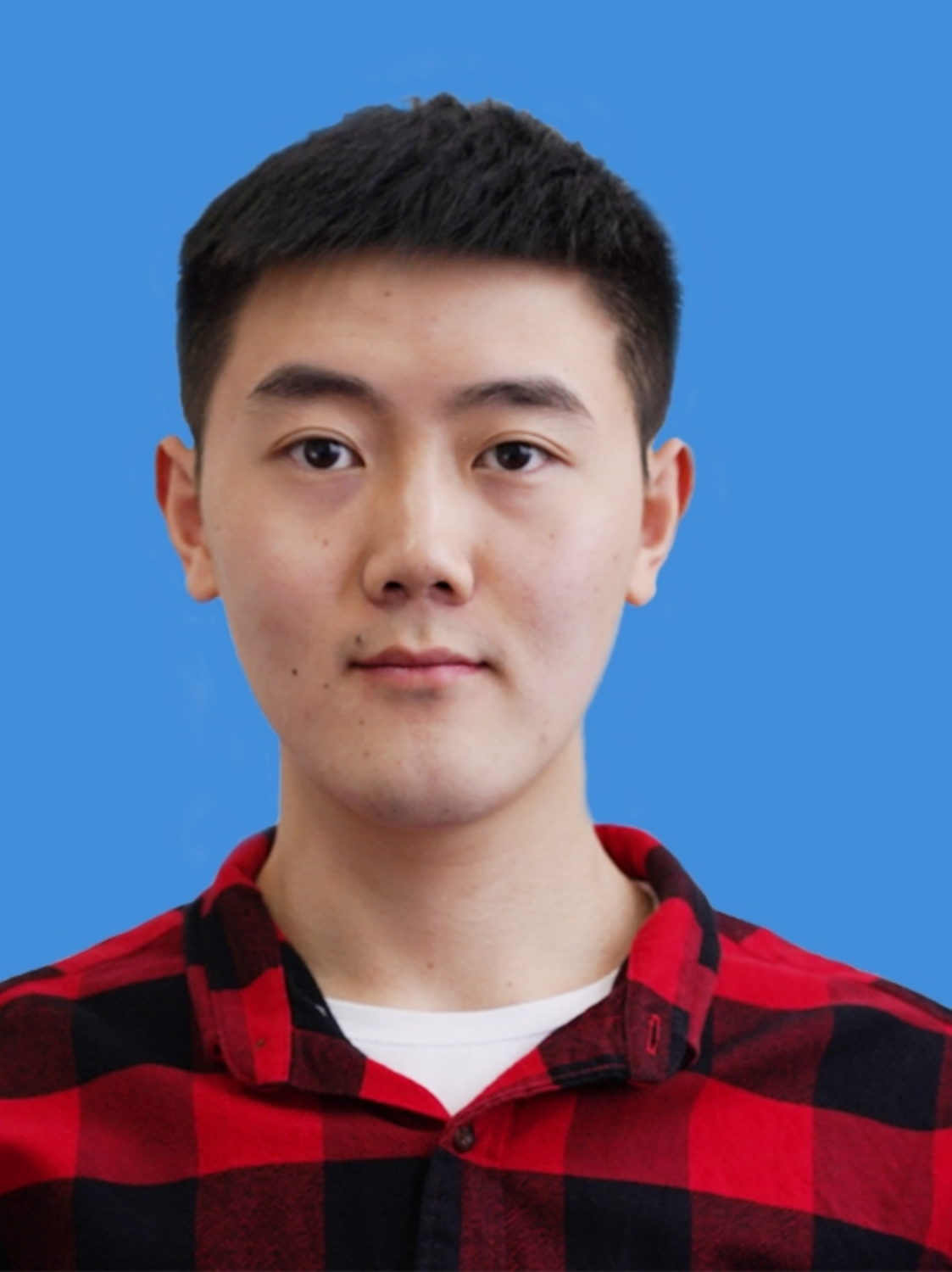}}]{Lutong Chen} received his Bachelor's Degree from the School of Cyber Science and Technology, University of Science and Technology of China in 2020. He is currently a Ph.D. student in School of Cyber Science and Technology. His research interests include quantum networking and network security.
\end{IEEEbiography}

\begin{IEEEbiography}[{\includegraphics[width=1in,height=1.25in,clip,keepaspectratio]{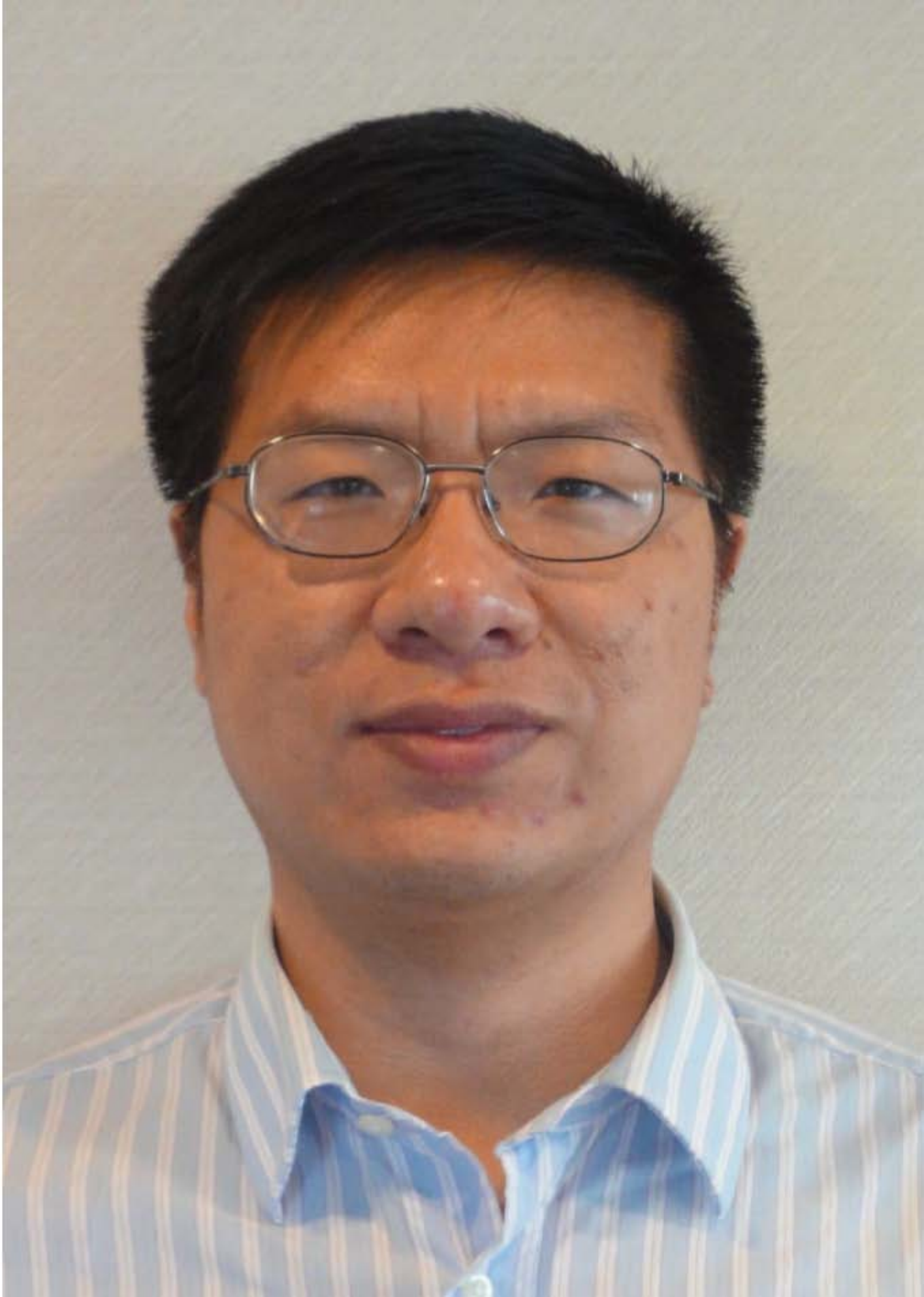}}]{Ruidong Li (SM'07)} received his bachelor's degree in engineering from  Zhejiang University, China, in 2001, and received a doctorate of engineering from the University of  Tsukuba in 2008. He is an associate professor in College of Science and Engineering, Kanazawa University, Japan. Before joining Kanazawa University, he was a senior researcher with the Network System Research Institute, National Institute of Information and Communications Technology (NICT). He is the founder and chair of the IEEE SIG on big data intelligent networking and IEEE SIG on intelligent Internet edge and the  secretary of IEEE Internet Technical Committee. He also serves as the chair for conferences and workshops, such as IWQoS 2021, MSN 2020, BRAINS 2020, ICC 2021 NMIC  symposium, ICCN 2019/2020, NMIC 2019/2020, and organized the special issues for the leading magazines and journals, such as IEEE Communications Magazine,  IEEE Network, IEEE IEEE Transactions on Network Science and Engineering (TNSE), etc. His current research interests include future networks, big data networking, blockchain, information-centric network, the internet of things, network security, wireless networks, and quantum  Internet. He is a senior member of the IEEE and a member of IEICE.
\end{IEEEbiography}
	
\begin{IEEEbiography}[{\includegraphics[width=1in,height=1.25in,clip,keepaspectratio]{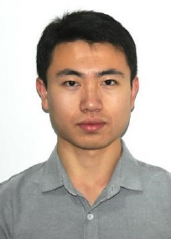}}]{Zhaoying Wang} received the B.S. degree in Information Security from the School of Cyber Science and Technology, University of Science and Technology of China in 2019. He is currently working toward the Ph.D degree from the School of Cyber Science and Technology. His research interests include Quantum Internet architecture and Quantum networking.
\end{IEEEbiography}	

\begin{IEEEbiography}[{\includegraphics[width=1in,height=1.25in,clip,keepaspectratio]{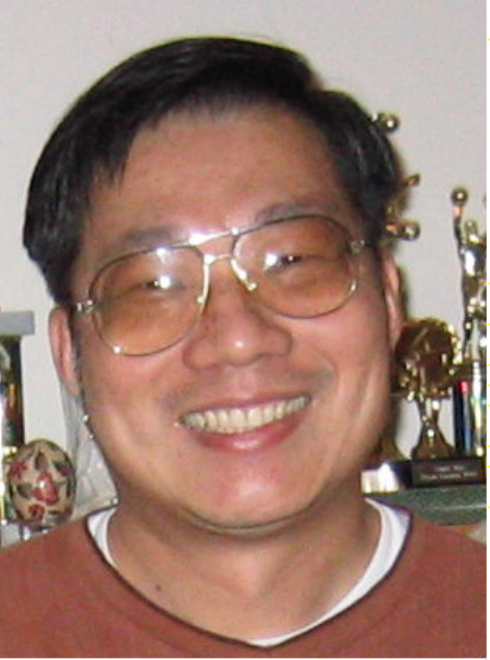}}]{David S.L. Wei}(SM'07) received his Ph.D. in Computer and Information Science from the University of Pennsylvania in 1991. Currently, he is a Full Professor in the Computer and Information Science Department at Fordham University. Dr. Wei has published over 140 technical papers in various archival journals and conference proceedings. He has served as a program committee member and a session chair for several well-known international conferences, including Infocom. Moreover, he was a lead guest editor or a guest editor for several special issues in the IEEE Journal on Selected Areas in Communications, the IEEE Transactions on Cloud Computing, and the IEEE Transactions on Big Data. Additionally, he served as an Associate Editor of IEEE Transactions on Cloud Computing from 2014 to 2018, an editor of IEEE J-SAC for the Series on Network Softwarization and Enablers from 2018 to 2020, and an Associate Editor of the Journal of Circuits, Systems, and Computers from 2013 to 2018. Dr. Wei's contributions to information security in wireless and satellite communications and cyber-physical systems were recognized with the IEEE Region 1 Technological Innovation Award (Academic) in 2020. He is a member of ACM and AAAS and holds life senior memberships in IEEE, IEEE Computer Society, and IEEE Communications Society. Currently, Dr. Wei's research is focused on cloud and edge computing, cybersecurity, and quantum computing and communications.
\end{IEEEbiography}
	
\begin{IEEEbiography}[{\includegraphics[width=1in,height=1.25in,clip,keepaspectratio]{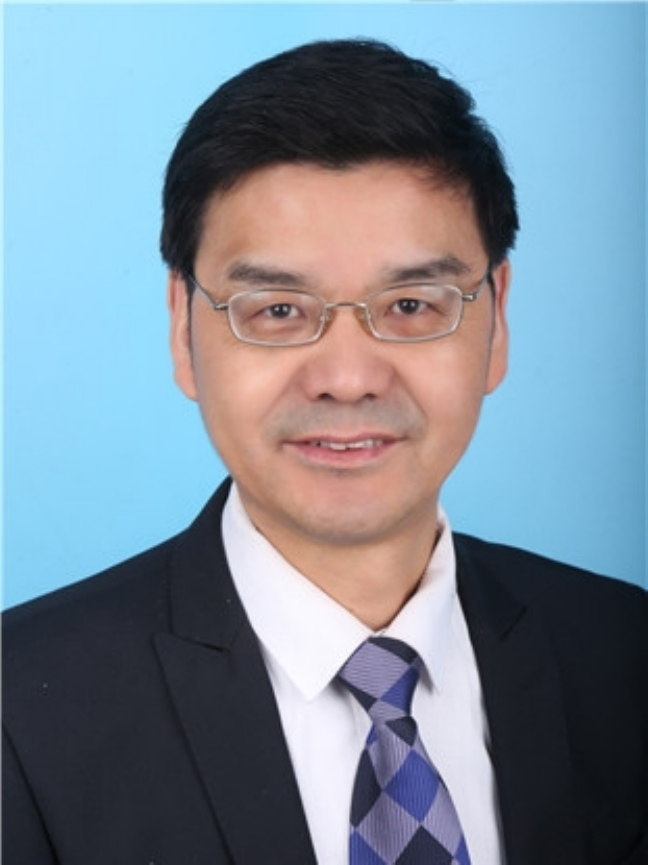}}]{Nenghai Yu} received the B.S. degree from the Nanjing University of Posts and Telecommunications, Nanjing, China, in 1987, the M.E. degree from Tsinghua University, Beijing, China, in 1992, and the Ph.D. degree from the Department of Electronic Engineering and Information Science (EEIS), University of Science and Technology of China (USTC), Hefei, China, in 2004. Currently, he is a Professor in the School of Cyber Science and Technology, and the School of Information Science and Technology, USTC. He is the Executive Dean of the School of Cyber Security, USTC, and the Director of the Information Processing Center, USTC. He has authored or co-authored more than 130 papers in journals and international conferences. His research interests include multimedia security, information hiding, and quantum networking.
\end{IEEEbiography}

\begin{IEEEbiography}[{\includegraphics[width=1in,height=1.25in,clip,keepaspectratio]{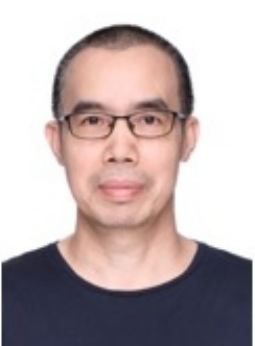}}]
{Qibin Sun (F'11)} received the Ph.D. degree from the Department of Electronic Engineering and Information Science (EEIS), University of Science and Technology of China (USTC), in 1997. He is currently a professor in the School of Cyber Science and Technology, USTC. His research interests include multimedia security, network intelligence and security, and so on. He has published more than 120 papers in international journals and conferences. He is a fellow of IEEE.
\end{IEEEbiography}

\begin{IEEEbiography}[{\includegraphics[width=1in,height=1.25in,clip,keepaspectratio]{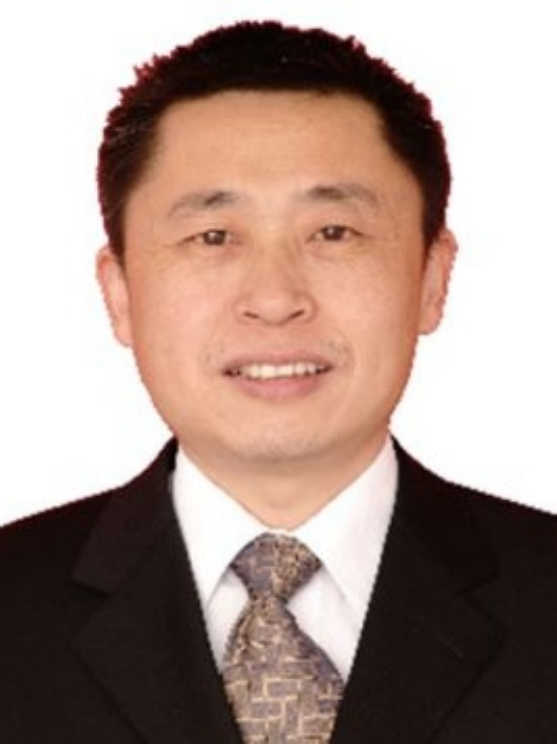}}]{Jun Lu} received his bachelor's degree from southeast university in 1985 and his master's degree from the Department of Electronic Engineering and Information Science (EEIS), University of Science and Technology of China (USTC), in 1988. Currently, he is a professor in the School of Cyber Science and Technology and the Department of EEIS, USTC. He is also the president of Jiaxing University. His research interests include theoretical research and system development in the field of integrated electronic information systems, network and information security. He is an Academician of the Chinese Academy of Engineering (CAE).
\end{IEEEbiography}
	
\vfill
\end{document}